\documentclass[12pt]{article}

\usepackage[title]{appendix}
\usepackage{lscape}
\usepackage{amsfonts}
\usepackage{amssymb}
\usepackage{amsmath}
\usepackage{amsthm}
\usepackage{bbm}
\usepackage{epstopdf}
\usepackage{float}
\usepackage[allcolors=blue,bookmarks=false,colorlinks=true]{hyperref}
\usepackage{natbib}
\usepackage[margin=.75in]{geometry}
\usepackage[hang]{caption}
\usepackage{color}
\usepackage{graphicx}
\usepackage{psfrag}
\usepackage{rotating}

\DeclareMathOperator{\arctanh}{arctanh}
\DeclareMathOperator{\argmin}{arg\,min}
\DeclareMathOperator{\Trace}{Trace}
\renewcommand{\qed}{\hfill $\blacksquare$}

\newtheorem{theorem}{Theorem}[section]

\newtheorem{corollary}{Corollary}[section]

\newtheorem{lemma}{Lemma}[section]

\newtheorem{proposition}{Proposition}[section]

\renewenvironment{proof}[1][Proof]{\begin{trivlist}
\item[\hskip \labelsep {\bfseries #1.}]}{\end{trivlist}}
\newcommand{\norm}[1]{\left\lVert#1\right\rVert}

\makeatletter
\newcommand*\rel@kern[1]{\kern#1\dimexpr\macc@kerna}
\newcommand*\widebar[1]{%
  \begingroup
  \def\mathaccent##1##2{%
    \rel@kern{0.8}%
    \overline{\rel@kern{-0.8}\macc@nucleus\rel@kern{0.2}}%
    \rel@kern{-0.2}%
  }%
  \macc@depth\@ne
  \let\math@bgroup\@empty \let\math@egroup\macc@set@skewchar
  \mathsurround\z@ \frozen@everymath{\mathgroup\macc@group\relax}%
  \macc@set@skewchar\relax
  \let\mathaccentV\macc@nested@a
  \macc@nested@a\relax111{#1}%
  \endgroup
}
\makeatother

\allowdisplaybreaks

\usepackage{multirow}
\usepackage{dcolumn}
\DeclareMathOperator*{\essinf}{ess\,inf}
\DeclareMathOperator*{\esssup}{ess\,sup}
\newcommand{\lleft}{\left}
\newcommand{\rright}{\right}
\newcommand{\llvert}{\vert}
\newcommand{\rrvert}{\vert}
\newcommand{\matfontas}[1]{#1}
\newcommand{\noxml}[1]{#1}
\newcommand{\upqed}{\qed}
\newenvironment{tabnotes}{\par\smallskip\begingroup\footnotesize}{\par\endgroup}
\newcommand{\legend}[1]{\noindent\textit{Note:} #1}

\newtheorem{assumptionj}{Assumption}
\newtheorem{assumptionv}{Assumption}
\newtheorem{assumptionc}{Assumption}
\newtheorem{assumptioncc}{Assumption}
\newtheorem{assumptionccc}{Assumption}
\newtheorem{assumptioncccc}{Assumption}
\newtheorem{assumptionccccc}{Assumption}
\newtheorem{assumptionnd}{Assumption}

\begin{document}
\title{A nonparametric test for diurnal variation \\ in spot correlation processes\thanks{We thank three anonymous reviewers at Quantitative Economics for insightful feedback that substantially improved the manuscript. We also thank Tim Bollerslev, Carsten Chong, Peter Reinhard Hansen, Yoann Potiron, Jun Yu, and participants in our sessions at the African Econometric Society Meeting in Nairobi, Kenya (June 2023), and the Financial Econometrics conference in Toulouse, France (May 2024), for helpful comments. We are indebted to Torben G. Andersen, Yingwen Tan, Zhensi Tan, Viktor Todorov, and Zhiyuan Zhang for identifying an inaccuracy, since corrected, in our application of the functional delta method. Christensen received funding from the Independent Research Fund Denmark (DFF 1028-00030B). He thanks the Rady School of Management, University of California, San Diego, for its hospitality while he worked on the paper. Zhi Liu was supported by MYRG-GRG2024-00190-FST-UMDF, MYRG-GRG2025-00093-FST, and the APAEM Seed Grant in Financial Econometrics from the University of Macau. This is the authors' version of the article published in \emph{Quantitative Economics}, 17 (2026), 1143--1202, \url{https://doi.org/10.3982/QE2690}.}}
\author{Kim Christensen\thanks{Aarhus University, Department of Economics and Business Economics, Universitetsbyen 51, 8000 Aarhus C, Denmark.} $^,$\thanks{Center for Research in Energy: Economics and Markets (CoRE), Aarhus University, Denmark.} $^,$\thanks{Research fellow at the Danish Finance Institute (DFI).}
\and Ulrich Hounyo\thanks{University at Albany -- State University of New York, Department of Economics, 1400 Washington Avenue, Albany, New York 12222, United States.} \and Zhi Liu\thanks{University of Macau, Department of Mathematics, Faculty of Science and Technology, Taipa, Macau, China.}}

\date{November 2026}

\maketitle

\begin{abstract}
The association between log-price increments of exchange-traded equities, as measured by their spot correlation estimated from high-frequency data, exhibits a pronounced upward-sloping and almost piecewise linear relationship at the intraday horizon. There is notably lower---on average less positive---correlation in the morning than in the afternoon. We develop a nonparametric testing procedure to detect such variation in a correlation process. The test statistic has a known distribution under the null hypothesis, whereas it diverges under the alternative. We run a Monte Carlo simulation to discover the finite sample properties of the test statistic, which are close to the large sample predictions, even for small sample sizes and realistic levels of diurnal variation. In an application, we implement the test on a high-frequency dataset covering the stock market over an extended period. The test leads to rejection of the null most of the time. This suggests diurnal variation in the correlation process is a nontrivial effect in practice. We show how conditioning information about macroeconomic news and corporate earnings announcements affects the intraday correlation curve.


\end{abstract}

\vspace*{-0.5cm}

\setlength{\baselineskip}{18pt}\setlength{\abovedisplayskip}{10pt} \belowdisplayskip\abovedisplayskip\setlength{\abovedisplayshortskip}{5pt} \abovedisplayshortskip\belowdisplayshortskip\setlength{\abovedisplayskip}{8pt} \belowdisplayskip\abovedisplayskip\setlength{\abovedisplayshortskip}{4pt}

\vfill

\thispagestyle{empty}

\pagebreak

\section{Introduction} \setcounter{page}{1}

\label{sec1}

Correlation percolates through financial economics. It is a critical ingredient in the determination of optimal portfolio weights in a \citet{markowitz:52a} mean-variance asset allocation problem, where the asset return correlations also determine a lower bound on diversification. Moreover, the correlation between the return of an asset and the return of the market portfolio is paramount in single- and multifactor capital asset pricing models \citep{fama-french:15a, sharpe:64a}, where it is used to calculate the so-called beta, which is an important driver of the premium over the risk-free rate earned as a compensation by investing in the risky asset. In addition, correlation is also employed in risk management and hedging.

It has long been recognized that correlations are time-varying, and the vast majority of parametric models to describe \textit{interday} correlation allow it to change dynamically (e.g., \cite{engle:02a,noureldin-shephard-sheppard:12a}). The properties of the correlation process have also been traversed in detail with nonparametric analysis from high-frequency data. This is typically done by studying a realized measure of the daily integrated covariance, which is mapped into a correlation estimate, for example, \citet{ait-sahalia-fan-xiu:10a} and \citet{boudt-cornelissen-croux:12a}.

Surprisingly, relatively little is known about the behavior of correlation at the \textit{intraday} horizon. This stands in sharp contrast to the volatility of individual equity returns that is known to evolve as a U- or reverse J-shaped curve with notably higher volatility near the opening and closing of the stock exchange than around noon (e.g., \cite{harris:86a,wood-mcinish-ord:85a}). Several estimators of the intraday volatility curve have emerged over the years, for example, \citet{andersen-bollerslev:97b, andersen-bollerslev:98b} propose a parametric model for periodicity in volatility, whereas \citet{boudt-croux-laurent:11b} and \citet{christensen-hounyo-podolskij:18a} develop nonparametric jump- and microstructure noise-robust estimators from high-frequency data that verify the existence of a pervasive structure in the intraday volatility.

The most common setup for describing the dynamic of spot volatility of an asset log-return at the interday and intraday horizon is a multiplicative time-series model:
\begin{equation}
\label{equation:spot-volatility} \sigma_{t} = \sigma_{sv,t} \sigma
_{u,t}, 
\end{equation}
where $\sigma_{sv,t}$ is a stationary process meant to capture stochastic volatility, whereas $\sigma_{u,t}$ is a deterministic component intended to capture diurnal variation and assumed to be a constant time-of-day factor (i.e., $\sigma_{u,t}= \sigma_{u,t-1}$).\footnote{In recent work, \citet{andersen-thyrsgaard-todorov:19a} suggest that the intraday volatility curve may be time-varying; see also \citet{andersen-su-todorov-zhang:24a}.}

In a bivariate setting, any systematic evolution in the volatility is automatically transferred to the covariance process, $c_{t}^{XY} = \sigma_{t}^{X} \sigma_{t}^{Y} \rho_{t}$, where $\sigma_{t}^{X}$ and $\sigma_{t}^{Y}$ represent the spot volatility of asset $X$ and $Y$, whereas $\rho_{t}$ is their correlation. If the individual return variations of $X$ and $Y$ follow \eqref{equation:spot-volatility}, the covariance inherits an ``imputed'' diurnal pattern:
\begin{equation}
\label{equation:total-covariance} c_{t}^{XY} = \underbrace{\sigma
_{sv,t}^{X} \sigma_{sv,t}^{Y}}_{\text{imputed stochastic covariance}} \times\underbrace{\sigma_{u,t}^{X} \sigma_{u,t}^{Y}}_{\text{imputed diurnal covariance}} \times\underbrace{\rho_{t}}_{\text{spot correlation}}. 
\end{equation}
However, observing \eqref{equation:total-covariance} suggests that there may be an additional source of diurnal variation in the covariance, since the dynamic of the spot correlation further affects it. As in \eqref{equation:spot-volatility}, we can capture a recurrent behavior in the spot correlation as follows:
\begin{equation}
\label{equation:spot-correlation} \rho_{t} = \rho_{sc,t}
k_{u,t}, 
\end{equation}
where $\rho_{sc,t}$ and $k_{u,t}$ are interpreted as above. In the modified setting of \eqref{equation:spot-correlation}, the breakdown of the covariance into its component parts is now given by
\begin{equation}
c_{t}^{XY} = \underbrace{\sigma_{sv,t}^{X} \sigma_{sv,t}^{Y} \rho_{sc,t}}_{\text{stochastic covariance}} \times\underbrace{\sigma_{u,t}^{X} \sigma_{u,t}^{Y} k_{u,t}}_{\text{diurnal covariance}}.
\label{eq4}
\end{equation}
To the extent that correlations vary systematically within a day, we should expect the actual and imputed diurnal covariance curve to deviate (see, e.g., \cite{bibinger-hautsch-malec-reiss:19a}, for initial evidence of this effect). To get a first impression of this, we begin with an inspection of panel A in Figure~\ref{figure:rho_d} in our empirical application in Section~\ref{section:empirical}, where we compare the average imputed and actual intraday covariance curve calculated pairwise for all constituents of the Dow Jones Industrial Average and a proxy for the market portfolio of aggregate movements in the U.S. equity market over the sample period 2010--2023. We observe a striking discrepancy between the two, most notably in the early morning and late afternoon. This provides strong evidence of this effect in the high-frequency data. Looking at it in terms of the correlation process in panel B of the figure, we locate a very significant upward-sloping intraday correlation curve, which increases monotonically during the trading session in an almost piecewise linear fashion. This is consistent with \citet{allez-bouchaud:11a} and concurrent work of \citet{hansen-luo:23a}. There are large jumps in the correlation around the release of macroeconomic information, which corresponds to an influx of systematic risk to the market.\footnote{The presence of diurnal variation in the correlation also has implications for the parametric modeling of intraday spot covariance. In particular, one has to account for this effect to extract the stationary component of the covariance process. A ``naive'' approach with the imputed diurnal covariance based on the idiosyncratic intraday volatility curve---amounting to asset-wise deflation---is insufficient to get a covariance free of systematic intraday evolution.}

In this paper, we construct a testing procedure to detect diurnal variation in a correlation process. It distills local estimates of the spot correlation, after the high-frequency return series has been devolatized to remove the effect of idiosyncratic volatility (both deterministic and stochastic), thus isolating the correlation process, while also controlling for potential price jump variation. If there are systematic changes in the spot correlation estimates, the test statistic grows large and rejects the null hypothesis of no diurnal variation. This is related to, but different from, previous work by \citet{reiss-todorov-tauchen:15a} for testing a constant beta. Overall, in our empirical high-frequency data, we implement the test statistic on a month-by-month basis and find that the proposed test statistic rejects the null hypothesis most of the time, thus confirming the circumstantial evidence from Figure~\ref{figure:rho_d}. Furthermore, we provide anecdotal evidence about how macroeconomic news and corporate earnings announcements affect the intraday correlation curve.

To highlight the exploitation of predictable dynamics in the correlation, we adopt the standpoint of a trader who hedges a long exposure in single stocks via the market portfolio. We report a nontrivial effect by incorporating diurnal correlation into the risk management process, relative to ignoring it, yielding a drop in combined portfolio variance close to 20\%. It also delivers a much more stable hedge ratio during the course of the trading day, helping to reduce transaction costs derived from warehousing the risk.

The roadmap of the paper is as follows. In Section~\ref{section:theory}, we present the model and list the assumptions required to extract an intraday correlation curve from a bivariate time series of high-frequency data. In Section~\ref{section:correlation}, we develop our point-in-time correlation estimator. In Section~\ref{section:test-procedure}, we propose a testing procedure, which can be employed to uncover the existence of diurnal variation in the correlation process. We derive the required asymptotic distribution theory, which is based on a functional central limit theorem. In Section~\ref{section:extension}, we elaborate on the relaxation of a crucial assumption. We also show how our framework can be extended to a conditional version that incorporates relevant information that may help to determine the functional form of the diurnal correlation curve. In Section~\ref{section:simulation}, we inspect the small sample attributes of our framework via Monte Carlo simulation. In Section~\ref{section:empirical}, we apply it to a large panel of equity data. In Section~\ref{section:conclusion}, we conclude. We relegate proofs and supplemental results to the Appendices.

\section{Theoretical setup}
\label{section:theory}

We suppose a filtered probability space $( \Omega, \mathcal{F}, (\mathcal{F}_{t})_{t \geq0}, P)$ describes a bivariate continu\-ous-time log-price process $Z = (X,Y)^{\top}$, where $( \mathcal{F}_{t})_{t \geq0}$ is a filtration and $^{\top}$ is the transpose operator.\footnote{Our analysis extends to $d$-dimensional processes in an obvious fashion.} $Z$ is observed on $[0,T]$, where $T$ is the number of days in the sample and the subinterval $[t-1,t]$ is the $t$th day, for $t = 1, \ldots, T$. We assume $Z$ is recorded discretely at the equidistant time points $t_{i} = t-1+i/n$, for $i=0,1, \ldots, n$, so a total of $nT$ increments are observed with a time gap of $\Delta_{n} = 1/n$. Throughout, the asymptotic theory is infill and long-span, that is, we look at limits in which the time gap between consecutive observations goes to zero ($ \Delta_{n} \rightarrow0$ or $n \rightarrow\infty$) and the sample period increases ($T \rightarrow\infty$).

In the absence of arbitrage (or rather a free lunch with vanishing risk), $Z$ is a semimartingale (e.g., \cite{delbaen-schachermayer:94a}). We suppose $Z$ is an It\^{o} process, so we can write the time $t$ value of $Z$ as follows:
\begin{equation}
\label{equation:Z} Z_{t} = Z_{0} +
\int_{0}^{t} a_{s} \,\mathrm{d}s + \int_{0}^{t} \sigma_{s} \,\mathrm{d}W_{s} + J_{t}, \quad t \geq0,
\end{equation}
where $Z_{0} = (X_{0},Y_{0})^{\top}$ is $\mathcal{F}_{0}$-measurable,
\begin{equation}
\label{equation:component} a_{t} = %
\begin{bmatrix}
a_{t}^{X} \\
a_{t}^{Y} \end{bmatrix} %
, \qquad\sigma_{t} = %
\begin{bmatrix} \sigma
_{t}^{X} & 0 \\
\rho_{t} \sigma_{t}^{Y} & \sqrt{1- \rho_{t}^{2}} \sigma_{t}^{Y}
\end{bmatrix} %
, \qquad W_{t} = %
\begin{bmatrix} W_{t}^{X}
\\
W_{t}^{Y} \end{bmatrix} %
, \quad\text{and}\quad J_{t} = %
\begin{bmatrix}
J_{t}^{X} \\
J_{t}^{Y} \end{bmatrix} %
, 
\end{equation}
where $(a_{t})_{t \geq0}$ is a predictable and locally bounded drift, $( \sigma_{t})_{t \geq0}$ is an adapted, c\`{a}dl\`{a}g volatility matrix, while $(W_{t})_{t \geq0}$ is a bivariate standard Brownian motion with\break$\langle W^{X}, W^{Y} \rangle_{t} = 0$, where $\langle\cdot, \cdot\rangle$ denotes the predictable part of the quadratic covariation process.

$J_{t}$ is a pure-jump process, for which we impose the following restriction.

\def\theassumptionj{(J)}
\begin{assumptionj}\label{ass(J)}
$J_{t}^{\wp}$ is such that
\begin{equation}
\label{equation:jump-process} J_{t}^{\wp} =
\int_{0}^{t} \int_{\mathbb{R}}x \mu^{\wp} ( \mathrm{d}s, \mathrm{d}x), 
\end{equation}
where $\mu^{\wp} $ is an integer-valued random measure on $\mathbb{R}_{+} \times\mathbb{R}$ with compensator $\nu^{\wp} ( \mathrm{d}t, \mathrm{d}x) = \chi_{t}^{\wp} \,\mathrm{d}t \otimes F^{\wp}( \mathrm{d}x)$, $\chi_{t}^{\wp}$ is an adapted, c\`{a}dl\`{a}g process, and $F^{\wp}$ is a measure on $\mathbb{R}$.
\end{assumptionj}

Here, and in the remainder of the article, the generic superscript $\wp$ notation is used to represent that the derived stochastic process is associated with $\wp$, where $\wp$ is either $X$ or $Y$.

We also assume that the stochastic volatility processes are It\^{o} semimartingales.

\def\theassumptionv{(V)}
\begin{assumptionv}\label{ass(V)}
$\sigma_{t}^{\wp}$ is of the form:
\begin{align}
\label{equation:stochastic-volatility-process} %
\begin{aligned} \sigma_{t}^{\wp}
&= \sigma_{0}^{\wp} + \int_{0}^{t} \tilde{a} _{s}^{\wp} \,\mathrm{d}s + \int_{0}^{t} \tilde{\sigma}_{s}^{\wp} \,\mathrm{d}W_{s} + \int_{0}^{t} \tilde{\nu}_{s}^{\wp} \,\mathrm{d} \tilde{W}_{s} + \int_{0}^{t} \int_{\mathbb{R}} x \tilde{\mu}^{\wp}( \mathrm{d}s, \mathrm{d}x), \end{aligned} 
\end{align}
where $( \tilde{a}_{t}^{\wp})_{t \geq0}$, $( \tilde{\sigma}_{t}^{\wp})_{t \geq0}$, $( \tilde{\nu}^{\wp})_{t \geq0}$ are adapted, c\`{a}dl\`{a}g stochastic processes, $\tilde{W}_{t} = ( \tilde{W}_{t}^{X},\allowbreak\tilde{W}_{t}^{Y})^{\top}$ is a bivariate standard Brownian motion, independent of $W$, but such that $\tilde{W}_{t}^{X}$ and $\tilde{W}_{t}^{Y}$ can be correlated. Lastly, $\tilde{\mu}^{\wp}( \mathrm{d}t, \mathrm{d}x)$ is the jump counting measure of $\sigma_{t}^{\wp}$ with compensator $\tilde{\chi}^{\wp}_{t} \,\mathrm{d}t \otimes\tilde{F}^{\wp}( \mathrm{d}x)$, where $\tilde{\chi}_{t}^{\wp}$ is an adapted, c\`{a}dl\`{a}g process, and $\tilde{F}^{\wp}$ is a measure on $\mathbb{R}$.
\end{assumptionv}

The above constitutes a more or less nonparametric framework for modeling arbitrage-free price processes, which accommodates most of the models employed in practice. Mainly, we exclude semimartingales that are not absolutely continuous, but this is not too restrictive.\footnote{An example of a continuous local martingale that has no stochastic integral representation is a Brownian motion time-changed with the Cantor function (or devil's staircase); see \citet{ait-sahalia-jacod:18a} and \citet{barndorff-nielsen-shephard:04a}.} Note that we integrate over the jump size distribution directly with respect to the Poisson random measure. Hence, we are assuming that the jump processes are of finite variation.\footnote{In general, the Poisson random measure needs to be compensated (i.e., converted to a martingale) for jump processes of infinite variation to ensure that the summation (over the small jumps) is convergent.} They may be infinitely active, but they should be absolutely summable. We add more regularity to the jump processes below. Furthermore, Assumption~(V) excludes the possibility that volatility can be rough, for example, that it is driven by a fractional Brownian motion with a Hurst exponent less than a half, which has been a recurrent theme in the recent literature (e.g., \cite{bolko-christensen-pakkanen-veliyev:23a,fukasawa-takabatake-westphal:22a,gatheral-jaisson-rosenbaum:18a,shi-yu:23a,wang-xiao-yu:23a}).

It is possible to expand our results to a more general setting. For instance, to cope with infinite variation jumps we can apply the Laplace transform-based estimator of \citet{liu-liu-liu:18a} or the debiased truncation-based estimator in \citet{boniece-figueroa-lopez-zhou:25a}. To handle roughness, we can rely on estimators of spot volatility that are robust to this assumption, such as the Fourier transform-based estimator from \citet{mancino-mariotti-toscano:25a} or the truncation-based estimator of \citet{christensen-thyrsgaard-veliyev:19a}. Then we can directly plug in such consistent estimators into our diurnal correlation framework. However, we do not pursue these extensions here.

In the maintained framework, the continuous part of the quadratic covariation process of $Z$ is absolutely continuous with respect to the Lebesgue measure, so it has a derivative:
\begin{equation}
\frac{\mathrm{d} \bigl\langle Z^{c}, Z^{c} \bigr\rangle_{t}}{\mathrm{d}t} = \sigma_{t} \sigma_{t}^{\top} = %
\begin{bmatrix} \bigl(
\sigma_{t}^{X} \bigr)^{2} & \sigma_{t}^{X} \sigma_{t}^{Y} \rho_{t} \\[0.10cm]
\sigma_{t}^{X} \sigma_{t}^{Y} \rho_{t} & \bigl( \sigma_{t}^{Y} \bigr)^{2}%
\end{bmatrix}
\equiv %
\begin{bmatrix} c_{t}^{X}
& c_{t}^{XY} \\[0.10cm]
c_{t}^{XY} & c_{t}^{Y}
\end{bmatrix} %
= c_{t}, \label{eq9}
\end{equation}
and instantaneous correlation:
\begin{equation}
\label{equation:correlation} \rho_{t} \equiv\frac{\mathrm{d} \bigl\langle
X^{c}, Y^{c} \bigr\rangle_{t}}{\sqrt{\mathrm{d} \bigl\langle X^{c}, X^{c} \bigr\rangle}_{t} \sqrt{\mathrm{d} \bigl\langle Y^{c},Y^{c} \bigr\rangle}_{t}}, 
\end{equation}
where $\wp^{c}$ is the continuous part of $\wp$.

We need to make some additional assumptions, starting with one for the correlation reminiscent of equation \eqref{equation:spot-volatility} for the stochastic volatility process.

\def\theassumptionc{(C1)}
\begin{assumptionc}\label{ass(C1)}
The spot correlation $\rho_{t}$ factors as
\begin{equation}
\label{equation:total-correlation} \rho_{t} = \rho_{sc,t}
k_{u,t}, 
\end{equation}
where $\rho_{sc,t}$ is a stochastic process and $k_{u,t}$ is a deterministic component.
\end{assumptionc}

In Assumption~\ref{ass(C1)}, only the left-hand side of \eqref{equation:total-correlation} is identified, so the scale of one of the terms on the right-hand side needs to be fixed. We add such an identification condition in Assumption~\ref{ass(C2)}. Furthermore, note that as the diurnal component is not a correlation in itself, there is nothing to stop it from venturing outside $(-1,1)$, so long as the overall product of the diurnal and stochastic component does not.

In view of \eqref{equation:spot-volatility} and \eqref{equation:total-correlation}, the spot covariance is the product of a stochastic process and a deterministic component, where the latter captures diurnal variation:
\begin{equation}
c_{t}^{XY} = \sigma_{t}^{X} \sigma_{t}^{Y} \rho_{t} = \sigma_{sv,t}^{X} \sigma_{u,t}^{X} \sigma_{sv,t}^{Y} \sigma_{u,t}^{Y} \rho_{sc,t} k_{u,t} = \underset{\underset{\text{stochastic covariance}}{= c_{sv,t}^{XY}}}{\underbrace{\sigma_{sv,t}^{X} \sigma_{sv,t}^{Y} \rho_{sc,t}}} \times\underset{\underset{\text{diurnal covariance}}{= c_{u,t}^{XY}}}{\underbrace{\sigma_{u,t}^{X} \sigma_{u,t}^{Y} k_{u,t}}}. \label{eq12}
\end{equation}
Note that for $X=Y$, $k_{u,t} = \rho_{sc,t} = 1$. Hence, our paper generalizes \citet{christensen-hounyo-podolskij:18a} to a multivariate context.

In view of Assumption~\ref{ass(C1)}, the spot covariance matrix factors as follows:
\begin{equation}
c_{t} = %
\begin{bmatrix} c_{t}^{X}
& c_{t}^{XY} \\[0.10cm]
c_{t}^{XY} & c_{t}^{Y}
\end{bmatrix} %
= %
\begin{bmatrix}
c_{u,t}^{X} & c_{u,t}^{XY} \\[0.10cm]
c_{u,t}^{XY} & c_{u,t}^{Y}
\end{bmatrix} %
\odot %
\begin{bmatrix}
c_{sv,t}^{X} & c_{sv,t}^{XY} \\[0.10cm]
c_{sv,t}^{XY} & c_{sv,t}^{Y}
\end{bmatrix} %
\equiv c_{u,t} \odot c_{sv,t},
\label{eq13}
\end{equation}
where $\odot$ denotes the Hadamard product.

We further impose the following.

\def\theassumptioncc{(C2)}
\begin{assumptioncc} \label{ass(C2)}
$( \sigma_{u,t}^{\wp})_{t \geq0}$ and $( k_{u,t})_{t \geq0}$ are bounded, Riemann integrable, one-\noxml{\break}periodic functions such that $\int_{t-1}^{t} \sigma_{u,s}^{X} \sigma_{u,s}^{Y} k_{u,s} \,\mathrm{d}s = 1$.
\end{assumptioncc}

\def\theassumptionccc{(C3)}
\begin{assumptionccc}\label{ass(C3)}
$\sigma_{sv,t}^{\wp} > 0$, $\sigma_{u,t}^{\wp} > 0$, $\rho_{sc,t} \neq0$ and $k_{u,t} \neq0$, for all $t \geq0$ except on a set with Lebesgue measure zero.
\end{assumptionccc}

Assumption~\ref{ass(C2)} adds some regularity on $\sigma_{u,t}^{\wp}$ and $k_{u,t}$. The requirement on the definite integral of the diurnal covariance function is a natural generalization from the univariate framework, where it reduces to the standard identification condition $\int_{t-1}^{t} ( \sigma_{u,s}^{\wp})^{2} \,\mathrm{d}s = 1$. We also suppose that $\sigma_{u}^{\wp}$ and $k_{u}$ are recurrent, that is, $\sigma_{u,t}^{\wp} = \sigma_{u,t-1}^{\wp}$ and $k_{u,t} = k_{u,t-1}$ for all $t\geq1$, so that these functions are consistently estimable from a long enough sample of high-frequency data. While the latter is not uncommon in the literature, it is a strong assumption that encounters problems in practice, since empirical evidence suggests that the intraday volatility curve may be time-varying \citep{andersen-thyrsgaard-todorov:19a}. We relax this part of the assumption in Section~\ref{section:extension} to allow for much more general dynamics in these processes. Assumption~\ref{ass(C3)} presupposes that both correlation components are nonzero, except on a set of Lebesgue measure zero, since we evidently cannot identify $\rho_{sc,t} \neq0$ if $k_{u,t} = 0$, and vice versa. The condition allows the correlation process to cross zero in a continuous fashion, provided it does not get ``stuck'' at the origin. For example, this holds if the driving force of the stochastic correlation is a Brownian motion, for which the zero set is uncountably infinite but of Lebesgue measure zero.

As our asymptotic theory is based on both $n \rightarrow\infty$ and $T \rightarrow\infty$, we cannot activate the localization procedure for high-frequency data described in \citet[Section~4.4.1]{jacod-protter:12a} to bound various processes, so instead we impose a related condition.

\def\theassumptioncccc{(C4)}
\begin{assumptioncccc}\label{ass(C4)}
The drift term $a^{\wp}$ is Lipschitz continuous (in mean square), that is, $E \bigl[|a_{t}^{\wp} - a_{s}^{\wp}|^{2} \bigr] \leq C|t-s|$, for any $s, t\in[0, \infty)$ and a positive constant $C$ (that does not depend on $s$ and $t$),
\begin{equation}
\sup_{t \in\mathbb{R}_{+}} E \bigl[ \exp\bigl( \bigl\llvert a_{t}^{\wp} \bigr\rrvert\bigr) \bigr] + \sup_{t \in\mathbb{R}_{+}} E \bigl[ \exp\bigl( \bigl\llvert\sigma_{t}^{\wp} \bigr\rrvert\bigr) \bigr] + \sup_{t \in\mathbb{R}_{+}} E \bigl[ \exp\bigl( \bigl\llvert\chi_{t}^{\wp} \bigr\rrvert\bigr) \bigr] < \infty.
\label{eq14}
\end{equation}
Moreover, $F^{\wp}( \mathbb{R}) < \infty$, $\tilde{F}^{\wp}( \mathbb{R}) < \infty$, $\int_{\mathbb{R}} |x|^{2} \tilde{F}^{\wp}( \mathrm{d}x) < \infty$, and
\begin{equation}
\sup_{t \in\mathbb{R}_{+}} E \bigl[ \bigl\llvert\tilde{a}_{t}^{\wp} \bigr\rrvert^{8} \bigr] + \sup_{t \in\mathbb{R}_{+}} E \bigl[ \bigl\llvert\tilde{\sigma}_{t}^{\wp} \bigr\rrvert^{8} \bigr] + \sup_{t \in\mathbb{R}_{+}} E \bigl[ \bigl\llvert\tilde{\nu}_{t}^{\wp} \bigr\rrvert^{8} \bigr] + \sup_{t \in\mathbb{R}_{+}} E \bigl[ \bigl\llvert\tilde{\chi}_{t}^{\wp} \bigr\rrvert^{8} \bigr] < \infty. \label{eq15}
\end{equation}
\end{assumptioncccc}

Assumption~\ref{ass(C4)} follows Assumption~I of \citet{andersen-su-todorov-zhang:24a} for the univariate case; see also Assumption~1 of \citet{andersen-tan-todorov-zhang:25a}. It restricts the jump processes to be of finite activity, but this can be relaxed, as shown in the Supplementary Appendix of their paper. Moreover, the moment conditions are also stricter than necessary.

The last set of assumptions concerns the stationarity and ergodicity of the stochastic volatility and correlation processes.

\def\theassumptionccccc{(C5)}
\begin{assumptionccccc}\label{ass(C5)}
For any positive integer $s$ and $\tau\in[0, 1)$, $\sigma_{sv,s-1+ \tau}^{\wp}$ and $\rho_{sc,s-1+ \tau}$ are functions (depending on $\tau$) of $M_{s-1+ \tau}$, where $(M_{t})_{t \geq0}$ is a multivariate Markov process, which is stationary, ergodic, and $\alpha$-mixing with mixing coefficient
\begin{equation}
\alpha_{s} = \sup_{t \geq0} \sup\bigl\{\bigl\llvert P(A \cap B) - P(A)P(B) \bigr\rrvert: A \in\mathcal{G}_{t}, B \in\mathcal{G}^{t+s} \bigr\}, \label{eq16}
\end{equation}
where $\mathcal{G}_{t} = \sigma(M_{u} \mid u \leq t)$ and $\mathcal{G}^{t} = \sigma(M_{u} \mid u \geq t)$ are the ``backward-'' and ``forward-looking'' $\sigma$-algebras, such that $\alpha_{s} = O(s^{-q- \ell})$ for some $q>0$ and an arbitrarily small constant $\ell> 0$.
\end{assumptionccccc}

Assumption~\ref{ass(C5)} follows Assumption~II of \citet{andersen-su-todorov-zhang:24a} and Assumption~$H_{0}$ in the recent contribution of \citet{andersen-tan-todorov-zhang:25a}. The astute indexation ensures that subsets of the volatility and correlation, separated by an integer-valued index set, can be time-dependent through a transformation of a multivariate Markov process. The remaining parts are standard regularity conditions for inference with weakly dependent processes. In particular, the decay rate $q$ of the sequence of mixing coefficients is restricted further to establish consistency and, more so, for a functional CLT.

Assumptions \ref{ass(C1)}--\ref{ass(C5)} are sufficient to identify both volatility and correlation components $\sigma_{sv,t}^{\wp}$, $\sigma_{u,t}^{\wp}$, $\rho_{sc,t}$, and $k_{u,t}$.

To construct our hypothesis, we partition the sample space $\Omega$ into
\begin{equation}
\label{equation:Omega-H0} \Omega_{\mathcal{H}_{0}}= \{\omega: k_{u,t} = 1,
t \geq0\}, 
\end{equation}
and $\Omega_{\mathcal{H}_{a}} = \Omega_{\mathcal{H}_{0}}^{\complement}$. The null is then defined as $\mathcal{H}_{0}: \omega\in\Omega_{\mathcal{H}_{0}}$, that is, it consists of paths with no diurnal correlation. The alternative is $\mathcal{H}_{a}: \omega\in\Omega_{\mathcal{H}_{a}}$. As usual in time-series analysis, the premise here is that we cannot repeat the experiment. We can access discrete high-frequency data from a single path. On this basis, the goal is to decide which subset our realization lies in. We note that an equivalent representation of the null hypothesis is the following: $\Omega_{\mathcal{H}_{0}} = \{\omega: \int_{0}^{1}( k_{u,t}-1)^{2} \,\mathrm{d}t = 0\}$.

\section{Spot correlation estimator}
\label{section:correlation}

To implement our testing procedure, we first need an estimator of the spot correlation coefficient, which we construct from a standard localized estimator of the continuous part of the quadratic covariation process.

We represent the log-price increments of $Z$ as follows:
\begin{equation}
\label{equation:increment} \Delta_{(t-1)n+i}^{n}Z \equiv
Z_{t-1+i/n} - Z_{t-1+(i-1)/n} = %
\begin{bmatrix} \Delta_{(t-1)n+i}^{n}X
\\[0.10cm]
\Delta_{(t-1)n+i}^{n}Y \end{bmatrix} %
,
\end{equation}
for $t = 1, \ldots,T$ and $i=1, \ldots, n$.

We split the sample into smaller blocks of $k_{n}$ log-price increments each. We suppose $k_{n}$ is a divisor of $n$ for notational convenience, so that there are $n/k_{n}$ blocks per day. We set $\tau_{0} \equiv0$, and over the $j$th block on day $t$ we define $\tau_{j} = \frac{j}{n/k_{n}}$. Then, we construct
\begin{equation}
\label{equation:realized-covariance} %
\begin{aligned}[b] \hat{c}_{t, \tau_{j}} &=
\frac{n}{k_{n}} \sum_{\ell= (j-1) {k}_{n}+1}^{jk_{n}} \bigl( \Delta_{(t-1)n+ \ell}^{n}Z \bigr) \bigl( \Delta_{(t-1)n + \ell}^{n}Z \bigr)^{\top} \odot %
\begin{bmatrix} \mathbbm{1}_{\mathcal{A}_{t, \tau_{j}}^{X,n}} & \mathbbm{1}_{
\mathcal{A}_{t, \tau_{j}}^{X,n} \cap\mathcal{A}_{t, \tau_{j}}^{Y,n}} \\
\mathbbm{1}_{\mathcal{A}_{t, \tau_{j}}^{X,n} \cap\mathcal{A}_{t, \tau_{j}}^{Y,n}} & \mathbbm{1}_{\mathcal{A}_{t, \tau_{j}}^{Y,n}} \end{bmatrix} \\
&\equiv %
\begin{bmatrix} \hat{c}_{t, \tau_{j}}^{X}
& \hat{c}_{t, \tau_{j}}^{XY} \\
\hat{c}_{t, \tau_{j}}^{XY} & \hat{c}_{t, \tau_{j}}^{Y}
\end{bmatrix}, \end{aligned} 
\end{equation}
for $t = 1, \dots, T$ and $j = 1, \dots, n/k_{n}$, $\mathcal{A}_{t, \tau_{j}}^{\wp, n}= \{| \Delta_{(t-1)n + \ell}^{n} \wp| \leq v_{n,t,j}^ {\wp} \}$, with
\begin{equation}
v_{n,t,j}^{\wp} = \alpha_{n,t,j}^{\wp}n^{- \varpi},
\label{eq20}
\end{equation}
where $\alpha_{n,t,j}^{\wp} = \alpha^{\wp} \sqrt{\operatorname{BV}_{n,t,j}^{\wp}}$ such that $\alpha^{\wp} > 0$, $\varpi\in(0,1/2)$, and
\begin{equation}
\operatorname{BV}_{n,t,j}^{\wp} = \frac{\pi}{2} \frac{1}{k_{n}-1} \sum_{\ell= ( j-1 ) k_{n}+2}^{jk_{n}} \bigl\llvert\sqrt{n} \Delta_{(t-1)n+ \ell-1}^{n} \wp\bigr\rrvert\bigl\llvert\sqrt{ n} \Delta_{(t-1)n+\ell}^{n} \wp\bigr\rrvert. \label{eq21}
\end{equation}
Equation \eqref{equation:realized-covariance} is the realized covariance of \citet{barndorff-nielsen-shephard:04a} upgraded with the truncation device of \citet{mancini:09a}. The latter removes returns that originate from the jump component of the log-price process. This ensures that $\hat{c}_{t, \tau_{j}}$ is consistent for the continuous part of the quadratic covariation, that is, integrated covariance, over the block. The threshold is a function of a localized bipower variation estimator \citep{barndorff-nielsen-shephard:04b}, so the truncation is time-varying and adapts to the level of intraday volatility. This is important, because failure to capture the dynamic of the volatility process can cause problems for inference (e.g., \cite{boudt-croux-laurent:11b}).

It is convenient to work with a statistic defined on the whole interval $[0,T]$, which we do by setting $\hat{c}_{t, \tau} \equiv\hat{c}_{t,\tau_{j}}$, for $\tau\in(\tau_{j-1}, \tau_{j}]$.

To proceed, we estimate the intraday curve in the spot covariance and transform this into an estimate of the diurnal component in the correlation process. We propose to scale an estimator targeting the average spot covariance at a particular time-of-day with another estimator of the unconditional covariance over the whole day, where the latter serves as a normalization to adhere to Assumption~\ref{ass(C2)}, that is,
\begin{equation}
\label{equation:diurnal-spot-covariance} \hat{c}_{u, \tau} = \tilde{c}_{u, \tau} \oslash
\bar{c}_{sv} \equiv %
\begin{bmatrix}
\hat{c}_{u, \tau}^{X} & \hat{c}_{u, \tau}^{XY} \\[0.10cm]
\hat{c}_{u, \tau}^{XY} & \hat{c}_{u, \tau}^{Y}
\end{bmatrix} %
, 
\end{equation}
with
\begin{equation}
\label{equation:time-series-average} \tilde{c}_{u, \tau_{j}} = \frac{1}{T} \sum
_{t=1}^{T} \hat{c}_{t, \tau_{j}} \equiv
\begin{bmatrix} \tilde{c}_{u, \tau_{j}}^{X}
& \tilde{c}_{u, \tau_{j}}^{XY} \\[0.10cm]
\tilde{c}_{u, \tau_{j}}^{XY} & \tilde{c}_{u, \tau_{j}}^{Y}
\end{bmatrix} %
\quad\text{and} \quad\widebar{c}_{sv} = \frac{1}{n/k_{n}} \sum_{j=1}^{n/k_{n}} \tilde{c}_{u, \tau_{j}} \equiv %
\begin{bmatrix}
\widebar{c}_{sv}^{X} & \widebar{c}_{sv}^{XY} \\[0.10cm]
\widebar{c}_{sv}^{XY} & \widebar{c}_{sv}^{Y}
\end{bmatrix} %
, 
\end{equation}
where $A \oslash B$ is the Hadamard division.

An estimator of the deterministic component of the intraday correlation is the following:
\begin{equation}
\hat{k}_{u, \tau} = \frac{\hat{c}_{u, \tau}^{XY}}{\sqrt{\hat{c}_{u, \tau}^{X}} \sqrt{\hat{c}_{u, \tau}^{Y}}}.
\label{eq24}
\end{equation}
It is worthwhile to note that $\hat{k}_{u, \tau}$ can equivalently be written as
\begin{equation}
\hat{k}_{u, \tau} = \frac{\tilde{k}_{u, \tau}}{\overline{\rho}_{sc}}, \label{eq25}
\end{equation}
where
\begin{equation}
\tilde{k}_{u, \tau} = \frac{\tilde{c}_{u, \tau}^{XY}}{\sqrt{\tilde{c}_{u, \tau}^{X}} \sqrt{\tilde{c}_{u, \tau}^{Y}}} \quad\text{and} \quad\widebar{\rho}_{sc} = \frac{\widebar{c}_{sv}^{XY}}{\sqrt{\widebar{c}_{sv}^{X}} \sqrt{\widebar{c}_{sv}^{Y}}}. \label{eq26}
\end{equation}
The next result derives the probability limit of the various estimators.

\begin{theorem}
\label{theorem:consistency}
Suppose that Assumptions \ref{ass(V)}, \ref{ass(J)}, and \ref{ass(C1)}--\ref{ass(C5)} (with $q = 1$ in Assumption~\ref{ass(C5)}) hold. As $n \rightarrow\infty$, $T \rightarrow\infty$, $k_{n} \rightarrow\infty$ such that $k_{n} / n \rightarrow0$, it holds that for $\tau\in[0,1]$,
\begin{equation}
\hat{c}_{u, \tau} \overset{p}{\longrightarrow} c_{u, \tau} \quad\text{and} \quad\widebar{c}_{sv} \overset{p}{\longrightarrow} \mathbb{E} \lleft( %
\begin{bmatrix}
c_{sv,1}^{X} & c_{sv,1}^{XY} \\[0.10cm]
c_{sv,1}^{XY} & c_{sv,1}^{Y}
\end{bmatrix} %
\rright). \label{eq27}
\end{equation}
Moreover,
\begin{equation}
\hat{k}_{u, \tau} \overset{p}{\longrightarrow} k_{u, \tau}, \qquad\tilde{k}_{u, \tau} \overset{p}{\longrightarrow} k_{u, \tau} E_{\bar{\rho}_{sc}}, \quad\text{and} \quad\bar{\rho}_{sc} \overset{p}{\longrightarrow} E_{\bar{\rho}_{sc}},
\label{eq28}
\end{equation}
where
\begin{equation}
E_{\bar{\rho}_{sc}} = \frac{\mathbb{E} \bigl(c_{sv,1}^{XY} \bigr)}{\sqrt{\mathbb{E} \bigl(c_{sv,1}^{X} \bigr)} \sqrt{\mathbb{E} \bigl(c_{sv,1}^{Y} \bigr)}}.
\label{eq29}
\end{equation}
\end{theorem}
The proof relies on a double asymptotic setting with $n \rightarrow\infty$ and $T \rightarrow\infty$. Intuitively, to retrieve the stationary expectation of the covariance process, the time horizon has to increase. In this regard, the requirement on the memory of the process is rather weak and merely states that the autocorrelation function has to be absolutely summable. As $n \rightarrow\infty$, on each block the realized covariance converges to the integrated covariance. The condition $k_{n} \rightarrow\infty$ with $n / k_{n} \rightarrow\infty$ says that we reduce the time span of such a block at a sufficiently slow rate so there is an accumulation of log-returns inside each estimation window. Taken together, this implies that realized covariance collapses to the latent point-in-time covariance and---after conversion---that our estimator of the diurnal component of the spot correlation process is consistent.

\section{Testing procedure}
\label{section:test-procedure}

In this section, we construct our testing procedure to discriminate between the null and alternative hypothesis. We develop a test statistic that accommodates the general setting for the spot covariance process (as outlined in Assumptions \ref{ass(C1)}--\ref{ass(C5)}).

\subsection{Test statistic}
\label{section:test-statistic}

We begin with a preliminary functional central limit theorem (CLT) concerning the asymptotic distribution of the diurnal covariance estimator from \eqref{equation:diurnal-spot-covariance}. We define the Hilbert space:
\begin{equation}
\mathcal{L}^{2} = \biggl\{g: [0,1] \rightarrow\mathbb{R} \Bigm| \int_{0}^{1} g(u)^{2} \,\mathrm{d}u< \infty\biggr\}, \label{eq30}
\end{equation}
equipped with the usual inner product $\langle\cdot\,, \cdot\rangle$ and the induced norm $ \lVert\cdot\rVert$. We use the notation $x_{n} \asymp y_{n}$ to represent that, as $n \rightarrow\infty$, $1/C \leq x_{n} / y_{n} \leq C$ for some positive constant~$C$.

\begin{theorem}
\label{theorem:functional-clt}
Suppose that Assumptions \ref{ass(V)}, \ref{ass(J)}, and \ref{ass(C1)}--\ref{ass(C5)} (with $q=3$ in Assumption~\ref{ass(C5)}) hold. As $n \rightarrow\infty$ and $T \rightarrow\infty$ such that $k_{n} \rightarrow\infty$, $k_{n}/n \rightarrow0$, $T \asymp n^{c}$, and $k_{n} \asymp n^{d}$, for some strictly positive exponents $c$ and $d$ that satisfy
\begin{equation}
\label{equation:growth-rate-T-n-k} 0 < c< 4 \varpi\quad\text{and} \quad1-4 \varpi< d <
1-c/2, 
\end{equation}
with $\varpi\in(0,1/2)$. Then it holds that
\begin{equation}
\label{equation:functional-clt} \sqrt{T} \begin{pmatrix} \hat{c}_{u, \tau}^{X}
- c_{u, \tau}^{X} \\[0.10cm]
\hat{c}_{u, \tau}^{XY} - c_{u, \tau}^{XY} \\[0.10cm]
\hat{c}_{u, \tau}^{Y} - c_{u, \tau}^{Y}
\end{pmatrix} \overset{d}{\longrightarrow} \mathcal{W}_{\tau},
\end{equation}
where $\mathcal{W} = ( \mathcal{W}_{1}, \mathcal{W}_{2}, \mathcal{W}_{3})^{\top}$, and the $\mathcal{W}_{i}$'s are $\mathcal{L}^{2}$-valued mean zero Gaussian processes with covariance matrix function between $\mathcal{W}_{\kappa}$ and $\mathcal{W}_{\tau}$ given by
\begin{align}
\Gamma_{\kappa, \tau} &=
\begin{bmatrix} \frac{1}{\mathbb{E}
\bigl(c_{sv,1}^{X}\bigr)^{2}} & \frac{1}{\mathbb{E} \bigl(c_{sv,1}^{X}\bigr) \mathbb{E}\bigl(c_{sv,1}^{XY} \bigr)} & \frac{1}{\mathbb{E}\bigl(c_{sv,1}^{X}\bigr) \mathbb{E}\bigl(c_{sv,1}^{Y}\bigr)} \\[10pt]
\frac{1}{\mathbb{E}\bigl(c_{sv,1}^{XY}\bigr) \mathbb{E} \bigl(c_{sv,1}^{X}\bigr)} & \frac{1}{\mathbb{E} \bigl(c_{sv,1}^{XY}\bigr)^{2}} & \frac{1}{\mathbb{E} \bigl(c_{sv,1}^{XY}\bigr) \mathbb{E}\bigl(c_{sv,1}^{Y} \bigr)} \\[10pt]
\frac{1}{\mathbb{E}\bigl(c_{sv,1}^{Y}\bigr) \mathbb{E} \bigl(c_{sv,1}^{X}\bigr)} & \frac{1}{\mathbb{E} \bigl(c_{sv,1}^{Y}\bigr) \mathbb{E}\bigl(c_{sv,1}^{XY} \bigr)} & \frac{1}{\mathbb{E}\bigl(c_{sv,1}^{Y} \bigr)^{2}} \end{bmatrix} %
\nonumber
\\
&\quad{}\odot\sum_{h=- \infty}^{\infty}
\begin{bmatrix} v_{\kappa, \tau}^{X, X}(h) &
v_{\kappa, \tau}^{X,XY}(h) & v_{\kappa, \tau}^{X,Y}(h) \\[8pt]
v_{\kappa, \tau}^{XY,X}(h) & v_{\kappa, \tau}^{XY, XY}(h) & v_{\kappa, \tau}^{XY,Y}(h) \\[8pt]
v_{\kappa, \tau}^{Y,X}(h) & v_{\kappa, \tau}^{Y,XY}(h) & v_{\kappa, \tau}^{Y, Y}(h) \end{bmatrix} %
.
\label{eq33}
\end{align}
Here, with $Z_{1}, Z_{2} \in\{ X, Y, XY \}$,
\begin{equation}
v_{\kappa, \tau}^{Z_{1}, Z_{2}}(h) = \operatorname{cov}\bigl(A_{1, \kappa}^{Z_{1}}, A_{1, \tau+h}^{Z_{2}}\bigr), \label{eq34}
\end{equation}
for $\kappa, \tau\in[0,1]$, and
\begin{equation}
A_{1, \kappa}^{Z} = c_{\kappa}^{Z} - c_{u, \kappa}^{Z} \int_{0}^{1}c_{s}^{Z} \,\mathrm{d}s. \label{eq35}
\end{equation}
\end{theorem}
This extends Theorem~1 of \citet{andersen-su-todorov-zhang:24a} from the univariate to the multivariate setting. Compared to Theorem~\ref{theorem:consistency}, we impose a faster rate of decay on the sequence of mixing coefficients.

In Assumption~\ref{ass(C5)}, we require the random component of the correlation process to follow the same stationarity condition imposed on the volatility process. Consequently, the random component of the covariance process satisfies this condition, allowing us to select the orders of $T$ and $k_{n}$ as in the univariate case. Condition \eqref{equation:growth-rate-T-n-k} restricts the growth of $T$ and $k_{n}$ relative to $n$. Such constraints also appear in closely related work on long-span estimation with high-frequency data; see, for example, equation (9) in \citet{andersen-su-todorov-zhang:24a} and equation (5) in \citet{andersen-tan-todorov-zhang:25a}. As they explain, when the truncation parameter $\varpi$ is close to $1/2$, the resulting bounds on $c$ and $d$ are weakest. That is, $c$ can be any number in $(0,2)$, making the length of the time period very flexible. Once $c$ is chosen, the optimal choice of $d$ is provided in \citet{andersen-su-todorov-zhang:24a}. In particular, when $c>1/2$, the optimal choice of $d$ is $(2-c)/3$. On the other hand, if we take the optimal convergence rate for spot volatility; namely, if $d$ is close to $1/2$, then $c \leq1$, indicating that $T$ cannot grow faster than $n$, implying that high-frequency sampling should increase at least as fast as the time span. We refer to Section~5 of \citet{andersen-su-todorov-zhang:24a} for a detailed discussion of the bias-variance tradeoff.

To derive the diurnal correlation process, we can apply the functional delta rule to \eqref{equation:functional-clt} with $g(x,y,z) = y(xz)^{-1/2}$. However, the correlation map is not Hadamard differentiable on the unrestricted product space $\mathcal{L}^{2}([0,1])^{3}$, as shown in \citet{andersen-tan-tan-todorov-zhang:26a}. In Lemma~\ref{lemma:covariance-cone} of Appendix~\ref{appendix:proof}, we establish Hadamard differentiability along a covariance-constrained estimator sequence, under an additional nondegeneracy condition stated as Assumption~\ref{ass(ND)}. Corollary~\ref{cor:stoch} converts this to the required $\mathcal{L}^{2}$ asymptotic linearization, and from this it follows that
\begin{equation}
\sqrt{T} ( \hat{k}_{u, \tau} - k_{u, \tau} ) \overset{d}{\longrightarrow} \nabla g \bigl(c_{u, \tau}^{X}, c_{u, \tau}^{XY}, c_{u, \tau}^{Y} \bigr) \cdot\mathcal{W}_{\tau}, \label{eq36}
\end{equation}
where
\begin{equation}
\nabla g \bigl(c_{u, \tau}^{X}, c_{u, \tau}^{XY}, c_{u, \tau}^{Y} \bigr) = \bigl(4 c_{u, \tau}^{X} c_{u, \tau}^{Y} \bigr)^{-1/2} \biggl(- \frac{c_{u, \tau}^{XY}}{c_{u, \tau}^{X}},2, - \frac{c_{u, \tau}^{XY}}{c_{u, \tau}^{Y}} \biggr). \label{eq37}
\end{equation}
Hence, under the null hypothesis (where $k_{u, \tau} \equiv1$):
\begin{equation}
\label{equation:S} S_{j} = \sqrt{T} ( \hat{k}_{u, \tau_{j}}-1 )
\overset{d}{\longrightarrow} \nabla g \bigl(c_{u, \tau_{j}}^{X}, c_{u, \tau_{j}}^{XY}, c_{u, \tau_{j}}^{Y} \bigr) \mathcal{W}_{\tau_{j}}. 
\end{equation}
Now, we propose our test statistic:
\begin{equation}
\label{equation:test-statistic} \mathcal{N} = \frac{1}{n/k_{n}}
\sum_{j=1}^{n/k_{n}} S_{j}^{2} = \frac{T}{n/k_{n}} \sum_{j=1}^{n/k_{n}} ( \hat{k}_{u, \tau_{j}}-1 )^{2}. 
\end{equation}
The next theorem helps to explain the behavior of $\mathcal{N}$.

\begin{theorem} \label{theorem:functional-clt-correlation}
Suppose that the assumptions of Theorem~\ref{theorem:functional-clt} are maintained and that Assumption~\ref{ass(ND)} in Appendix~\ref{appendix:proof} holds.

\begin{enumerate}
\item[(a)] In general,\vspace*{-2pt}
\begin{equation}
\frac{1}{n/k_{n}} \sum_{j=1}^{n/k_{n}} ( \hat{k}_{u, \tau_{j}}-1 )^{2} \overset{p}{\longrightarrow} \int_{0}^{1} ( k_{u,t}-1 )^{2} \,\mathrm{d}t. \label{eq40}
\end{equation}
\item[(b)] In restriction to $\Omega_{\mathcal{H}_{0}}$,\vspace*{-2pt}
\begin{equation}
\mathcal{N} \overset{d}{\longrightarrow} \bigl\lVert\nabla g \bigl(c_{u, \tau}^{X}, c_{u, \tau}^{XY}, c_{u, \tau}^{Y} \bigr) \cdot\mathcal{W}_{\tau} \bigr\rVert^{2} \equiv\lVert\mathcal{H} \rVert^{2}.
\label{eq41}
\end{equation}
\end{enumerate}
\end{theorem}

Theorem~\ref{theorem:functional-clt-correlation} implies that $\mathcal{N} \rightarrow\infty$ under $\mathcal{H}_{a}$, so a test based on it is consistent. Note that part (a) holds irrespective of whether $k_{u,t} = 1$ (i.e., there is no diurnal variation in the correlation) or not.

The asymptotic variance matrix is latent and has to be replaced with an estimator. Note that $\mathcal{H}$ is a mean zero Gaussian process with covariance kernel:\vspace*{-2pt}
\begin{equation}
\begin{aligned}[b] C( \kappa, \tau) &= \nabla g \bigl(c_{u, \kappa}^{X},
c_{u, \kappa}^{XY}, c_{u, \kappa}^{Y} \bigr) \Gamma_{\kappa, \tau} \nabla g \bigl(c_{u, \tau}^{X}, c_{u, \tau}^{XY}, c_{u, \tau}^{Y} \bigr)^{\top} \\
&= \frac{1}{4} \bigl(c_{u, \kappa}^{X} c_{u, \kappa}^{Y} c_{u, \tau}^{X} c_{u, \tau}^{Y} \bigr)^{-1/2} \biggl( - \frac{c_{u, \kappa}^{XY}}{c_{u, \kappa}^{X}},2, - \frac{c_{u, \kappa}^{XY}}{c_{u, \kappa}^{Y}} \biggr) \Gamma_{\kappa, \tau} \biggl( -\frac{c_{u, \tau}^{XY} }{c_{u, \tau}^{X}},2, - \frac{c_{u, \tau}^{XY}}{c_{u, \tau}^{Y}} \biggr)^{\top}. \end{aligned} \label{eq42}
\end{equation}
According to Theorem~\ref{theorem:consistency}, we can estimate $c_{u, \tau}^{X}$, $c_{u, \tau}^{XY}$, and $c_{u, \tau}^{Y}$ with $\hat{c}_{u, \tau}^{X}$, $\hat{c}_{u, \tau}^{XY}$, and $\hat{c}_{u, \tau}^{Y}$, respectively, and likewise for terms with index $\kappa$. We propose a standard HAC-based estimator of $\Gamma_{\kappa, \tau}$:\vspace*{-2pt}
\begin{equation}
\widehat{\Gamma}_{\kappa, \tau} = %
\begin{bmatrix}
\frac{1}{\bigl( \widebar{c}_{sv}^{X}\bigr)^{2}} & \frac{1}{\widebar{c}_{sv}^{X} \widebar{c}_{sv}^{XY}} & \frac{1}{\widebar{c}_{sv}^{X} \widebar{c}_{sv}^{Y}} \\[10pt]
\frac{1}{\widebar{c}_{sv}^{XY} \widebar{c}_{sv}^{X}} & \frac{1}{\bigl( \widebar{c}_{sv}^{XY} \bigr)^{2}} & \frac{1}{\widebar{c}_{sv}^{XY} \widebar{c}_{sv}^{Y}} \\[10pt]
\frac{1}{\widebar{c}_{sv}^{Y} \widebar{c}_{sv}^{X}} & \frac{1}{\widebar{c}_{sv}^{Y} \widebar{c}_{sv}^{XY}} & \frac{1}{\bigl( \widebar{c}_{sv}^{Y} \bigr)^{2}} \end{bmatrix} %
\odot\Biggl( \hat{v}_{\kappa, \tau}(0) + \sum_{h=1}^{H_{T}} \omega\biggl( \frac{h}{H_{T}} \biggr) \bigl( \hat{v}_{\kappa, \tau}(h) + \hat{v}_{\kappa, \tau}(-h) \bigr) \Biggr), \label{eq43}
\end{equation}
where\vspace*{-2pt}
\begin{equation}
\hat{v}_{\kappa, \tau}(h) = \frac{1}{T} \sum_{t=1}^{T}
\begin{bmatrix} \hat{A}_{t, \kappa}^{X}
\\[10pt]
\hat{A}_{t, \kappa}^{XY} \\[10pt]
\hat{A}_{t, \kappa}^{Y} \end{bmatrix} %
\begin{bmatrix} \hat{A}_{t, \tau+h}^{X}
\\[10pt]
\hat{A}_{t, \tau+h}^{XY} \\[10pt]
\hat{A}_{t, \tau+h}^{Y} \end{bmatrix} %
^{\top},
\label{eq44}
\end{equation}
$H_{T}$ is the lag length, $\omega$ is a kernel (see, e.g., \cite{andrews:91a}), and\vspace*{-2pt}
\begin{align}
\begin{aligned} \hat{A}_{t, \kappa}^{X} &=
\hat{c}_{t-1+\kappa}^{X} - \hat{c}_{u, \kappa}^{X} \sum_{\ell=1}^{n} \bigl( \Delta_{(t-1)n+\ell}^{n}X \bigr)^{2} \mathbbm{1}_{\mathcal{A}_{t, \tau_{j}}^{X,n}}, \\
\hat{A}_{t, \kappa}^{XY} &= \hat{c}_{t-1+\kappa}^{XY} - \hat{c}_{u, \kappa}^{XY} \sum_{\ell=1}^{n} \bigl( \Delta_{(t-1)n+\ell}^{n}X \Delta_{(t-1)n+\ell}^{n}Y \bigr) \mathbbm{1}_{\mathcal{A}_{t, \tau_{j}}^{X,n} \cap\mathcal{A}_{t, \tau_{j}}^{Y,n}}, \\
\hat{A}_{t, \kappa}^{Y} &= \hat{c}_{t-1+\kappa}^{Y} -\hat{c}_{u, \kappa}^{Y} \sum_{\ell=1}^{n} \bigl( \Delta_{(t-1)n+\ell}^{n}Y \bigr)^{2} \mathbbm{1}_{\mathcal{A}_{t, \tau_{j}}^{Y,n}}. \end{aligned} \label{eq45}
\end{align}
It should be noted that the expectation of $A_{t, \kappa+h}^{Z}$ is zero for $h = 0, \dots, H_{T}$. The following result then gives the consistency of $\widehat{\Gamma}_{\kappa, \tau}$.

\begin{proposition}
\label{proposition:hac}
Suppose that the assumptions of Theorem~\ref{theorem:functional-clt} are maintained. Then, if $H_{T} \rightarrow\infty$ such that $H_{T}/ \sqrt{T} \rightarrow0$, it further holds that
\begin{equation}
\widehat{\Gamma}_{\kappa, \tau} \overset{p}{\longrightarrow} \Gamma_{\kappa, \tau}. \label{eq46}
\end{equation}
\end{proposition}

Hence, we arrive at the following estimator of $C( \kappa, \tau)$:
\begin{equation}
\label{equation:kernel-estimator} \widehat{C}( \kappa, \tau) = \frac{1}{4} \bigl(
\hat{c}_{u, \kappa}^{X} \hat{c}_{u, \kappa}^{Y} \hat{c}_{u, \tau}^{X} \hat{c}_{u, \tau}^{Y} \bigr)^{-1/2} \biggl( - \frac{\hat{c}_{u, \kappa}^{XY}}{\hat{c}_{u, \kappa}^{X}},2, - \frac{\hat{c}_{u, \kappa}^{XY}}{\hat{c}_{u, \kappa}^{Y}} \biggr) \widehat{\Gamma}_{\kappa,\tau} \biggl( - \frac{\hat{c}_{u, \tau}^{XY}}{\hat{c}_{u, \tau}^{X}},2, - \frac{\hat{c}_{u, \tau}^{XY}}{\hat{c}_{u, \tau}^{Y}} \biggr)^{\top}. 
\end{equation}
Now, define $\widehat{\mathcal{H}}$ to be an $\mathcal{F}$-conditional $\mathcal{L}^{2}$-valued mean zero Gaussian process with covariance kernel $\widehat{C}$, as defined in \eqref{equation:kernel-estimator}. We can then show that $\widehat{\mathcal{H}}$ converges in law to $\mathcal{H}$ (in~$\mathcal{L}^{2}$).

\begin{theorem}
\label{theorem:process-estimator}
Suppose that the assumptions of Theorem~\ref{theorem:functional-clt} are maintained (with $q = 4$ in Assumption~\ref{ass(C5)}), $c+d > 1-16 \varpi/ 7$ and $d > (3-8 \varpi)/3$. In addition, if $\int_{\mathbb{R}}x^{8} \tilde{F}( \mathrm{d}x) < \infty$ and $H_{T} \asymp n^{\gamma}$ for a strictly positive exponent $\gamma$ that satisfies
\begin{equation}
\gamma< \min\bigl\{d/2, (1-d)/4, 2 \varpi-(1-d)/2, c/2, 2 \varpi-7/8+7(c+d)/8 \bigr\}. \label{eq48}
\end{equation}
Then it holds that
\begin{equation}
\widehat{\mathcal{H}} \overset{d}{\longrightarrow} \mathcal{H}.
\label{eq49}
\end{equation}
\end{theorem}
The CLT in Theorem~\ref{theorem:process-estimator} again generalizes the associated Theorem~6 in \citet{andersen-su-todorov-zhang:24a} to the multivariate case. Compared to Theorem~\ref{theorem:functional-clt}, it imposes the additional rate conditions $c + d > 1-16 \varpi/ 7$ and $d > (3-8 \varpi)/3$. Requiring $d>(3-8 \varpi)/3$ is stronger than $d > 1-4\varpi$ in Theorem~\ref{theorem:functional-clt}, but for $\varpi\geq3/8$ it is automatically satisfied. This observation also applies to $c + d > 1-16 \varpi/ 7$, which is automatic for $\varpi\geq7/16$.

We can approximate the asymptotic distribution of the test statistic, $ \lVert\mathcal{H} \rVert^{2}$, using simulations. We partition the interval $[0,1]$ into $m$ subintervals of equal length, where $m = n/k_{n}$, and consider an $m$-dimensional normal random vector $( \widehat{\mathcal{H}}_{\tau_{1}}, \dots, \widehat{\mathcal{H}}_{\tau_{m}})^{\top}$ with mean zero and conditional covariance matrix $\widehat{C} = ( \widehat{C}_{\tau_{i}, \tau_{j}})_{1 \leq i,j \leq m}$, where $\tau_{j} = j/m$ for $j=1, \dots, m$. Next, observe that
\begin{equation}
\label{equation:approximate} \widehat{\mathcal{Z}} = \frac{1}{m} \sum
_{j=1}^{m} \widehat{\mathcal{H}}_{\tau_{j}}^{2} \overset{d} = \frac{1}{m} \sum_{j=1}^{m} \lambda_{j} \chi_{j}^{2}, 
\end{equation}
where $( \lambda_{j})_{j=1}^{m}$ are the eigenvalues of $\widehat{C}$ and $( \chi_{j}^{2})_{j=1}^{m}$ are independent $\chi^{2}(1)$-distributed random variates, defined on an extension of the original probability space and independent from $\mathcal{F}$. Since $\widehat{C}$ is an estimate of a covariance matrix, it can possess negative eigenvalues in practice. We therefore follow \citet{andersen-su-todorov-zhang:24a} and retain only those terms in \eqref{equation:approximate} that are associated with positive eigenvalues. The above process delivers one possible outcome and can be repeated as many times as necessary to get an acceptable approximation to the law of $ \lVert\mathcal{H} \rVert^{2}$.

\section{Extensions}
\label{section:extension}

\subsection{Stochastic diurnal correlation}
\label{section:stochastic}

In Assumption~\ref{ass(C1)}, we restricted the intraday curve in the correlation process to be deterministic. To allow for a more general structure that incorporates stochastic diurnal correlation, we follow \citet{andersen-su-todorov-zhang:24a} and suppose instead that for $Z \in\{ X, Y, XY \}$,
\begin{equation}
\label{equation:total-covariance-random} \mathbb{E}\bigl[c_{t}^{Z}\bigr] =
c_{u, t- \lfloor t \rfloor}^{Z}. 
\end{equation}
In contrast to before, \eqref{equation:total-covariance-random} only restricts the calendar effect in correlation to be present in expectation. This implies that, on average, the function is periodic as in Assumption~\ref{ass(C2)}, such that it remains consistently estimable in the infill and long-span limit by ergodicity.

The diurnal correlation function is now given by\vspace*{-2pt}
\begin{equation}
\label{equation:spot-correlation-random} \frac{\mathbb{E}\bigl[c_{t}^{XY}
\bigr]}{\sqrt{\mathbb{E}\bigl[c_{t}^{X}\bigr] \mathbb{E} \bigl[c_{t}^{Y}\bigr]}} = \frac{c_{u, t- \lfloor t \rfloor}^{XY}}{\sqrt{c_{u,t- \lfloor t \rfloor}^{X} c_{u,t- \lfloor t \rfloor}^{Y}}} \equiv k_{u,t- \lfloor t \rfloor}. 
\end{equation}
We can construct a test of the hypothesis\vspace*{-2pt}
\begin{equation}
\mathcal{H}_{0}: k_{u, t- \lfloor t \rfloor} = k_{u} \quad\text{against} \quad\mathcal{H}_{a}: k_{u, t- \lfloor t \rfloor} \neq k_{u}. \label{eq53}
\end{equation}
In this setting, it readily follows that\vspace*{-2pt}
\begin{equation}
\hat{k}_{u, \tau} = \frac{\tilde{c}_{u, \tau}^{XY}}{\sqrt{\tilde{c}_{u, \tau}^{X}} \sqrt{\tilde{c}_{u, \tau}^{Y}}} \overset{p}{\longrightarrow} k_{u, \tau}. \label{eq54}
\end{equation}
Moreover, the following result establishes a functional CLT.
\begin{theorem}
\label{theorem:functional-clt-random}
Suppose that the assumptions of Theorem~\ref{theorem:functional-clt} are maintained. Then it holds that\vspace*{-2pt}
\begin{equation}
\label{equation:functional-clt-random} \sqrt{T} \begin{pmatrix} \tilde{c}_{u, \tau}^{X}
- \mathbb{E}\bigl[c_{\tau}^{X}\bigr] \\[6pt]
\tilde{c}_{u, \tau}^{XY} - \mathbb{E}\bigl[c_{\tau}^{XY} \bigr] \\[6pt]
\tilde{c}_{u, \tau}^{Y} - \mathbb{E}\bigl[c_{\tau}^{Y} \bigr] \end{pmatrix} \overset{d}{\longrightarrow} \mathcal{W}_{\tau},
\end{equation}
where $\mathcal{W} = ( \mathcal{W}_{1}, \mathcal{W}_{2}, \mathcal{W}_{3})^{\top}$, and the $\mathcal{W}_{i}$'s are $\mathcal{L}^{2}$-valued mean zero Gaussian processes with covariance matrix function between $\mathcal{W}_{\kappa}$ and $\mathcal{W}_{\tau}$ given by\vspace*{-2pt}
\begin{equation}
\Gamma_{\kappa, \tau} = \sum_{h=- \infty}^{\infty}
\begin{bmatrix} v_{\kappa, \tau}^{X, X}(h) &
v_{\kappa, \tau}^{X,XY}(h) & v_{\kappa, \tau}^{X,Y}(h) \\[6pt]
v_{\kappa, \tau}^{XY,X}(h) & v_{\kappa, \tau}^{XY, XY}(h) & v_{\kappa, \tau}^{XY,Y}(h) \\[6pt]
v_{\kappa, \tau}^{Y,X}(h) & v_{\kappa, \tau}^{Y,XY}(h) & v_{\kappa, \tau}^{Y, Y}(h) \end{bmatrix} %
.
\label{eq56}
\end{equation}
Here, with $Z_{1}, Z_{2} \in\{ X, Y, XY \}$,\vspace*{-2pt}
\begin{equation}
v_{\kappa, \tau}^{Z_{1}, Z_{2}}(h) = \operatorname{cov}\bigl(c_{\kappa}^{Z_{1}}, c_{\tau+ h}^{Z_{2}}\bigr), \label{eq57}
\end{equation}
for $\kappa, \tau\in[0,1]$.\vadjust{\goodbreak}
\end{theorem}

We propose the following test statistic:
\begin{equation}
\label{equation:test-statistic-random} \widetilde{\mathcal{N}} =
\frac{T}{n/k_{n}} \sum_{j=1}^{n/k_{n}} ( \hat{k}_{u, \tau_{j}}-\bar{k}_{u} )^{2},
\end{equation}
where $\hat{k}_{u,\tau}$ is defined in \eqref{eq54} and $\bar{k}_{u} = \frac{1}{n/k_{n}} \sum_{j=1}^{n/k_{n}} \hat{k}_{u, \tau_{j}}$. We can then prove the following.

\begin{theorem}
\label{theorem:functional-clt-correlation-random}
Suppose that the assumptions of Theorem~\ref{theorem:functional-clt} are maintained and that Assumption~\ref{ass(ND)} in Appendix~\ref{appendix:proof} holds.
\begin{enumerate}
\item[(a)] In general,
\begin{equation}
\frac{1}{n/k_{n}} \sum_{j=1}^{n/k_{n}} ( \hat{k}_{u, \tau_{j}}- \bar{k}_{u} )^{2} \overset{p}{\longrightarrow} \int_{0}^{1} \biggl( k_{u,t}- \int_{0}^{1} k_{u,s} \,\mathrm{d}s \biggr)^{2} \,\mathrm{d}t. \label{eq59}
\end{equation}
\item[(b)] In restriction to $\Omega_{\mathcal{H}_{0}}$,
\begin{equation}
\begin{aligned}[b]
\widetilde{\mathcal{N}} & \overset{d}{\longrightarrow} \biggl\lVert\nabla g \bigl( \mathbb{E}\bigl[c_{\tau}^{X}\bigr], \mathbb{E}\bigl[c_{\tau}^{XY}\bigr], \mathbb{E} \bigl[c_{\tau}^{Y}\bigr] \bigr) \cdot\mathcal{W}_{\tau} \\
&\quad{}- \int_{0}^{1} \nabla g \bigl( \mathbb{E} \bigl[c_{t}^{X}\bigr], \mathbb{E}\bigl[c_{t}^{XY} \bigr], \mathbb{E}\bigl[c_{t}^{Y}\bigr] \bigr) \cdot\mathcal{W}_{t} \,\mathrm{d}t \biggr\rVert^{2} \\
&\overset{d}{=} \int_{0}^{1} \bigl( \nabla g \bigl( \mathbb{E}\bigl[c_{t}^{X}\bigr], \mathbb{E} \bigl[c_{t}^{XY}\bigr], \mathbb{E}\bigl[c_{t}^{Y} \bigr] \bigr) \cdot\mathcal{W}_{t} \bigr)^{2} \,\mathrm{d}t \\
&\quad{}- \biggl( \int_{0}^{1} \nabla g \bigl( \mathbb{E} \bigl[c_{t}^{X}\bigr], \mathbb{E}\bigl[c_{t}^{XY} \bigr], \mathbb{E}\bigl[c_{t}^{Y}\bigr] \bigr) \cdot\mathcal{W}_{t} \,\mathrm{d}t \biggr)^{2}.%
\end{aligned} %
\label{eq60}
\end{equation}
\end{enumerate}
\end{theorem}
Again, we can design a standard HAC-based estimator of $\Gamma_{\kappa, \tau}$:
\begin{equation}
\widehat{\Gamma}_{\kappa, \tau} = \hat{v}_{\kappa, \tau}(0) + \sum_{h=1}^{H_{T}} \omega\biggl( \frac{h}{H_{T}} \biggr) \bigl( \hat{v}_{\kappa, \tau}(h) + \hat{v}_{\kappa, \tau}(-h) \bigr) , \label{eq61}
\end{equation}
where
\begin{equation}
\hat{v}_{\kappa, \tau}(h) = \frac{1}{T} \sum_{t=1}^{T}
\begin{bmatrix} \hat{c}_{t-1+\kappa}^{X}
- \tilde{c}_{u, \kappa}^{X} \\[6pt]
\hat{c}_{t-1+\kappa}^{XY} - \tilde{c}_{u, \kappa}^{XY} \\[6pt]
\hat{c}_{t-1+\kappa}^{Y} - \tilde{c}_{u, \kappa}^{Y}
\end{bmatrix} %
\begin{bmatrix} \hat{c}_{t-1+h+ \tau}^{X}
- \tilde{c}_{u, \tau}^{X} \\[6pt]
\hat{c}_{t-1+h+ \tau}^{XY} - \tilde{c}_{u, \tau}^{XY} \\[6pt]
\hat{c}_{t-1+h+ \tau}^{Y} - \tilde{c}_{u, \tau}^{Y}
\end{bmatrix} %
^{\top}, \label{eq62}
\end{equation}
with $H_{T}$ and $\omega$ defined as above. Accordingly, we can construct an estimator of the covariance kernel as follows:
\begin{equation}
\label{equation:kernel-estimator-random} \widehat{C}( \kappa, \tau) = \frac{1}{4} \bigl(
\tilde{c}_{u, \kappa}^{X} \tilde{c}_{u, \kappa}^{Y} \tilde{c}_{u, \tau}^{X} \tilde{c}_{u, \tau}^{Y} \bigr)^{-1/2} \biggl( - \frac{\tilde{c}_{u, \kappa}^{XY}}{\tilde{c}_{u, \kappa}^{X}},2, - \frac{\tilde{c}_{u, \kappa}^{XY}}{\tilde{c}_{u, \kappa}^{Y}} \biggr) \widehat{\Gamma}_{\kappa,\tau} \biggl( - \frac{\tilde{c}_{u, \tau}^{XY}}{\tilde{c}_{u, \tau}^{X}},2, - \frac{\tilde{c}_{u, \tau}^{XY}}{\tilde{c}_{u, \tau}^{Y}} \biggr)^{\top}.
\end{equation}
As before, we can simulate the distribution of the test statistic by partitioning the interval $[0,1]$ into $m$ subintervals of equal length, where $m = n/k_{n}$. We generate the $m$-dimensional normal random vector $( \widehat{\mathcal{H}}_{\tau_{1}}, \dots, \widehat{\mathcal{H}}_{\tau_{m}})$ with mean zero and conditional covariance matrix $\widehat{C} = ( \widehat{C}_{\tau_{i}, \tau_{j}})_{1 \leq i,j \leq m}$, now based on \eqref{equation:kernel-estimator-random}, where $\tau_{j} = j/m$ for $j=1, \dots, m$. Next, we set
\begin{equation}
\label{equation:approximate-random} \widehat{\mathcal{Z}} = \frac{1}{m} \sum
_{j=1}^{m} \widehat{\mathcal{H}}_{\tau_{j}}^{2} - \Biggl( \frac{1}{m} \sum_{j=1}^{m} \widehat{\mathcal{H}}_{\tau_{j}} \Biggr)^{2}. 
\end{equation}
In the context of random diurnal volatility and correlation, the decompositions in \eqref{equation:spot-volatility} and \eqref{equation:spot-correlation} are lost, and the identification condition in Assumption~\ref{ass(C2)} becomes meaningless. So, the hypothesis $k_{u,t} \equiv k_{u}$ (a constant) for $t \in(0,1)$ does not imply that $k_{u,t} \equiv1$. Therefore, we employ the equivalent condition of $k_{u,t} \equiv k_{u}$ for $t \in(0,1)$; namely, $k_{u,t} - \int_{0}^{1} k_{u,s} \,\mathrm{d}s \equiv0$, to create the modified test statistic in \eqref{equation:test-statistic-random}, which is different from the previous one. That being said, although the new test statistic is of course also available for testing with a deterministic diurnal correlation function, it is not identical to \eqref{equation:test-statistic}, because the old version of the test statistic incorporates the extra information provided by the identification condition in Assumption~\ref{ass(C2)}.

\subsection{Incorporating conditioning information}
\label{section:conditional}

In this section, we follow \citet{andersen-thyrsgaard-todorov:19a} by showing how our theoretical framework can be generalized to a conditional version that incorporates some additional information that may help to explain the form of the diurnal correlation function, such as the release of important news announcements; an idea that we explore further in the empirical application. To this end, we redefine the random variables $\tilde{c}_{u, \tau_{j}}$ and $\widebar{c}_{sv}$ in \eqref{equation:time-series-average} as follows:
\begin{equation}
\begin{aligned}
\tilde{c}^{\mathcal{B}}_{u, \tau_{j}} &= \frac{1}{T} \sum_{t=1}^{T} \mathbbm{1}_{\mathcal{B}_{t-1}} \hat{c}_{t, \tau_{j}} \equiv %
\begin{bmatrix} \tilde{c}_{u, \tau_{j}}^{X, \mathcal{B}} &
\tilde{c}_{u, \tau_{j}}^{XY, \mathcal{B}} \\[6pt]
\tilde{c}_{u, \tau_{j}}^{XY, \mathcal{B}} & \tilde{c}_{u, \tau_{j}}^{Y, \mathcal{B}}
\end{bmatrix} %
\quad\text{and} \\
\widebar{c}_{sv}^{\mathcal B} &= \frac{1}{n/k_{n}} \sum_{j=1}^{n/k_{n}} \tilde{c}^{{\mathcal B}}_{u, \tau_{j}} \equiv %
\begin{bmatrix} \widebar{c}_{sv}^{X, {\mathcal B}} &
\widebar{c}_{sv}^{XY, {\mathcal B}} \\[6pt]
\widebar{c}_{sv}^{XY, {\mathcal B}} & \widebar{c}_{sv}^{Y, {\mathcal B}}
\end{bmatrix} %
,
\end{aligned}
\label{eq65}
\end{equation}
where $\mathcal{B}_{t-1}$ is an $\mathcal{F}_{t-1}$-adapted random set. Provided appropriate stationarity, ergodicity, and mixing conditions hold, we can deduce a straightforward extension of Theorem~\ref{theorem:functional-clt}. Namely, letting $\hat{c}_{u, \tau}^{\mathcal{B}} = \tilde{c}_{u, \tau}^{\mathcal{B}} \oslash\bar{c}_{sv}^{\mathcal{B}}$, we deduce that
\begin{equation}
\label{equation:functional-clt-conditional} \sqrt{T} \begin{pmatrix} \hat{c}_{u, \tau}^{X, \mathcal{B}}
- c_{u, \tau}^{X, \mathcal{B}} \\[6pt]
\hat{c}_{u, \tau}^{XY, \mathcal{B}} - c_{u, \tau}^{XY, \mathcal{B}} \\[6pt]
\hat{c}_{u, \tau}^{Y, \mathcal{B}} - c_{u, \tau}^{Y, \mathcal{B}}
\end{pmatrix} \overset{d}{\longrightarrow} \mathcal{W}^{\mathcal{B}}_{
\tau},
\end{equation}
where $\mathcal{W}^{\mathcal{B}} = ( \mathcal{W}^{\mathcal{B}}_{1}, \mathcal{W}_{2}^{\mathcal{B}}, \mathcal{W}^{\mathcal{B}}_{3})^{\top}$, and the $\mathcal{W}^{\mathcal{B}}_{i}$'s are $\mathcal{L}^{2}$-valued mean zero Gaussian processes with covariance matrix function:
\begin{align}
&\operatorname{cov}\bigl( \mathcal{W}^{\mathcal{B}}_{\kappa}, \mathcal{W}^{\mathcal{B}}_{\tau}\bigr) \nonumber\\
&\quad= \Gamma_{\kappa, \tau}^{\mathcal{B}}\nonumber\\
&\quad= %
\matfontas{\begin{bmatrix} \frac{1}{\mathbb{E} \bigl(c_{sv,t}^{X} \mathbbm{1}_{\mathcal{B}_{t-1}}\bigr)^{2}} & \frac{1}{\mathbb{E}\bigl(c_{sv,t}^{X} \mathbbm{1}_{\mathcal{B}_{t-1}}\bigr) \mathbb{E}\bigl(c_{sv,t}^{XY} \mathbbm{1}_{\mathcal{B}_{t-1}}\bigr)} & \frac{1}{\mathbb{E}\bigl(c_{sv,t}^{X} \mathbbm{1}_{\mathcal{B}_{t-1}}\bigr) \mathbb{E}\bigl(c_{sv,t}^{Y} \mathbbm{1}_{\mathcal{B}_{t-1}}\bigr)} \\[10pt]
\frac{1}{\mathbb{E}\bigl(c_{sv,t}^{XY} \mathbbm{1}_{\mathcal{B}_{t-1}}\bigr) \mathbb{E}\bigl(c_{sv,t}^{X} \mathbbm{1}_{\mathcal{B}_{t-1}}\bigr)} & \frac{1}{\mathbb{E} \bigl(c_{sv,t}^{XY} \mathbbm{1}_{\mathcal{B}_{t-1}}\bigr)^{2}} & \frac{1}{\mathbb{E}\bigl(c_{sv,t}^{XY} \mathbbm{1}_{\mathcal{B}_{t-1}}\bigr) \mathbb{E}\bigl(c_{sv,t}^{Y} \mathbbm{1}_{\mathcal{B}_{t-1}}\bigr)} \\[10pt]
\frac{1}{\mathbb{E}\bigl(c_{sv,t}^{Y} \mathbbm{1}_{\mathcal{B}_{t-1}}\bigr) \mathbb{E}\bigl(c_{sv,t}^{X} \mathbbm{1}_{\mathcal{B}_{t-1}}\bigr)} & \frac{1}{\mathbb{E}\bigl(c_{sv,t}^{Y} \mathbbm{1}_{\mathcal{B}_{t-1}}\bigr) \mathbb{E}\bigl(c_{sv,t}^{XY} \mathbbm{1}_{\mathcal{B}_{t-1}}\bigr)} & \frac{1}{\mathbb{E} \bigl(c_{sv,t}^{Y} \mathbbm{1}_{\mathcal{B}_{t-1}}\bigr)^{2}}
\end{bmatrix}} \nonumber
\\
&\quad\quad{}\odot\sum_{h=- \infty}^{\infty} %
\begin{bmatrix} v_{\kappa, \tau}^{X, X, \mathcal{B}}(h) &
v_{\kappa, \tau}^{X,XY, \mathcal{B}}(h) & v_{\kappa, \tau}^{X,Y, \mathcal{B}}(h) \\[6pt]
v_{\kappa, \tau}^{XY,X, \mathcal{B}}(h) & v_{\kappa, \tau}^{XY, XY, \mathcal{B}}(h) & v_{\kappa, \tau}^{XY,Y, \mathcal{B}}(h) \\[6pt]
v_{\kappa, \tau}^{Y,X, \mathcal{B}}(h) & v_{\kappa, \tau}^{Y,XY, \mathcal{B}}(h) & v_{\kappa, \tau}^{Y, Y, \mathcal{B}}(h) \end{bmatrix} %
.\label{eq67}
\end{align}
Here, with $Z_{1}, Z_{2} \in\{ X, Y, XY \}$,
\begin{equation}
v_{\kappa, \tau}^{Z_{1}, Z_{2}, \mathcal{B}}(h) = \operatorname{cov}\bigl(A_{1, \kappa}^{Z_{1}, \mathcal{B}}, A_{1, \tau+h}^{Z_{2}, \mathcal{B}}\bigr), \label{eq68}
\end{equation}
for $\kappa, \tau\in[0,1]$, and
\begin{equation}
A_{t, \kappa}^{Z, \mathcal{B}} = \mathbbm{1}_{\mathcal{B}_{t-1}} \cdot\biggl(c_{t-1+ \kappa}^{Z} - c_{u, \kappa}^{Z} \int_{0}^{1}c_{t-1+s}^{Z} \,\mathrm{d}s\biggr). \label{eq69}
\end{equation}
Thus, we can proceed as above to construct both point estimates of $k_{u,t}$ and the test statistic. We omit a formal proof of this result, as it follows directly from the above.

\section{Small sample comparisons}
\label{section:simulation}

In the above, we developed a procedure to detect diurnal variation in a correlation process. We continue with a Monte Carlo exploration to gauge the finite sample properties of the proposed test statistic in a controlled environment.

We simulate a bivariate jump-diffusion process on the time interval $[0,T]$. It has a continuous part, which is given by
\begin{align}
\begin{aligned} \mathrm{d}X_{t}^{\mathrm{c}} &= \sigma
_{t}^{X} \,\mathrm{d} W_{t}^{X}, \\
\mathrm{d}Y_{t}^{\mathrm{c}} &= \sigma_{t}^{Y} \bigl( \rho_{t} \,\mathrm{d}W_{t}^{X} + \sqrt{1 - \rho_{t}^{2}} \,\mathrm{d}W_{t}^{Y} \bigr), \end{aligned} \label{eq70}
\end{align}
where $W_{t}^{\wp}$ is a standard Brownian motion.\footnote{Throughout this section, the driving stochastic processes are assumed to be mutually independent, unless explicitly stated otherwise.} This implies a conditional spot covariance $\mathbb{E} ( \mathrm{d}X_{t}^{\mathrm{c}} \,\mathrm{d} Y_{t}^{\mathrm{c}} \mid\mathcal{F}_{t} ) = \sigma_{t}^{X} \sigma_{t}^{Y} \rho_{t} \,\mathrm{d}t$ with correlation $\rho_{t}$.

The idiosyncratic volatility $\sigma_{t}^{\wp} = \sigma_{sv,t}^{\wp} \sigma_{u,t}$ is modeled as
\begin{align}
\label{equation:idiosyncratic-volatility} %
\begin{aligned} \mathrm{d}c_{sv,t}^{\wp}
&= \lambda(c_{0} - c_{sv,t}^{\wp})\,\mathrm{d}t + \xi\sqrt{c_{sv,t}^{\wp}} \,\mathrm{d}B_{t}^{\wp}, \\
\sigma_{u,t} &= \sqrt{C + A \bigl\llvert t - \lfloor t \rfloor- 0.5 \bigr\rrvert}, \end{aligned} 
\end{align}
where $c_{sv,t}^{\wp} \equiv( \sigma_{sv,t}^{\wp})^{2}$.

$\sigma_{sv,t}$ has a \citet{heston:93a}-type dynamic. As in \citet{christensen-thyrsgaard-veliyev:19a}, we set $\lambda= 0.05$, $c_{0} = 1$, and $\xi= 0.2$. We allow for a leverage effect by taking $\operatorname{corr}( \mathrm{d}W_{t}^{\wp}, \mathrm{d}B_{t}^{\wp}) = -\sqrt{0.5}$. Furthermore, in line with our empirical work the intraday volatility curve is V-shaped. We take $C = 0.5$ and $A = 2.0$, which renders volatility close to twice as large at the start and end of the unit interval as in the middle.\footnote{We also inspected a superposition of exponential functions: $\sigma_{u,t} = C + A e^{-a_{1}t} + B e^{-a_{2} (1-t)}$, where $A = 0.75$, $B = 0.25$, $C = 0.88929198$, and $a_{1} = a_{2} = 10$ (e.g., \cite{andersen-dobrev-schaumburg:12a,hasbrouck:99a}). The odd value of $C$ is such that $\int_{0}^{1} \sigma_{u,t}^{2} \,\mathrm{d}t = 1$. This delivers an inverse J-shaped curve, which agrees better with panel A of Figure~\ref{figure:rho_d} in our empirical application. However, the results are basically unchanged compared to those we report here and are available upon request.}

As required by Assumption~\ref{ass(C1)}, we decompose $\rho_{t} = \rho_{sc,t} k_{u,t}$, where the diurnal correlation component $k_{u,t}$ is an affine function of $t$:
\begin{equation}
\label{equation:diurnal-correlation} k_{u,t} = a + b \bigl(t - \lfloor t \rfloor\bigr).
\end{equation}
We assume that $b = 2(1-a)$.\footnote{Taken together, the functional forms of $\sigma_{u,t}$ and $k_{u,t}$ imply that $\int_{0}^{1} \sigma_{u,t}^{2} \,\mathrm{d}t = \int_{0}^{1} k_{u,t} \,\mathrm{d}t = \int_{0}^{1} c_{u,t}^{XY} \,\mathrm{d}t = 1$.} As such, the null hypothesis of no diurnal variation in $\rho_{t}$ is equivalent to the restriction $\mathcal{H}_{0}: a = 1$, whereas the alternative is $\mathcal{H}_{a}: a \neq1$. We examine $a = (1.00, 0.95, \dots, 0.80)$. Apart from being convenient, the nondecreasing linear form is also a decent description of the diurnal pattern observed in the correlation processes investigated in Section~\ref{section:empirical}. Our parametric model further fixes $k_{u,0.5} = 1$, which is consistent with prevailing evidence in panel B of Figure~\ref{figure:rho_d} in that section. The domain of $a$ is also shown in the figure. The lowest value $a = 0.8$---or $b = 0.4$---is small relative to the slope $\hat{b} = 0.8062$ estimated from that dataset, so our results should be conservative.

The stochastic correlation process follows:
\begin{equation}
\label{equation:stochastic-correlation} \frac{\mathrm{d} \rho_{sc,t}}{1 - \rho
_{sc,t}^{2}} = \kappa( \rho- \rho_{sc,t}) \,\mathrm{d}t + \sigma\,\mathrm{d} \tilde{B}_{t}, 
\end{equation}
with $\rho_{sc,0} \in(-1,1)$.

\begin{figure}[t!]
\caption{Illustration of stochastic correlation process.}
\label{figure:scp}
\begin{center}
\begin{tabular}{cc}
{\small{Panel A: Stationary density of $\rho_{sc,t}$}} & {\small{Panel B: Sample path of $\rho_{sc,t}$ and $\rho_{t}$.}} \\
\includegraphics[height=8cm,width=0.48\textwidth]{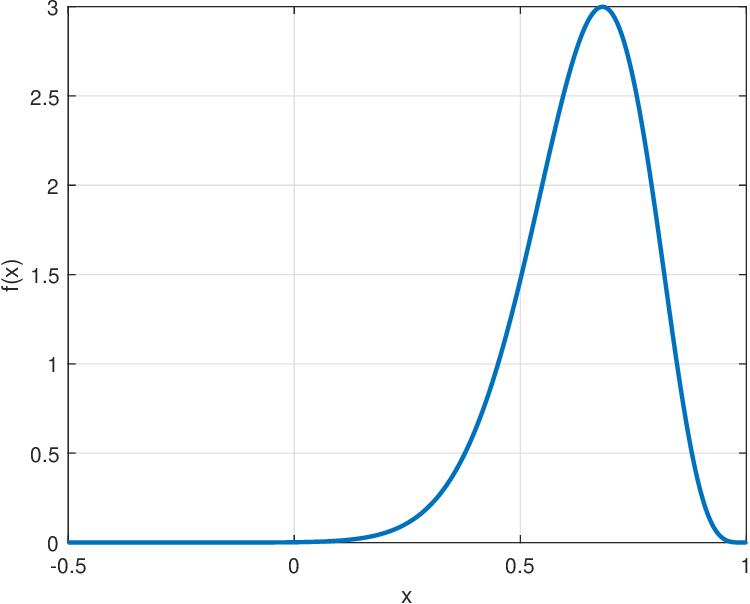} &
\includegraphics[height=8cm,width=0.48\textwidth]{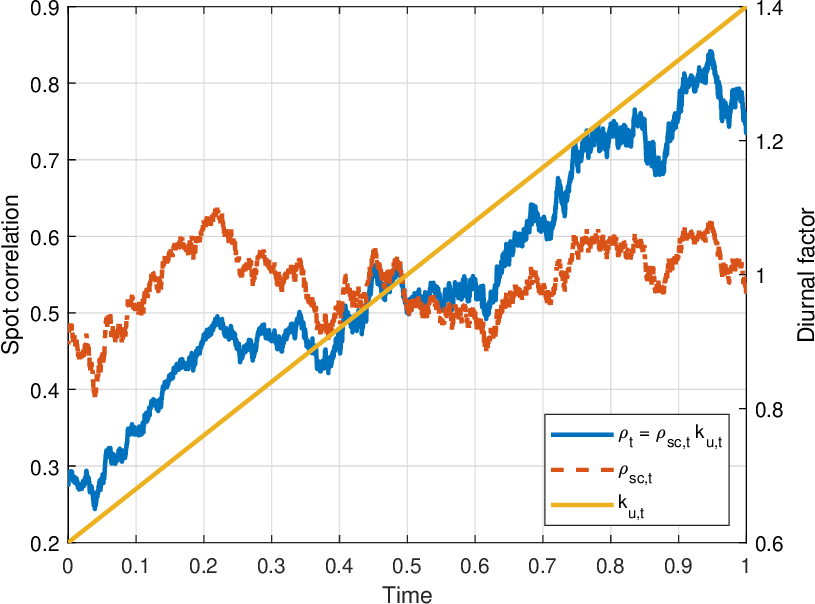}
\end{tabular}
\begin{scriptsize}
\parbox{\textwidth}{\emph{Note.} In panel A, we plot the stationary distribution of $\rho_{sc,t}$ implied by the stochastic correlation model in \eqref{equation:stochastic-correlation}. The parameter vector is $(\kappa, \rho, \sigma) = (1.5,0.6,0.3)$. In panel B, we show a sample path of this process with $t \in[0,1]$ and $\mathrm{d}t = 1/23{,}400$. We further plot the diurnal correlation function $k_{u,t}$ from \eqref{equation:diurnal-correlation} with $(a,b) = (0.6,0.8)$, which together form the spot correlation $\rho_{t} = \rho_{sc,t} k_{u,t}$.}
\end{scriptsize}
\end{center}
\end{figure}

The above SDE can be constructed via a Fisher transformation of $\rho_{sc,t}$ (e.g., \cite{teng-ehrhardt-gunther:16a}): $P_{sc,t} = \arctanh( \rho_{sc,t}) = \frac{1}{2} \ln( \frac{1 + \rho_{sc,t} }{1 - \rho_{sc,t}} )$. Suppose $P_{sc,t}$ is a modified Gaussian Ornstein--Uhlenbeck process $\mathrm{d} P_{sc,t} = - \tilde{\kappa} ( \tanh(P_{sc,t} ) - \tilde{\rho} ) \,\mathrm{d}t + \tilde{\sigma} \,\mathrm{d} \tilde{B}_{t}$ with $\tilde{\kappa}, \tilde{\sigma} > 0$ and $\tilde{\rho} \in(-1,1)$. An application of It\^{o}'s lemma to the inverse $\rho_{sc,t} = \tanh(P_{sc,t}) = \frac{\exp(2P_{sc,t})-1}{\exp(2P_{sc,t})+1}$ then delivers \eqref{equation:stochastic-correlation} with $\sigma^{2} = \tilde{\sigma}^{2}$, $\kappa= \tilde{\kappa} + \tilde{\sigma}^{2}$ and $\rho= \frac{\tilde{\kappa}}{\tilde{\kappa} + \tilde{\sigma}^{2}} \tilde{\rho}$. If the parameters satisfy the Feller-type condition $\kappa> \frac{\sigma^{2}}{1 \pm\rho}$, $\rho_{sc,t}$ is stationary with state space $(-1,1)$, that is, the probability mass at the boundary goes sufficiently fast to zero as $\rho_{sc,t} \rightarrow\pm1$, such that the barriers are not attainable (nor attractive). This is suitable for a dynamic correlation model.

We set $\kappa= 1.5$, $\sigma= 0.3$, and $\rho= (0.2, 0.4, 0.6)$, so that the above condition is fulfilled in every scenario. This implies an unconditional mean correlation of $\mathbb{E}(\rho_{sc,t}) = \frac{\kappa}{\kappa- \sigma^{2}} \rho$, or about $(0.213,0.426,0.638)$. Our choices of $\rho$ incur a weak to strong positive association between $X$ and $Y$ in line with the descriptive statistics of the unconditional sample correlation coefficient presented in Table~\ref{table:djia-descriptive} in the empirical investigation. On the one hand, the intermediate and largest value of $\rho$ are in line with what we observe there, whereas the lowest value of $\rho$ is beneath the first quartile of the sample correlation between every asset pair. On the other hand, the former rarely lead to a negative spot correlation, whereas the unconditional distribution of the latter has a nontrivial amount of probability mass below zero (i.e., the chance of observing a negative correlation is around 0.1 for $\rho= 0.2$, whereas it is close to zero otherwise). This is intended to show the impact of weak correlation on our test statistic, since in this case $\rho_{sc,t}$ can linger about zero. Moreover, examining a smaller numeric value of $\rho$ is relevant for other asset classes. In any case, we draw the initial condition $\rho_{sc,0}$ at random from the stationary distribution of $\rho_{sc,t}$, which is illustrated in panel A of Figure~\ref{figure:scp} for $\rho= 0.6$.\footnote{The stationary density is given by $f_{\rho}(x) = \frac{m}{2^{c}}(1+x)^{a+b}(1-x)^{a-b}$, for $x \in(-1,1)$, where $a = \frac{\kappa- 2 \sigma^{2}}{\sigma^{2}}$, $b = \frac{\kappa\rho}{\sigma^{2}}$, and $c = \frac{\kappa}{\sigma^{2}}$. $m$ is a normalizing constant, such that $\int_{-1}^{1} f_{\rho}(x) \,\mathrm{d}x = 1$, which can be expressed analytically via the hypergeometric and gamma function.} A~realization of the full-blown continuous-time dynamics of $\rho_{sc,t}$ in this case is shown in panel B.\footnote{We employ full truncation to enforce that $\rho_{t}$ remains in $(-1,1)$.}

We add a pure-jump component to the continuous sample path of log-price, which is simulated as a compound Poisson process:
\begin{equation*}
\mathrm{d} J_{t}^{\wp} = q_{t}^{\wp} \,\mathrm{d}N_{t}^{\wp},
\end{equation*}
where $q_{t}^{\wp}$ is the jump size at time $t$ and $N_{t}^{\wp}$ is a Poisson process with intensity $\lambda_{J}$. We draw $q_{t}^{\wp} \sim N(0, \sigma_{J}^{2})$ with $\sigma_{J} = \sqrt{\frac{1}{\lambda_{J}} \frac{p_{J}}{1-p_{J}} c_{0}}$, so the quadratic jump variation is proportional to the average diffusive variance. $p_{J}$ controls how much of the second-order variation in the log-price process is due to the jump component. We assume $\lambda_{J} = 0.2$ and $p_{J} = 0.1$, such that a jump is observed every fifth day, while accounting for 10\% of the quadratic variation, on average. This conforms with empirical evidence on jump testing (e.g., \cite{ait-sahalia-jacod-li:12a,ait-sahalia-xiu:16a,bajgrowicz-scaillet-treccani:16a}).

We discretize the system with an Euler scheme and a baseline step of $\mathrm{d}t = 1/23{,}400$. This represents the ``continuous-time'' foundation from which we extract a coarser sample of size $n = $ 26, 39, 78, 390, 780, 1{,}560, and 4{,}680, equidistant log-price increments over each interval $[t-1,t]$, for $t = 1, \dots, T$ and $T = $ 5, 22, and 66. The former can be interpreted as discretely sampling a process every 900, 600, 300, 60, 30, 15, and 5 seconds, while the latter corresponds to observing such high-frequency data over a week, month, and quarter.\footnote{In practice, recording a price at 5- or 15-second intervals induces a nontrivial amount of microstructure noise in the estimation. Hence, $n = 1{,}560$ or $n = 4{,}680$ is a much larger sampling frequency than we feel comfortable with in the empirical application. It is mainly added to illustrate the convergence properties of our test.}

In practice, high-frequency estimation of the correlation between asset returns is known to diminish as the sampling frequency goes up, because the observed data are asynchronous, that is, lack alignment in time (e.g., \cite{epps:79a}). To gauge the importance of this, we also consider a scenario, where $X$ and $Y$ are observed at irregularly spaced sampling times. We simulate the number of observations on day $t$ as $n_{t}^{\wp} \sim\operatorname{Poi}( \lambda_{n})$, where $\lambda_{n} = 4{,}680$, such that the average daily number of data points is equal to the largest value of $n$ from the equidistant setting.\footnote{To put this in perspective, the choice of $\lambda_{n}$ is merely a quarter of the average daily number of trades in the least liquid asset considered in our empirical application (17{,}920 for TRV, as shown in Table~\ref{table:djia-descriptive}), so it exacerbates the degree of asynchronicity we encounter there.} Conditional on $n_{t}$, we select the observation grid as a random sample without replacement of size $n_{t}$ from $0, \mathrm{d}t, \dots, 1$ and proceed as above, but using previous-tick interpolation to construct an equidistant (and synchronous) sample of size $n$.\footnote{The refresh time approach of \cite{barndorff-nielsen-hansen-lunde-shephard:11a} was another option.}

A total of 10{,}000 replications are made. As described in Section~\ref{section:correlation}, in each simulation we divide the available high-frequency data $(\Delta_{(t-1)n+i}^{n} X)_{i = 1, \dots,n \text{ and } t=1, \dots, T}$ and $(\Delta_{(t-1)n+i}^{n} Y)_{i = 1, \dots,n \text{ and } t=1, \dots, T}$ into nonoverlapping subsets of size $k_{n} = $ 13, 13, 26, 130, 195, 390, and 936, corresponding to $n/k_{n} = $ 2, 3, 3, 3, 4, 4, and 5, so the number of blocks is rising slowly with $n$, as required by the rate condition from Theorem~\ref{theorem:functional-clt}. Indeed, because the testing procedure explores the properties of the covariance process, a casual robustness check suggests that it is preferable with a smaller number of blocks consisting of a larger number of increments, than vice versa, as it is important to get a good approximation of its intraday dynamic.

We calculate the jump-robust bipower variation and relieve log-returns from the jump component by blockwise truncation of increments that are numerically above $v_{n} = q \sqrt{\operatorname{BV}}n^{- \varpi}$ with $q = 5$ and $\varpi= 0.49$. Hence, our procedure labels a log-return as a jump if it exceeds about five diffusive standard deviations.

To compute the test statistic, we implement the HAC estimator of the asymptotic covariance matrix with a lag length $H_{T} = [T^{1/3}]$ and a Parzen kernel to ensure positive semidefiniteness.\footnote{We also experimented with a Bartlett kernel, but that did not lead to substantial changes.} The results are robust to the concrete choice of lag length, so long as it is not exceedingly large. To evaluate the test statistic, we draw $9{,}999$ realizations of $\widehat{\mathcal{Z}}$ and extract an appropriate quantile from the induced empirical distribution function.

\begin{table}
\tabcolsep=0pt

\caption{Rejection rate of the test statistic for diurnal variation in the
correlation process at the $\alpha= 0.01$ significance level and with $\rho= 0.60$.}
\label{table:sim-diurnal-correlation-t=N-r=0.60-q=0.01.tex}
\begin{tabular*}{\textwidth}{@{\extracolsep{\fill}}D{.}{.}{5.0}D{.}{.}{3.0}cccccccccc@{}}
\hline
\multicolumn{1}{l}{\multirow{3}{*}[-17pt]{$n$}}& \multicolumn{1}{c}{\multirow{3}{*}[-17pt]{$k_{n}$}} &
\multicolumn{5}{c}{Equidistant Sampling} & \multicolumn{5}{c}{Irregular Sampling}
\\
\cline{3-7}\cline{8-12}
& & \multicolumn{5}{c}{$a = $} & \multicolumn{5}{c}{$a = $} \\
\cline{3-7}\cline{8-12}
& & 1.000 & 0.950 & 0.900 & 0.850 & 0.800 & 1.000 & 0.950 & 0.900 & 0.850 & 0.800 \\
\hline
\multicolumn{12}{l}{Panel A: $T = 5$}\\
26 & 13 & 0.080 & 0.091 & 0.134 & 0.186 & 0.254 & 0.076 & 0.088 & 0.128 & 0.176 & 0.241 \\
39 & 13 & 0.054 & 0.063 & 0.101 & 0.156 & 0.229 & 0.052 & 0.064 & 0.100 & 0.151 & 0.216 \\
78 & 26 & 0.056 & 0.087 & 0.178 & 0.295 & 0.422 & 0.056 & 0.090 & 0.165 & 0.273 & 0.402 \\
390 & 130 & 0.066 & 0.248 & 0.565 & 0.775 & 0.881 & 0.057 & 0.180 & 0.467 & 0.690 & 0.833 \\
780 & 195 & 0.045 & 0.327 & 0.699 & 0.870 & 0.942 & 0.048 & 0.158 & 0.478 & 0.744 & 0.874 \\
1{,}560 & 390 & 0.043 & 0.504 & 0.849 & 0.951 & 0.982 & 0.048 & 0.154 & 0.487 & 0.772 & 0.904 \\
4{,}680 & 936 & 0.035 & 0.745 & 0.955 & 0.990 & 0.996 & 0.032 & 0.076 & 0.245 & 0.539 & 0.779 \\[6pt]
\multicolumn{12}{l}{Panel B: $T = 22$}\\
26 & 13 & 0.023 & 0.047 & 0.124 & 0.254 & 0.390 & 0.022 & 0.052 & 0.120 & 0.256 & 0.387 \\
39 & 13 & 0.015 & 0.038 & 0.110 & 0.253 & 0.417 & 0.017 & 0.036 & 0.112 & 0.248 & 0.407 \\
78 & 26 & 0.015 & 0.090 & 0.328 & 0.587 & 0.780 & 0.017 & 0.086 & 0.313 & 0.581 & 0.763 \\
390 & 130 & 0.020 & 0.536 & 0.881 & 0.964 & 0.988 & 0.020 & 0.423 & 0.832 & 0.948 & 0.982 \\
780 & 195 & 0.015 & 0.694 & 0.948 & 0.985 & 0.996 & 0.015 & 0.467 & 0.889 & 0.971 & 0.992 \\
1{,}560 & 390 & 0.015 & 0.855 & 0.984 & 0.996 & 0.998 & 0.015 & 0.469 & 0.918 & 0.983 & 0.995 \\
4{,}680 & 936 & 0.011 & 0.964 & 0.996 & 0.999 & 0.999 & 0.013 & 0.224 & 0.817 & 0.964 & 0.992 \\[6pt]
\multicolumn{12}{l}{Panel C: $T = 66$}\\
26 & 13 & 0.014 & 0.063 & 0.237 & 0.499 & 0.706 & 0.015 & 0.062 & 0.247 & 0.501 & 0.717 \\
39 & 13 & 0.010 & 0.049 & 0.250 & 0.544 & 0.767 & 0.010 & 0.049 & 0.251 & 0.540 & 0.763 \\
78 & 26 & 0.009 & 0.213 & 0.673 & 0.893 & 0.964 & 0.013 & 0.201 & 0.679 & 0.891 & 0.963 \\
390 & 130 & 0.012 & 0.831 & 0.980 & 0.995 & 0.998 & 0.014 & 0.760 & 0.973 & 0.995 & 0.998 \\
780 & 195 & 0.011 & 0.922 & 0.994 & 0.998 & 0.999 & 0.014 & 0.830 & 0.987 & 0.997 & 0.999 \\
1{,}560 & 390 & 0.011 & 0.977 & 0.999 & 0.999 & 1.000 & 0.011 & 0.871 & 0.993 & 0.999 & 0.999 \\
4{,}680 & 936 & 0.009 & 0.996 & 1.000 & 1.000 & 1.000 & 0.010 & 0.714 & 0.986 & 0.998 & 0.999 \\
\hline\end{tabular*}
\begin{tabnotes}
\legend{We simulate a bivariate jump-diffusion model with diurnal variation in the correlation coefficient, such that
$\rho_{t} = \rho_{sc,t} k_{u,t}$, where $\rho_{sc,t}$ is a stochastic process and $k_{u,t} = a + bt$ with $b = 2(1-a)$ captures the deterministic component. The hypothesis $\mathcal{H}_{0}: \int_{0}^{1} ( k_{u,t} - 1)^{2} \,\mathrm{d}t = 0$ is tested against $\mathcal{H}_{a}: \int_{0}^{1} ( k_{u,t} - 1)^{2} \,\mathrm{d}t \neq0$. In the model, the null is equivalent to $a = 1$, whereas the alternative corresponds to $a \neq1$. The table reports rejection rates of the test statistic from \eqref{equation:test-statistic} at significance level $\alpha= 0.01$. $n$ is the number of intraday observations over a sample period of $T$ days, while $k_{n}$ is the number of log-price increments used to compute the blockwise realized covariance estimator.}
\end{tabnotes}%
\end{table}

\begin{table}
\tabcolsep=0pt

\caption{Rejection rate of the test statistic for diurnal variation in the
correlation process at the $\alpha= 0.01$ significance level and with $\rho= 0.40$.}
\label{table:sim-diurnal-correlation-t=N-r=0.40-q=0.01.tex}
\begin{tabular*}{\textwidth}{@{\extracolsep{\fill}}D{.}{.}{5.0}D{.}{.}{3.0}cccccccccc@{}}
\hline
\multicolumn{1}{l}{\multirow{3}{*}[-17pt]{$n$}}& \multicolumn{1}{c}{\multirow{3}{*}[-17pt]{$k_{n}$}} &
\multicolumn{5}{c}{Equidistant Sampling} & \multicolumn{5}{c}{Irregular Sampling}
\\
\cline{3-7}\cline{8-12}
& & \multicolumn{5}{c}{$a = $} & \multicolumn{5}{c}{$a = $} \\
\cline{3-7}\cline{8-12}
& & 1.000 & 0.950 & 0.900 & 0.850 & 0.800 & 1.000 & 0.950 & 0.900 & 0.850 & 0.800 \\
\hline
\multicolumn{12}{l}{Panel A: $T = 5$}\\
26 & 13 & 0.071 & 0.074 & 0.084 & 0.098 & 0.125 & 0.068 & 0.071 & 0.084 & 0.100 & 0.123 \\
39 & 13 & 0.051 & 0.049 & 0.063 & 0.078 & 0.104 & 0.047 & 0.053 & 0.059 & 0.080 & 0.095 \\
78 & 26 & 0.050 & 0.063 & 0.087 & 0.120 & 0.170 & 0.056 & 0.066 & 0.085 & 0.117 & 0.167 \\
390 & 130 & 0.066 & 0.099 & 0.233 & 0.389 & 0.529 & 0.056 & 0.095 & 0.198 & 0.328 & 0.475 \\
780 & 195 & 0.043 & 0.111 & 0.304 & 0.500 & 0.649 & 0.045 & 0.079 & 0.194 & 0.369 & 0.528 \\
1{,}560 & 390 & 0.042 & 0.177 & 0.468 & 0.670 & 0.800 & 0.047 & 0.086 & 0.226 & 0.417 & 0.604 \\
4{,}680 & 936 & 0.033 & 0.335 & 0.693 & 0.846 & 0.916 & 0.031 & 0.049 & 0.118 & 0.265 & 0.450 \\[6pt]
\multicolumn{12}{l}{Panel B: $T = 22$} \\
26 & 13 & 0.022 & 0.029 & 0.053 & 0.098 & 0.163 & 0.022 & 0.032 & 0.051 & 0.098 & 0.154 \\
39 & 13 & 0.017 & 0.023 & 0.045 & 0.094 & 0.165 & 0.018 & 0.025 & 0.044 & 0.087 & 0.150 \\
78 & 26 & 0.020 & 0.036 & 0.109 & 0.234 & 0.387 & 0.019 & 0.036 & 0.099 & 0.230 & 0.363 \\
390 & 130 & 0.021 & 0.172 & 0.513 & 0.716 & 0.838 & 0.019 & 0.133 & 0.440 & 0.668 & 0.801 \\
780 & 195 & 0.015 & 0.262 & 0.654 & 0.829 & 0.904 & 0.016 & 0.152 & 0.516 & 0.746 & 0.860 \\
1{,}560 & 390 & 0.013 & 0.448 & 0.804 & 0.911 & 0.950 & 0.016 & 0.178 & 0.600 & 0.809 & 0.898 \\
4{,}680 & 936 & 0.011 & 0.710 & 0.921 & 0.962 & 0.977 & 0.013 & 0.092 & 0.453 & 0.730 & 0.861 \\[6pt]
\multicolumn{12}{l}{Panel C: $T = 66$} \\
26 & 13 & 0.013 & 0.031 & 0.095 & 0.216 & 0.360 & 0.016 & 0.032 & 0.098 & 0.216 & 0.354 \\
39 & 13 & 0.011 & 0.025 & 0.091 & 0.231 & 0.402 & 0.012 & 0.025 & 0.096 & 0.227 & 0.400 \\
78 & 26 & 0.012 & 0.067 & 0.283 & 0.528 & 0.698 & 0.011 & 0.064 & 0.273 & 0.523 & 0.699 \\
390 & 130 & 0.011 & 0.407 & 0.783 & 0.904 & 0.943 & 0.012 & 0.326 & 0.747 & 0.890 & 0.939 \\
780 & 195 & 0.011 & 0.572 & 0.881 & 0.946 & 0.969 & 0.012 & 0.415 & 0.830 & 0.925 & 0.960 \\
1{,}560 & 390 & 0.012 & 0.757 & 0.940 & 0.971 & 0.984 & 0.011 & 0.483 & 0.876 & 0.946 & 0.970 \\
4{,}680 & 936 & 0.009 & 0.903 & 0.975 & 0.985 & 0.992 & 0.009 & 0.335 & 0.821 & 0.929 & 0.962 \\
\hline\end{tabular*}
\begin{tabnotes}
\legend{We simulate a bivariate jump-diffusion model with diurnal variation in the correlation coefficient, such that
$\rho_{t} = \rho_{sc,t} k_{u,t}$, where $\rho_{sc,t}$ is a stochastic process and $k_{u,t} = a + bt$ with $b = 2(1-a)$ captures the deterministic component. The hypothesis $\mathcal{H}_{0}: \int_{0}^{1} ( k_{u,t} - 1)^{2} \,\mathrm{d}t = 0$ is tested against $\mathcal{H}_{a}: \int_{0}^{1} ( k_{u,t} - 1)^{2} \,\mathrm{d}t \neq0$. In the model, the null is equivalent to $a = 1$, whereas the alternative corresponds to $a \neq1$. The table reports rejection rates of the test statistic from \eqref{equation:test-statistic} at significance level $\alpha= 0.01$. $n$ is the number of intraday observations over a sample period of $T$ days, while $k_{n}$ is the number of log-price increments used to compute the blockwise realized covariance estimator.}
\end{tabnotes}%
\end{table}

\begin{table}
\tabcolsep=0pt

\caption{Rejection rate of the test statistic for diurnal variation in the
correlation process at the $\alpha= 0.01$ significance level and with $\rho= 0.20$.}
\label{table:sim-diurnal-correlation-t=N-r=0.20-q=0.01.tex}
\begin{tabular*}{\textwidth}{@{\extracolsep{\fill}}D{.}{.}{5.0}D{.}{.}{3.0}cccccccccc@{}}
\hline
\multicolumn{1}{l}{\multirow{3}{*}[-17pt]{$n$}}& \multicolumn{1}{c}{\multirow{3}{*}[-17pt]{$k_{n}$}} &
\multicolumn{5}{c}{Equidistant Sampling} & \multicolumn{5}{c}{Irregular Sampling}
\\
\cline{3-7}\cline{8-12}
& & \multicolumn{5}{c}{$a = $} & \multicolumn{5}{c}{$a = $} \\
\cline{3-7}\cline{8-12}
& & 1.000 & 0.950 & 0.900 & 0.850 & 0.800 & 1.000 & 0.950 & 0.900 & 0.850 & 0.800 \\
\hline
\multicolumn{12}{l}{Panel A: $T = 5$}\\
26 & 13 & 0.050 & 0.049 & 0.052 & 0.056 & 0.065 & 0.045 & 0.049 & 0.052 & 0.052 & 0.061 \\
39 & 13 & 0.031 & 0.031 & 0.036 & 0.041 & 0.050 & 0.032 & 0.032 & 0.037 & 0.043 & 0.045 \\
78 & 26 & 0.037 & 0.042 & 0.048 & 0.057 & 0.069 & 0.041 & 0.044 & 0.048 & 0.057 & 0.066 \\
390 & 130 & 0.054 & 0.059 & 0.093 & 0.148 & 0.206 & 0.049 & 0.062 & 0.090 & 0.126 & 0.179 \\
780 & 195 & 0.038 & 0.053 & 0.106 & 0.188 & 0.271 & 0.038 & 0.045 & 0.079 & 0.132 & 0.201 \\
1{,}560 & 390 & 0.041 & 0.069 & 0.178 & 0.293 & 0.410 & 0.040 & 0.048 & 0.090 & 0.161 & 0.248 \\
4{,}680 & 936 & 0.032 & 0.108 & 0.301 & 0.465 & 0.591 & 0.030 & 0.034 & 0.053 & 0.109 & 0.167 \\[6pt]
\multicolumn{12}{l}{Panel B: $T = 22$}\\
26 & 13 & 0.013 & 0.015 & 0.020 & 0.033 & 0.049 & 0.014 & 0.016 & 0.022 & 0.034 & 0.045 \\
39 & 13 & 0.012 & 0.013 & 0.019 & 0.031 & 0.041 & 0.012 & 0.012 & 0.020 & 0.028 & 0.040 \\
78 & 26 & 0.015 & 0.018 & 0.033 & 0.066 & 0.112 & 0.014 & 0.019 & 0.030 & 0.067 & 0.104 \\
390 & 130 & 0.017 & 0.047 & 0.159 & 0.304 & 0.446 & 0.018 & 0.041 & 0.131 & 0.265 & 0.383 \\
780 & 195 & 0.011 & 0.060 & 0.247 & 0.413 & 0.556 & 0.013 & 0.040 & 0.165 & 0.334 & 0.468 \\
1{,}560 & 390 & 0.013 & 0.130 & 0.390 & 0.574 & 0.678 & 0.012 & 0.052 & 0.223 & 0.406 & 0.537 \\
4{,}680 & 936 & 0.009 & 0.284 & 0.593 & 0.721 & 0.798 & 0.010 & 0.032 & 0.144 & 0.326 & 0.469 \\[6pt]
\multicolumn{12}{l}{Panel C: $T = 66$}\\
26 & 13 & 0.008 & 0.015 & 0.030 & 0.061 & 0.108 & 0.011 & 0.014 & 0.032 & 0.063 & 0.110 \\
39 & 13 & 0.008 & 0.014 & 0.030 & 0.065 & 0.121 & 0.009 & 0.013 & 0.031 & 0.065 & 0.120 \\
78 & 26 & 0.011 & 0.023 & 0.070 & 0.165 & 0.275 & 0.010 & 0.024 & 0.066 & 0.159 & 0.270 \\
390 & 130 & 0.009 & 0.108 & 0.353 & 0.532 & 0.640 & 0.010 & 0.084 & 0.313 & 0.495 & 0.618 \\
780 & 195 & 0.008 & 0.176 & 0.479 & 0.643 & 0.729 & 0.010 & 0.114 & 0.397 & 0.589 & 0.686 \\
1{,}560 & 390 & 0.010 & 0.319 & 0.619 & 0.749 & 0.806 & 0.009 & 0.158 & 0.472 & 0.648 & 0.735 \\
4{,}680 & 936 & 0.009 & 0.533 & 0.764 & 0.844 & 0.875 & 0.006 & 0.099 & 0.397 & 0.601 & 0.705 \\
\hline
\end{tabular*}
\begin{tabnotes}
\legend{We simulate a bivariate jump-diffusion model with diurnal variation in the correlation coefficient, such that
$\rho_{t} = \rho_{sc,t} k_{u,t}$, where $\rho_{sc,t}$ is a stochastic process and $k_{u,t} = a + bt$ with $b = 2(1-a)$ captures the deterministic component. The hypothesis $\mathcal{H}_{0}: \int_{0}^{1} ( k_{u,t} - 1)^{2} \,\mathrm{d}t = 0$ is tested against $\mathcal{H}_{a}: \int_{0}^{1} ( k_{u,t} - 1)^{2} \,\mathrm{d}t \neq0$. In the model, the null is equivalent to $a = 1$, whereas the alternative corresponds to $a \neq1$. The table reports rejection rates of the test statistic from \eqref{equation:test-statistic} at significance level $\alpha= 0.01$. $n$ is the number of intraday observations over a sample period of $T$ days, while $k_{n}$ is the number of log-price increments used to compute the blockwise realized covariance estimator.}
\end{tabnotes}%
\end{table}

The outcome of the exercise is presented in Tables~\ref{table:sim-diurnal-correlation-t=N-r=0.60-q=0.01.tex}--\ref{table:sim-diurnal-correlation-t=N-r=0.20-q=0.01.tex}, which show rejection rates of the testing procedure at the $\alpha= 0.01$ level of significance.\footnote{The corresponding analysis at the $\alpha= 0.05$ and $\alpha= 0.10$ significance levels is reported in Appendix~\ref{appendix:monte-carlo}.} The various intraday sample sizes appear in rows and diurnal correlation slopes in columns, while the different values of $T$ are reported in panels A--C, respectively. In addition, the left-hand (right-hand) side of each table is for the equidistant (irregular) sampling scheme.

The column headings with $a = 1.00$ refer to the null hypothesis and we look at those to begin with. We observe that for $T = 5$ the test is somewhat oversized, as the rejection rates are higher than the nominal level. With such a small $T$, the time-averaged blockwise realized covariance is inevitably going to be a very crude measure of the associated time-of-day spot covariance, which introduces some distortion. At $T = 22$, the rejection rates have already settled around the anticipated value at the 1\% nominal level, but we still see a slight overrejection. The latter can arise from discrepancies between the sampling distribution of the test statistic for a finite number of blocks and that predicted by the asymptotic theory. Of course, it can potentially also be attributed to our choice of tuning parameters. By and large, however, the numbers line up with the asymptotic distribution theory under the null. We therefore leave the pursuit of more optimal tuning parameters to a future endeavor.

Moving to the right toward columns with $a \neq1$, which defines our alternative, we observe a monotonic rise in the rejection rates as $a$ gets smaller, which steepens the slope of the intraday correlation curve, and as the sample size increases (either $n$ or $T$). This is as prescribed by the asymptotic theory from Section~\ref{section:test-procedure}. Note that for commonly employed intraday sample sizes (e.g., $n = 78$ or $n = 390$) and a month's worth of high-frequency data (i.e., $T = 22$), the power is often rather good. This is compelling, since our naive configuration with a straight line understates the evolution of the nonlinear curve observed in practice.

To gauge the effect of changing $\rho$, that is, the average degree of asset return correlation, we note that a lower value leads to a decrease in the rejection rates. That is, weak correlation is detrimental to both the size and power of the test statistic. This effect is rather substantial for $\rho= 0.20$ compared to $\rho= 0.60$, but as expected we do observe a sustained and significant improvement with increasing $n$ and $T$ or a reduction in $a$.

Lastly, we inspect the robustness of the test statistic to random sampling times. Consistent with the analysis for varying $\rho$ in the previous paragraph, we learn that irregularly spaced data reduces the rejection rates relative to the equidistant setting. This can be ascribed to the Epps effect, which induces an attenuation bias in the estimated level of the correlation process. Indeed, the discrepancy gets more pronounced as the sampling frequency $n$ is increased relative to the intensity of the counting process $\lambda_{n}$, which causes a gradual worsening of the synchronization problem. However, whereas the drop in power remains present even with larger $n$ so long as we look at a small value of $T$, the effect is much less pronounced for data stretching over even a modest time period. This suggests that this problem should not be a big concern in practice.

In summary, the test statistic has acceptable size control and decent power in most of the settings that are relevant to our empirical application, which we turn to next.

\section{Empirical application}
\label{section:empirical}

We conduct an assessment of the presence of diurnal variation in the empirical correlation process by studying a vast dataset covering an extended time frame and a broad selection of companies from the large-cap segment of the US stock market.

\subsection{Data description}
\label{sec7.1}

At our disposal are high-frequency data from the members of the Dow Jones Industrial Average index, as of the August 31, 2020, recomposition. In addition, we include the SPDR (formerly known as Standard \& Poor's Depository Receipts) S\&P 500 trust, listed under the ticker symbol SPY. The latter is an exchange-traded fund that aims to replicate the total return of the S\&P 500 index (before expenses). Its price development is therefore representative of market-wide changes in the valuation of US equities.

We downloaded a time series of transaction and quotation data for each security from the \citet{nyse:26a} Trade and Quote (TAQ) database for the sample period January 4, 2010, to April 28, 2023. Prior to our investigation, we preprocessed the raw high-frequency data with a standard filtering algorithm to remove outliers (see, e.g., \cite{barndorff-nielsen-hansen-lunde-shephard:09a,christensen-oomen-podolskij:14a}).

The US stock market is open for trading from 9:30 a.m. to 4:00 p.m. on normal business days. However, on a regular basis most venues halt trading at an earlier time in observance of upcoming holidays. This is, for example, done before Independence Day, Thanksgiving, and Christmas Eve. In such instances, the trading session is shortened and the exchanges close at 1:00 p.m. As the diurnal correlation pattern on those days can be expected to deviate substantially from that on a regular business day with a usual trading schedule, we remove them from the sample. Furthermore, we purge the Flash Crash of May 6, 2010, due to its highly irregular volatility that exerts a disproportionate effect on our estimation procedure. As a result, the empirical investigation is based on the $T = 3{,}325$ days remaining in our sample.

In Table~\ref{table:djia-descriptive}, we present a list of ticker symbols and descriptive statistics of the associated high-frequency data.

\begin{sidewaystable}[p!]
\setlength{ \tabcolsep}{0.15cm}
\begin{center}
\caption{Descriptive statistics of TAQ high-frequency data.}
\label{table:djia-descriptive}
\vspace*{-0.25cm}
\begin{small}
\begin{tabular}{lrcccccccccccc}
\hline \hline
& & & & & \multicolumn{4}{c}{Versus SPY (point estimate)} & & \multicolumn{4}{c}{Versus rest (interquartile range)} \\ \cline{6-9} \cline{11-14}
Ticker & $N$ & $\bar{ \sigma}$ & H & & $\bar{ \rho}$ & $\hat{a}$ & $\hat{b}$ & $\hat{P}( \mathcal{N} > q_{1-\alpha/\#T})$ & & $\bar{ \rho}$ & $\hat{a}$ & $\hat{b}$ & $\hat{P}( \mathcal{N} > q_{1-\alpha/\#T})$ \\
\hline
AAPL & 279{,}133 & 0.214 & 0.026 &  & 0.701 & 0.876 & 0.248 & 0.531 &  & $[0.354; 0.457]$ & $[0.529; 0.687]$ & $[0.626; 0.942]$ & $[0.400; 0.506]$ \\
AMGN & 33{,}223 & 0.214 & 0.044 &  & 0.534 & 0.726 & 0.548 & 0.631 &  & $[0.331; 0.380]$ & $[0.524; 0.590]$ & $[0.820; 0.952]$ & $[0.375; 0.487]$ \\
AXP & 38{,}304 & 0.213 & 0.033 &  & 0.674 & 0.838 & 0.325 & 0.544 &  & $[0.383; 0.486]$ & $[0.552; 0.729]$ & $[0.542; 0.896]$ & $[0.425; 0.531]$ \\
BA & 64{,}796 & 0.268 & 0.030 &  & 0.594 & 0.837 & 0.325 & 0.531 &  & $[0.334; 0.423]$ & $[0.507; 0.736]$ & $[0.528; 0.986]$ & $[0.375; 0.469]$ \\
CAT & 41{,}780 & 0.233 & 0.020 &  & 0.668 & 0.820 & 0.359 & 0.675 &  & $[0.379; 0.488]$ & $[0.524; 0.729]$ & $[0.542; 0.952]$ & $[0.469; 0.544]$ \\
CRM & 46{,}594 & 0.281 & 0.025 &  & 0.603 & 0.829 & 0.342 & 0.606 &  & $[0.310; 0.414]$ & $[0.451; 0.683]$ & $[0.635; 1.098]$ & $[0.350; 0.438]$ \\
CSCO & 108{,}339 & 0.201 & 0.057 &  & 0.666 & 0.862 & 0.275 & 0.312 &  & $[0.392; 0.458]$ & $[0.609; 0.715]$ & $[0.571; 0.782]$ & $[0.344; 0.412]$ \\
CVX & 62{,}546 & 0.213 & 0.023 &  & 0.611 & 0.793 & 0.414 & 0.694 &  & $[0.351; 0.442]$ & $[0.502; 0.689]$ & $[0.622; 0.997]$ & $[0.438; 0.544]$ \\
DIS & 72{,}947 & 0.202 & 0.031 &  & 0.662 & 0.838 & 0.323 & 0.606 &  & $[0.374; 0.458]$ & $[0.542; 0.701]$ & $[0.599; 0.916]$ & $[0.419; 0.512]$ \\
DOW & 36{,}603 & 0.234 & 0.037 &  & 0.619 & 0.814 & 0.371 & 0.519 &  & $[0.361; 0.468]$ & $[0.531; 0.745]$ & $[0.510; 0.938]$ & $[0.375; 0.456]$ \\
GS & 36{,}885 & 0.231 & 0.031 &  & 0.648 & 0.835 & 0.330 & 0.594 &  & $[0.359; 0.466]$ & $[0.532; 0.713]$ & $[0.574; 0.937]$ & $[0.425; 0.525]$ \\
HD & 50{,}243 & 0.197 & 0.030 &  & 0.666 & 0.806 & 0.388 & 0.694 &  & $[0.407; 0.463]$ & $[0.572; 0.661]$ & $[0.679; 0.857]$ & $[0.475; 0.575]$ \\
HON & 29{,}750 & 0.191 & 0.034 &  & 0.704 & 0.849 & 0.303 & 0.662 &  & $[0.415; 0.498]$ & $[0.618; 0.740]$ & $[0.519; 0.764]$ & $[0.494; 0.575]$ \\
IBM & 39{,}641 & 0.169 & 0.029 &  & 0.672 & 0.778 & 0.443 & 0.750 &  & $[0.420; 0.486]$ & $[0.579; 0.667]$ & $[0.665; 0.841]$ & $[0.519; 0.606]$ \\
INTC & 135{,}872 & 0.226 & 0.051 &  & 0.656 & 0.853 & 0.293 & 0.344 &  & $[0.361; 0.444]$ & $[0.549; 0.707]$ & $[0.585; 0.902]$ & $[0.338; 0.412]$ \\
JNJ & 58{,}556 & 0.149 & 0.034 &  & 0.580 & 0.687 & 0.627 & 0.787 &  & $[0.365; 0.438]$ & $[0.465; 0.597]$ & $[0.805; 1.070]$ & $[0.419; 0.537]$ \\
JPM & 112{,}419 & 0.217 & 0.022 &  & 0.689 & 0.842 & 0.316 & 0.625 &  & $[0.385; 0.488]$ & $[0.553; 0.732]$ & $[0.537; 0.894]$ & $[0.469; 0.562]$ \\
KO & 66{,}613 & 0.149 & 0.051 &  & 0.571 & 0.694 & 0.612 & 0.637 &  & $[0.364; 0.421]$ & $[0.503; 0.618]$ & $[0.764; 0.995]$ & $[0.388; 0.506]$ \\
MCD & 37{,}803 & 0.157 & 0.035 &  & 0.574 & 0.718 & 0.563 & 0.769 &  & $[0.367; 0.412]$ & $[0.534; 0.586]$ & $[0.828; 0.932]$ & $[0.425; 0.519]$ \\
MMM & 28{,}154 & 0.177 & 0.038 &  & 0.668 & 0.801 & 0.398 & 0.637 &  & $[0.416; 0.482]$ & $[0.589; 0.689]$ & $[0.623; 0.823]$ & $[0.500; 0.569]$ \\
MRK & 64{,}628 & 0.177 & 0.036 &  & 0.552 & 0.702 & 0.595 & 0.600 &  & $[0.351; 0.406]$ & $[0.502; 0.601]$ & $[0.798; 0.997]$ & $[0.369; 0.469]$ \\
MSFT & 199{,}454 & 0.205 & 0.038 &  & 0.721 & 0.862 & 0.275 & 0.556 &  & $[0.388; 0.475]$ & $[0.517; 0.668]$ & $[0.664; 0.966]$ & $[0.419; 0.506]$ \\
NKE & 44{,}202 & 0.204 & 0.026 &  & 0.639 & 0.797 & 0.406 & 0.662 &  & $[0.373; 0.446]$ & $[0.525; 0.663]$ & $[0.675; 0.949]$ & $[0.450; 0.525]$ \\
PG & 56{,}081 & 0.153 & 0.044 &  & 0.543 & 0.636 & 0.728 & 0.781 &  & $[0.343; 0.414]$ & $[0.422; 0.566]$ & $[0.869; 1.156]$ & $[0.425; 0.531]$ \\
TRV & 17{,}920 & 0.182 & 0.061 &  & 0.577 & 0.729 & 0.543 & 0.750 &  & $[0.361; 0.442]$ & $[0.534; 0.645]$ & $[0.710; 0.933]$ & $[0.431; 0.537]$ \\
UNH & 39{,}641 & 0.218 & 0.042 &  & 0.552 & 0.724 & 0.551 & 0.688 &  & $[0.350; 0.383]$ & $[0.530; 0.574]$ & $[0.852; 0.941]$ & $[0.388; 0.481]$ \\
V & 56{,}107 & 0.201 & 0.041 &  & 0.637 & 0.814 & 0.372 & 0.613 &  & $[0.365; 0.439]$ & $[0.540; 0.673]$ & $[0.654; 0.920]$ & $[0.438; 0.500]$ \\
VZ & 78{,}106 & 0.161 & 0.037 &  & 0.512 & 0.670 & 0.660 & 0.637 &  & $[0.330; 0.383]$ & $[0.501; 0.588]$ & $[0.824; 0.998]$ & $[0.344; 0.425]$ \\
WBA & 41{,}062 & 0.221 & 0.043 &  & 0.532 & 0.742 & 0.515 & 0.512 &  & $[0.343; 0.383]$ & $[0.585; 0.656]$ & $[0.689; 0.830]$ & $[0.338; 0.406]$ \\
WMT & 60{,}565 & 0.156 & 0.034 &  & 0.538 & 0.664 & 0.673 & 0.769 &  & $[0.350; 0.398]$ & $[0.453; 0.557]$ & $[0.887; 1.094]$ & $[0.400; 0.500]$ \\
SPY & 414{,}927 & 0.130 & 0.010 &  & -- & -- & -- & -- &  & $[0.571; 0.668]$ & $[0.724; 0.838]$ & $[0.325; 0.551]$ & $[0.556; 0.694]$ \\
\hline \hline
\end{tabular}
\end{small}
\smallskip
\begin{scriptsize}
\parbox{0.98\textwidth}{\emph{Note.} Ticker is the stock symbol. $N$ is the number of transaction data before previous-tick interpolation to a 60-second sampling frequency. $\bar{ \sigma}$ is the truncated realized variance of \cite{mancini:09a} converted to an annualized standard deviation. H is the rejection rate of the Hausman test for microstructure noise described in \cite{ait-sahalia-xiu:19a}. $\bar{ \rho}$ is the sample correlation coefficient. $(\hat{a}, \hat{b})$
are OLS estimates of the parametric diurnal correlation function from \eqref{equation:diurnal-correlation}. We implement the test statistic from \eqref{equation:test-statistic} of no diurnal correlation $\mathcal{H}_{0}: \int _{0}^{1} ( k_{u,t}-1)^{2} \,\mathrm{d}t = 0$ each month. The sample period is January 4, 2010 to April 28, 2023. $\hat{P}( \mathcal{N} > q_{1-\alpha /\#T})$ is the fraction of the test statistics that exceed the $(1-\alpha /\#T)$-quantile of the simulated distribution function, as described in the main text, where $\alpha = 0.01$ is the overall significance level and $\#T = 160$ is the total number of hypotheses tested. We employ a Bonferroni correction to control the family-wise error rate.}
\end{scriptsize}
\end{center}
\end{sidewaystable}

We construct a 60-second equidistant transaction price series from the cleaned high-frequency data using the previous-tick rule of \citet{wasserfallen-zimmermann:85a}, so we collect $n = 390$ high-frequency returns per day for each asset. Although the asymptotic theory requires $n \rightarrow\infty$ and the amount of tick-by-tick data is an order of magnitude larger---as evident from column ``$N$'' in Table~\ref{table:djia-descriptive}---a 60-second window is the smallest time gap at which the data can be perceived noise-free, as gauged by the \citet{ait-sahalia-xiu:19a} Hausman test for microstructure noise. We compute their test statistic at the daily horizon and report the rejection rate in the ``H'' column in Table~\ref{table:djia-descriptive}.\footnote{Thanks to Dacheng Xiu for making Matlab code to implement the test available at his website.} This should be compared to a 1\% level of significance. Apart from a few stocks, the rejection rate is typically close to the nominal level, showing that noise is not a major concern. Meanwhile, raising the sampling frequency further raises the rejection rate materially (unreported, but available upon request) and is not recommended, unless a noise-robust approach is adopted.\footnote{One option is to pre-average the available high-frequency data; see, for example, \citet{jacod-li-mykland-podolskij-vetter:09a, podolskij-vetter:09a, podolskij-vetter:09b}. While this facilitates an increase in sampling frequency, one should be aware that noise-robust estimators converge at a very slow rate and may be less efficient than noise-free estimators in practice if the data are at the margin of being noisy. Still, pre-averaging can potentially improve the power of the test statistic, but we leave this extension for future research.}

\begin{figure}[t!]
\caption{A representative diurnal covariance and correlation function.}
\label{figure:rho_d}
\begin{center}
\begin{tabular}{cc}
{\small{Panel A: Diurnal covariance}}. & {\small{Panel B: Diurnal correlation}}. \\
\includegraphics[height=8cm,width=0.48\textwidth]{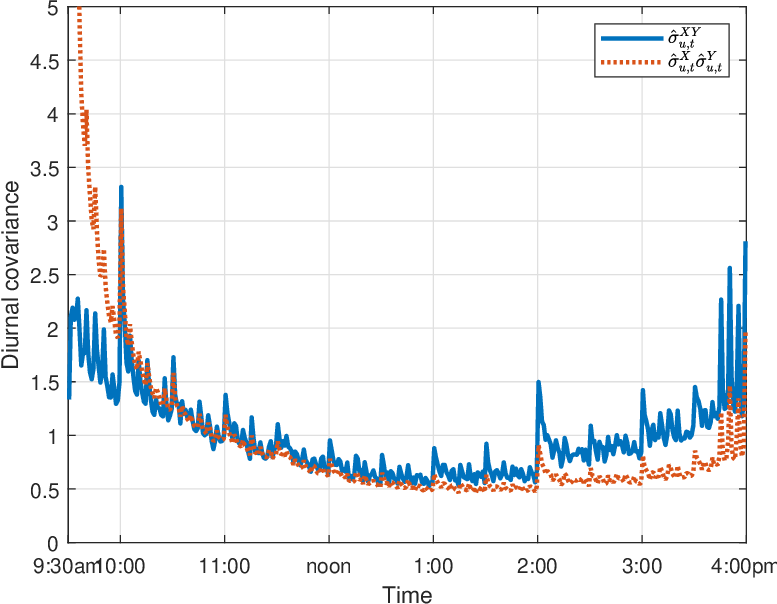} &
\includegraphics[height=8cm,width=0.48\textwidth]{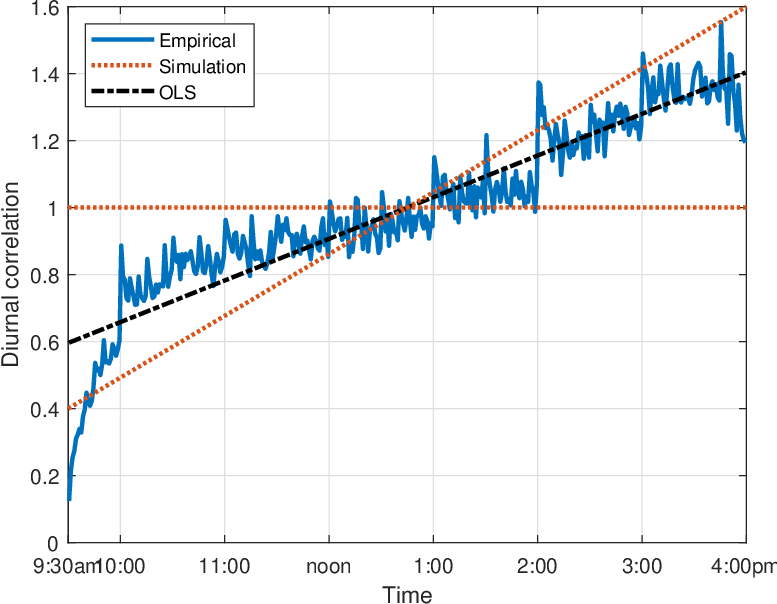}
\end{tabular}
\begin{scriptsize}
\parbox{\textwidth}{\emph{Note.} In panel A, we report a jump-robust estimator of the diurnal covariance function, $\hat{\sigma}_{u,t}^{XY}$, and compare it to $\hat{\sigma}_{u,t}^{X} \hat{\sigma}_{u,t}^{Y}$, where the latter is the imputed diurnal
covariation in the absence of deterministic variation in the intraday correlation coefficient, $k_{u,t}$. The estimator $\hat{k}_{u,t}$ is reported in panel B. ``OLS'' is the fitted least squares regression $\hat{k}_{u,t} = \hat{a} + \hat{b}t$ [with the restriction $b = 2(1-a)$]. ``Simulation'' shows the range of $a$ and $b$ values that are inspected in the Monte Carlo analysis.}
\end{scriptsize}
\end{center}
\end{figure}

\subsection{The diurnal pattern in correlation}
\label{sec7.2}

In panel A of Figure~\ref{figure:rho_d}, we plot a representative example of the diurnal covariance pattern inherent in our data. We follow \citet{christensen-hounyo-podolskij:18a} and compute it as the 0.5\% trimmed mean realized covariance estimate (after jump-truncation) at a fixed 60-second time-of-day slot, where the average is taken across the $T = 3{,}325$ days in the sample and $\frac{d(d-1)}{2} = 465$ pairwise combinations of the number of included equities, $d$. We contrast this to the geometric mean of the idiosyncratic diurnal variance, $\hat{\sigma}_{u,t}^{X} \hat{\sigma}_{u,t}^{Y}$ (everything is normalized as in Assumption~\ref{ass(C2)} to be comparable). Since $c_{u,t}^{XY} = \sigma_{u,t}^{X} \sigma_{u,t}^{Y} k_{u,t}$, the latter can be interpreted as the imputed diurnal covariance pattern present with no periodicity in the intraday correlation (i.e., $k_{u,t} = 1$). In agreement with prior literature (e.g., \cite{andersen-bollerslev:97b,bibinger-hautsch-malec-reiss:19a,christensen-hounyo-podolskij:18a}), $\hat{\sigma}_{u,t}^{X} \hat{\sigma}_{u,t}^{Y}$ resembles a ``tilted J.'' In contrast, we observe the actual diurnal covariance, $\hat{c}_{u,t}^{XY}$, is almost symmetric and much closer to U-shaped. This is anecdotal evidence that $k_{u,t}$ is not always equal to one.\footnote{Interestingly, there also appears to be a subperiodic structure in the diurnal covariance pattern at the whole- and half-hourly horizon.}

Next, we map each 60-second pairwise realized covariance matrix into a correlation estimate and repeat the above averaging procedure. The ensuing time-of-day correlation measure---portrayed in panel B of Figure~\ref{figure:rho_d}---should be randomly distributed around one under the null of no diurnal variation. Instead, we observe a pronounced upward-sloping and almost piecewise linear curve. There is notably lower (on average less positive) correlation in the morning than in the afternoon, which is in accord with \citet{allez-bouchaud:11a} and \citet{hansen-luo:23a}. These findings are further corroborated by estimating the equation $k_{u,t} = a + bt$ in \eqref{equation:diurnal-correlation} from the empirical high-frequency data. The OLS parameter estimates, subject to the maintained restriction $b = 2(1-a)$, are $\hat{a} = 0.5969$ and $\hat{b} = 0.8062$ with the fitted regression line inserted into the figure as a reference point. In practice, of course, $k_{u,t}$ evolves in a much more nonlinear and discontinuous fashion. We notice a positive jump at 10:00 a.m., arguably caused by the publication of macroeconomic information. There is another upsurge around 2:00 p.m., corresponding to the release of statements from Federal Open Market Committee \citep{fomc:26a} meetings.

The right-hand side of Table~\ref{table:djia-descriptive} has further descriptive statistics on diurnal correlation. It also reports the outcome of our testing procedure. We proceed as in Section~\ref{section:simulation} in terms of tuning parameters, that is, for $n = 390$ we take $k_{n} = 130$. We calculate the test statistic each month (of which there are 160 in total) with a Parzen kernel and lag length $H_{T_{m}} = [T_{m}^{1/3}]$, where $T_{m}$ is the number of days in month $m$ (with $T_{m} = 21$ on average). The analysis is then divided into two: We correlate individual members of the DJIA index against the SPY (``versus SPY'') and summarize with the interquartile range the results of pairing each stock---including the SPY---against all the thirty remaining ones (``versus rest'').

Glancing at the ``versus SPY'' part, several interesting findings emerge. First, every asset in our sample is positively related to the stock market portfolio exhibiting a typical level of correlation $\bar{\rho} = 0.556$. Second, on an individual stock basis the estimated $a$ and $b$ parameters are broadly in line with the aggregate figures reported above and remarkably consistent over the cross-section of equities. In the end, it translates into an average rejection rate of around two out of three with our proposed test statistic. Apart from a few instances, these rates are remarkably close for the vast majority of the assets.

Switching to the ``versus rest'' part, single names display a weaker association with each other than with the market. This is further reflected in the tendency for the intraday correlation to exhibit a more upward-sloping linear association with $a$ being lower and $b$ being higher. Interestingly, there is a somewhat larger discrepancy between the rejection rates of the test statistic for individual assets tested against each other. These rejection rates are notably lower than those for the market index, but remain far above the nominal level.

Overall, our results suggest diurnal variation in the correlation process is a nontrivial effect, which is present most of the months in our sample.

\subsection{Conditioning information}
\label{sec7.3}

To delve deeper into our empirical results, we follow the guidance from Section~\ref{section:conditional} and extend the previous analysis by investigating whether and how conditioning information helps to determine the functional form of the intraday correlation curve.

First, we gauge the impact of macroeconomic news in the form of monetary policy decisions made by the FOMC, which in the majority of our sample are released at 2:00~p.m. followed by a press conference at 2:30 p.m.\footnote{Earlier, the FOMC statements were released at the conclusion of the meeting, which gradually converged toward 2:15 p.m. The current format was adopted beginning in 2011, and hence, covers nearly our entire sample.} There are eight regularly scheduled meetings during the year. We acquired historical announcement dates from the Federal Reserve Board's website. Second, we analyze the influence of quarterly earnings announcements (QEA) issued by the individual companies in our stock universe. Here, the historical announcement dates were extracted from the Center for Research in Security Prices database \citep{crsp:26a}. We only include earnings announcements released either in the after-hours session on the previous day or during pre-market trading on the same day, such that the earliest opportunity to react to the news for the general public is at the commencement of exchange trading at 9:30 a.m. Thus, whereas the former application centers around market-wide systematic announcements released during active trading that are likely to affect the stock market in its entirety, the latter concerns largely idiosyncratic news---at least within the domain of the equities we look at---that are released prior to the opening of the stock exchange.\footnote{Financial information from a company can trigger price changes in related firms and the broader market (e.g., \cite{patton-verardo:12a,savor-wilson:16a}). However, as shown by \citet{christensen-timmermann-veliyev:25a}, the spillover effect is often small in magnitude.}

\begin{figure}[t!]
\caption{Conditional diurnal correlation function.}
\label{figure:conditional}
\begin{center}
\begin{tabular}{cc}
{\small{Panel A: FOMC}}. & {\small{Panel B: QEA}}. \\
\includegraphics[height=8cm,width=0.48\textwidth]{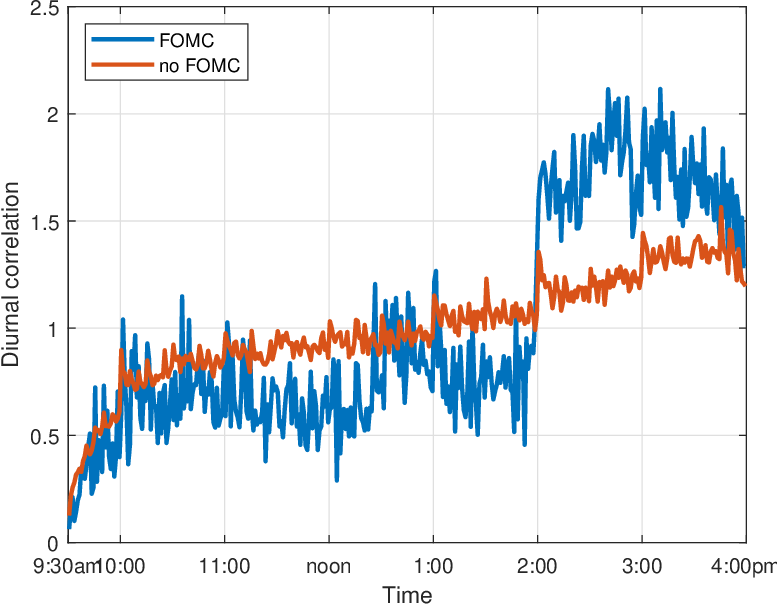} &
\includegraphics[height=8cm,width=0.48\textwidth]{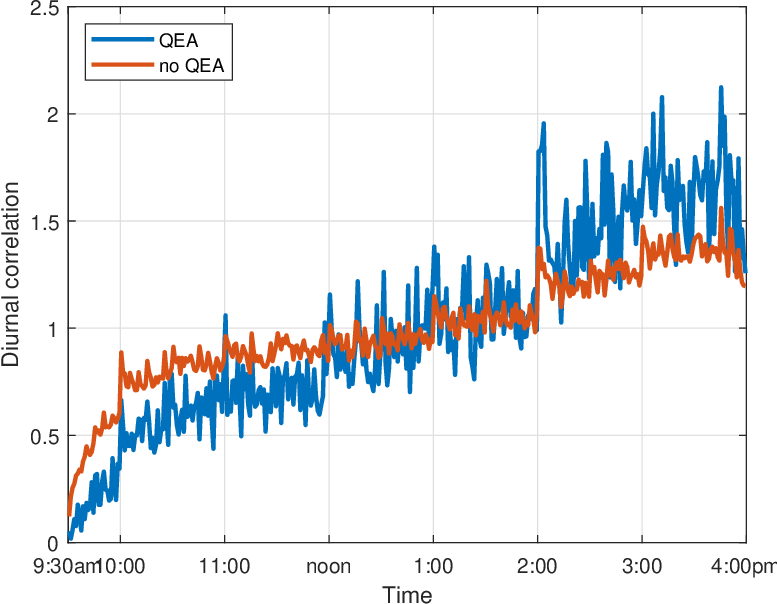} \\
\end{tabular}
\begin{scriptsize}
\parbox{0.925\textwidth}{\emph{Note.} In this figure, we show how conditioning information alters the intraday correlation curve. In panel A, we split the sample based on macroeconomic announcements, while in panel B we do it based on whether or not the stock in question made an earnings announcement. In both panels, the ``no'' curve refers to the no announcement sample.}
\end{scriptsize}
\end{center}
\end{figure}

The outcome of this analysis is presented in Figure~\ref{figure:conditional}. In panel A, we show the results for the macroeconomic news announcements, while panel B reports the associated results for earnings releases. The ``no'' curve refers to the complement based on the no announcement sample. In both cases, the latter is very close to the unconditional curve from panel B in Figure~\ref{figure:rho_d}, although the jump at 2:00 p.m. is slightly smaller in panel A of Figure~\ref{figure:conditional} than previously. Furthermore, we should note that since the announcement sample is much smaller than the no announcement sample, the reported point estimates are subject to considerable measurement error. However, the overall evolution can still be deciphered.

The results are compelling. In particular, the typical FOMC announcement leads to a distinct positive jump in the diurnal correlation pattern, which is much larger than above. As anticipated, the influx of a market-wide news component leads to a systematic response in the prices of most equities, which temporarily reinforces their intraday return correlation, before it starts to recede and taper off toward the no announcement curve at the closing of the stock exchange at 4:00 p.m. Turning our attention to panel B for the earnings announcements, the results are also rather intuitive. Specifically, an earnings announcement causes the security price of the issuing company to be largely detached from the market during the early phases of trading while the price discovery process is being completed and portfolio holdings being updated, before the intraday correlation curve reconnects with the no announcement sample around noon.\looseness=1\footnote{In unreported results, we also examined whether stock characteristics can help to explain the pattern in the diurnal correlation process. In particular, we studied the influence of liquidity and industry connectedness. First, we sorted our stocks based on liquidity, as defined by the ``$N$'' column in Table~\ref{table:djia-descriptive}. We selected the ten most liquid and least liquid companies, while leaving out the middle portion of the sample, and calculated a separate intraday correlation curve for each subsample. However, there was no discernible difference between them. This is possibly because we are only considering large-cap stocks that are highly liquid in \textit{absolute} terms, even if some are \textit{relatively} illiquid. Second, we split the stocks based on industry proximity, as defined by the ``closeness'' of their SIC codes (see, e.g., \cite{christensen-timmermann-veliyev:25a,wang-zajac:07a}). This showed that more distant companies are less correlated in the morning; a finding that mirrors our results for the QEA. Intuitively, when a company announces its earnings results, its security price also trades relatively ``distant'' to the market, being driven mainly by the idiosyncratic contents of the announcement in the short term. The details are available upon request.}

\subsection{Implications for risk management}
\label{section:hedging}

In closing, we highlight the importance of incorporating diurnal variation in the correlation process as exemplified via the operations of a trading desk. We suppose a dealer is long one stock from the DJIA index. The risk is offset with a dynamic short position in the market index (SPY in our context). We assume the trader employs a conventional 5-minute frequency and updates the hedge at the end of each time interval---based on available information---in order to minimize the expected variance of the combined portfolio during the next 5-minute window. The minimum variance hedge ratio, denoted $\phi_{i \mid i-1}^{n}$, is an adapted discrete-time stochastic process that is selected at the beginning of the $i$th interval $[(i-1)/n,i/n]$ via the following optimization problem:
\begin{equation}
\phi_{i \mid i-1}^{n} = \underset{\phi}{\argmin} \ \operatorname{var} \bigl( \Delta_{i}^{n} X - \phi\Delta_{i}^{n} Y \mid\mathcal{F}_{\frac{i-1}{n}} \bigr). \label{eq74}
\end{equation}
The solution is given by
\begin{equation}
\phi_{i \mid i-1}^{n} = \frac{\operatorname{cov} \bigl( \Delta_{i}^{n} X, \Delta_{i}^{n} Y \mid\mathcal{F}_{\frac{i-1}{n}} \bigr)}{\operatorname{var} \bigl( \Delta_{i}^{n} Y \mid\mathcal{F}_{\frac{i-1}{n}} \bigr)},
\label{eq75}
\end{equation}
for $i = 1, \ldots, n$, where $\Delta_{i}^{n} X$ is the subsequent 5-minute log-return on the underlying asset and $\Delta_{i}^{n} Y$ is the associated SPY log-return (note that in this subsection we set $n = 78$ to represent a 5-minute frequency for notational convenience).

The trading policy depends on the conditional covariance matrix:
\begin{equation}
\Sigma_{i \mid i-1}^{n} = \operatorname{var} \lleft(
\begin{matrix} \Delta_{i}^{n}
X \\
\Delta_{i}^{n} Y \end{matrix} %
\Bigm| \mathcal{F}_{\frac{i-1}{n}} \rright). \label{eq76}
\end{equation}
In practice, $\Sigma_{i \mid i-1}^{n}$ is not known in advance and has to be modeled. However, we do not pursue this approach here. Instead, we assume that an estimator of $\Sigma_{i \mid i-1}^{n}$ is accessible via the 5-minute ex post realized covariance matrix of $X$ and $Y$ (calculated from the 60-second high-frequency data extracted above).

$\phi_{i \mid i-1}^{n}$ is then selected as
\begin{equation}
\phi_{i \mid i-1}^{n} = \hat{\rho}_{[i-1,i]} \frac{\hat{\sigma}_{[i-1,i]}^{X}}{\hat{\sigma}_{[i-1,i]}^{Y}},
\label{eq77}
\end{equation}
with $\hat{\sigma}_{[i-1,i]}^{X}$ and $\hat{\sigma}_{[i-1,i]}^{Y}$ being the square-root realized variance of $X$ and $Y$ on the $i$th interval, whereas $\hat{\rho}_{[i-1,i]}$ is the realized correlation.

In other words, $\phi_{i \mid i-1}^{n}$ is the ex post minimum variance hedge ratio, conditional on knowing the subsequent realized covariance matrix over that window. It follows that $(\phi_{i \mid i-1}^{n})_{i=1}^{n}$ adapts to intraday periodicity in both the variance and correlation processes. Suppose that the stochastic correlation component is constant within a day, that is, $\rho_{t} = \rho_{sc, \lfloor t \rfloor} k_{u,t}$, where $\rho_{sc, \lfloor t \rfloor}$ is determined at the start of day $t$. This assumption is common in the discrete-time multivariate stochastic volatility literature, and it is a decent approximation to the dynamic of the stochastic correlation process in view of its persistence. In this case, the high-frequency correlation estimate can be decomposed as $\hat{\rho}_{[i-1,i]} = \hat{k}_{u, [i-1,i]} \hat{\rho}_{sc}$, where $\hat{\rho}_{sc} = n^{-1} \sum_{i=1}^{n} \hat{\rho}_{[i-1,i]}$ is the average realized correlation over the whole day and $\hat{k}_{u,[i-1,i]}$ is the diurnal coefficient. This further implies that
\begin{equation}
\phi_{i \mid i-1}^{n} = \hat{k}_{u,[i-1,i]} \hat{\rho}_{sc} \frac{\hat{\sigma}_{[i-1,i]}^{X}}{\hat{\sigma}_{[i-1,i]}^{Y}} = \hat{k}_{u, [i-1,i]} \bar{\phi}_{i \mid i-1}^{n}, \label{eq78}
\end{equation}
where ${\bar{\phi}}_{i \mid i-1}^{n}$ is the optimal ex post hedge ratio, when the local correlation estimate is replaced by an average for the entire day, all else equal. Hence, ${\bar{\phi}}_{i \mid i-1}^{n}$ adapts to diurnal variation in the variance but not the correlation.

\begin{figure}[t!]
\caption{The distribution of the minimum variance hedge ratio.}
\label{figure:minimum_variance}
\begin{center}
\begin{tabular}{cc}
{\small{Panel A: Intraday evolution}}. & {\small{Panel B: Unconditional distribution}}. \\
\includegraphics[height=8cm,width=0.48\textwidth]{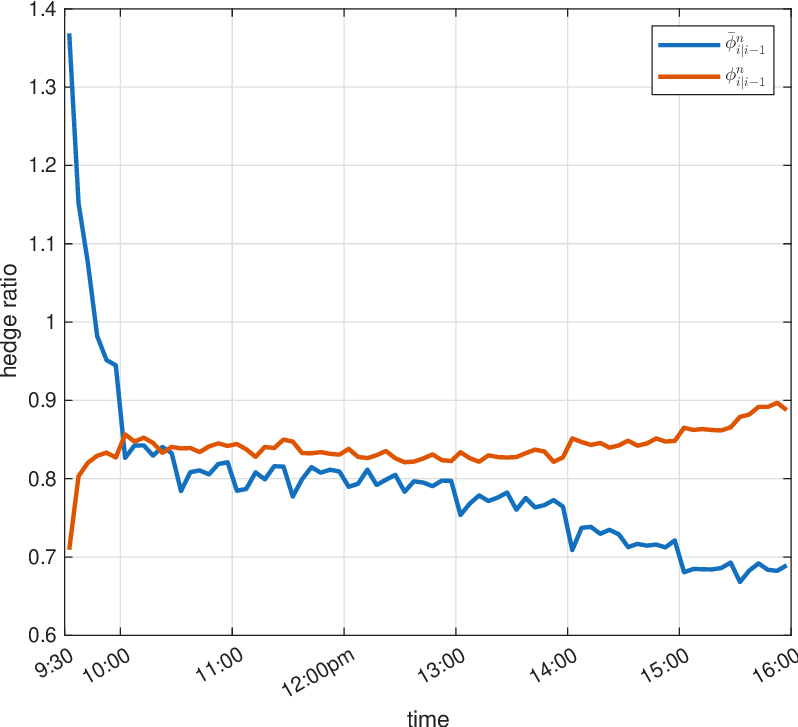} &
\includegraphics[height=8cm,width=0.48\textwidth]{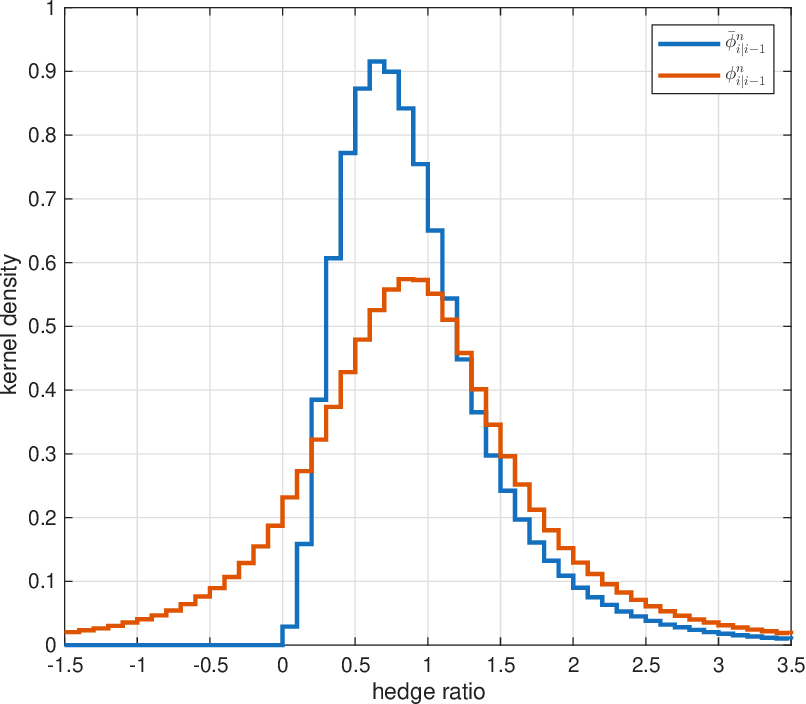}
\end{tabular}
\begin{scriptsize}
\parbox{0.925\textwidth}{\emph{Note.} In panel A, we plot the evolution of the average intraday minimum variance hedge ratio, that is, $\phi_{i \mid i-1}^{n}$ and $\bar{\phi}_{i \mid i-1}^{n}$. In panel B, we show the unconditional distribution of $\phi_{i \mid i-1}^{n}$ and $\bar{\phi}_{i \mid i-1}^{n}$. }
\end{scriptsize}
\end{center}
\end{figure}

We compare $\phi_{i \mid i-1}^{n}$ and $\bar{\phi}_{i \mid i-1}^{n}$ to illustrate the effect on risk management. The minimum variance hedge ratio is computed as described above across the components of the DJIA index and for each 5-minute interval in the sample. Figure~\ref{figure:minimum_variance} reports the results. In panel A, we plot the intraday profile of $\phi_{i \mid i-1}^{n}$ and $\bar{\phi}_{i \mid i-1}^{n}$. The optimal $\phi_{i \mid i-1}^{n}$ is around 0.7--0.9. In contrast, there is pronounced periodicity in $\bar{\phi}_{i \mid i-1}^{n}$. The latter fails to acknowledge that lower correlation in the morning has a detrimental impact on the diversification effect, causing a reduced hedge ratio (and vice versa in the afternoon). Interestingly, this means there are fewer transaction costs associated with managing a portfolio based on $\phi_{i \mid i-1}^{n}$. In panel B, we see the unconditional distribution of $\phi_{i \mid i-1}^{n}$ is more symmetric and has mass below zero, as it automatically adapts to brief lapses of low-to-negative correlation. In contrast, the histogram of $\bar{\phi}_{i \mid i-1}^{n}$ is floored at zero, because the daily correlation with the stock index tends to be positive.

The variance ratio of the full sample ex post portfolio return:
\begin{equation}
\frac{\widehat{\operatorname{var}}\bigl( \Delta_{i}^{n} X - \phi_{i \mid i-1}^{n} \Delta_{i}^{n} Y\bigr)}{\widehat{\operatorname{var}}\bigl( \Delta_{i}^{n} X - \bar{\phi}_{i \mid i-1}^{n} \Delta_{i}^{n} Y\bigr)} = 0.824, \label{eq79}
\end{equation}
suggesting it is possible to achieve a highly nontrivial reduction in risk exposure close to 20\% in a risk management model that controls for diurnal variation in correlation.

\section{Conclusion}
\label{section:conclusion}

We develop a nonparametric test of the hypothesis that there is no diurnal variation in a correlation process. The proposed test statistic has a known distribution under the null, whereas it diverges under an alternative with deterministic variation in the correlation. In a simulation study, the testing procedure aligns closely with the theoretical predictions and it attains a good rejection rate for moderate sample sizes and realistic shapes in the diurnal correlation process. In our empirical application, we document pervasive intraday correlation dynamics in the US equity market. Consistent with \citet{allez-bouchaud:11a}, \citet{bibinger-hautsch-malec-reiss:19a}, and \citet{hansen-luo:23a}, we find that correlations are low in the morning and rise systematically during the trading session. We further show how conditioning information about macroeconomic news and corporate earnings announcements affects the evolution of the intraday correlation curve.

\citet{andersen-thyrsgaard-todorov:19a} test whether the intraday volatility curve is changing over time (see \cite{andersen-su-todorov-zhang:24a}, for related work). They exploit an assumed stationarity of the stochastic volatility and compare the unconditional distribution of different time-of-day high-frequency returns. As in this paper, their results are derived based on a combination of infill and long-span analysis. It may be possible to adapt that setting to our framework by feeding their test statistic with devolatized high-frequency returns. We leave this idea for inspiration.

\clearpage

\begin{appendices}

\section{Proofs} \label{appendix:proof}

In this Appendix, we prove the theoretical results presented in the main text. To facilitate the derivations, we denote the continuous part of $X$ and $Y$ by
\begin{align*}
X_{t}^{c} &\equiv X_{0} + \int_{0}^{t} a^{X} \,\mathrm{d}s + \int_{0}^{t} \sigma_{s}^{X} \,\mathrm{d}W_{s}^{X} \quad\text{and} \\
Y_{t}^{c} &\equiv Y_{0} + \int_{0}^{t} a^{Y} \,\mathrm{d}s + \int_{0}^{t} \sigma_{s}^{Y} \bigl( \rho_{s} \,\mathrm{d}W_{s}^{X} + \sqrt{1-\rho_{s}^{2}} \,\mathrm{d}W_{s}^{Y} \bigr).
\end{align*}
We set $e^{XY}_{t, \tau} = \hat{c}_{t, \tau}^{XY} - \hat{c}_{t, \tau}^{X^{c}Y^{c}}$ for $t=1, \dots, T$ and $\tau\in(0, 1]$, which is the blockwise difference between the realized covariance calculated on the whole process and only its continuous component. We also denote $i_{t,\ell} = t-1+(\ell-1) \Delta_{n}$ and write $U_{t,\ell} = U_{i_{t,\ell}}$ for any stochastic process $U$.

Furthermore, we define
\begin{equation*}
\begin{aligned} \zeta
_{1}^{n}(t, \tau_{j}) &= \frac{n}{k_{n}} \sum_{\ell= (j-1)k_{n}+1}^{jk_{n}} \int_{i_{t, \ell}}^{i_{t, \ell+1}} a_{s}^{X} \,\mathrm{d}s \cdot\int_{i_{t, \ell}}^{i_{t, \ell+1}} a_{s}^{Y} \,\mathrm{d}s, \\
\zeta_{2}^{n}(t, \tau_{j}) &= \frac{n}{k_{n}} \sum_{\ell= (j-1)k_{n}+1}^{jk_{n}} \int_{i_{t, \ell}}^{i_{t, \ell+1}} a_{s}^{X} \,\mathrm{d}s \cdot\int_{i_{t, \ell}}^{i_{t, \ell+1}} \sigma_{s}^{Y} \bigl( \rho_{s} \,\mathrm{d}W_{s}^{X} + \sqrt{1- \rho_{s}^{2}} \,\mathrm{d}W_{s}^{Y} \bigr), \\
\zeta_{3}^{n}(t, \tau_{j}) &= \frac{n}{k_{n}} \sum_{\ell= (j-1)k_{n}+1}^{jk_{n}} \int_{i_{t, \ell}}^{i_{t, \ell+1}} a_{s}^{Y} \,\mathrm{d}s \cdot\int_{i_{t, \ell}}^{i_{t, \ell+1}} \sigma_{s}^{X} \,\mathrm{d}W_{s}^{X}, \\
\zeta_{4}^{n}(t, \tau_{j}) &= \frac{n}{k_{n}} \sum_{\ell= (j-1)k_{n}+1}^{jk_{n}} \int_{i_{t, \ell}}^{i_{t, \ell+1}} \bigl( \sigma_{s}^{X} - \sigma_{t,j}^{X} \bigr) \,\mathrm{d}W_{s}^{X} \cdot\int_{i_{t, \ell}}^{i_{t, \ell+1}} \sigma_{s}^{Y} \bigl( \rho_{s} \,\mathrm{d}W_{s}^{X} + \sqrt{1- \rho_{s}^{2}} \,\mathrm{d}W_{s}^{Y} \bigr), \\
\zeta_{5}^{n}(t, \tau_{j}) &= \frac{n}{k_{n}} \sum_{\ell= (j-1)k_{n}+1}^{jk_{n}} \int_{i_{t, \ell}}^{i_{t, \ell+1}} \sigma_{s}^{X} \,\mathrm{d}W^{X}_{s} \cdot\int_{i_{t, \ell}}^{i_{t, \ell+1}} \bigl( \sigma_{s}^{Y} - \sigma_{t,j}^{Y} \bigr) \bigl( \rho_{s} \,\mathrm{d}W_{s}^{X} + \sqrt{1- \rho_{s}^{2}} \,\mathrm{d}W_{s}^{Y} \bigr), \\
\zeta_{6}^{n}(t, \tau_{j}) &= \frac{n}{k_{n}} \sum_{\ell= (j-1)k_{n}+1}^{jk_{n}} \int_{i_{t, \ell}}^{i_{t, \ell+1}} \bigl( \sigma_{s}^{X} - \sigma_{t,j}^{X} \bigr) \,\mathrm{d}W_{s}^{X} \\
&\quad{}\cdot\int_{i_{t, \ell}}^{i_{t, \ell+1}} \bigl( \sigma_{s}^{Y} - \sigma_{t,j}^{Y} \bigr) \bigl( \rho_{s} \,\mathrm{d}W_{s}^{X} + \sqrt{1- \rho_{s}^{2}} \,\mathrm{d}W_{s}^{Y} \bigr).%
\end{aligned} %
\end{equation*}
We are going to need a couple of auxiliary lemmas.

\begin{lemma}\label{lem1}
Suppose the boundedness condition in Assumption~\ref{ass(C4)} holds. Then, for $i = 1, \dots, 3$,
\begin{equation*}
\mathrm{\mathbb{E}} \Biggl( \Biggl\llvert\frac{1}{T} \sum_{t=1}^{T} \zeta_{i}^{n}(t, \tau_{j}) \Biggr\rrvert^{m} \Biggr) \leq\frac{C}{n^{m/2}},
\end{equation*}
for any $m \geq2$ and $\tau_{j} \in[0,1]$.
\end{lemma}

\begin{proof}
The term $\frac{1}{T} \sum_{t=1}^{T} \zeta_{1}^{n}(t, \tau_{j})$ is handled with Jensen's inequality and the $C_{r}$ inequality:\vspace*{-2pt}
\begin{equation*}
\begin{aligned}
&\mathbb{E} \Biggl( \Biggl\llvert\frac{1}{T} \sum_{t=1}^{T} \zeta_{1}^{n}(t, \tau_{j}) \Biggr\rrvert^{m} \Biggr) \\
&\quad\leq\frac{n^{m}}{Tk_{n}} \sum_{t=1}^{T} \sum_{\ell= (j-1)k_{n}+1}^{jk_{n}} \mathbb{E} \biggl[ \biggl( \int_{i_{t, \ell}}^{i_{t, \ell+1}} a_{s}^{X} \,\mathrm{d}s \cdot\int_{i_{t, \ell}}^{i_{t, \ell+1}} a_{s}^{Y} \,\mathrm{d}s \biggr)^{m} \biggr] \\
&\quad\leq\frac{C_{m}n^{m}}{Tk_{n}} \sum_{t=1}^{T} \sum_{\ell=(j-1)k_{n}+1}^{jk_{n}} \mathbb{E} \biggl[ \biggl( \int_{i_{t, \ell}}^{i_{t, \ell+1}} a_{s}^{X} \,\mathrm{d}s \biggr)^{2m} + \biggl( \int_{i_{t, \ell}}^{i_{t, \ell+1}} a_{s}^{Y} \,\mathrm{d}s \biggr)^{2m} \biggr] \\
&\quad\leq\frac{C_{m}}{Tk_{n}n^{m-1}} \sum_{t=1}^{T} \sum_{\ell= (j-1)k_{n}+1}^{jk_{n}} \biggl( \int_{i_{t, \ell}}^{i_{t, \ell+1}} \mathbb{E} \bigl(a_{s}^{X} \bigr)^{2m} \,\mathrm{d}s + \int_{i_{t, \ell}}^{i_{t, \ell+1}} \mathbb{E} \bigl(a_{s}^{Y} \bigr)^{2m} \,\mathrm{d}s \biggr) \\
&\quad\leq\frac{C}{n^{m}},%
\end{aligned} %
\end{equation*}
where the first line in the array is based on the trivial inequality $ab \leq a^{2} + b^{2}$ and the last line is due to the boundedness condition.

The treatment of the second and third term is nearly identical, so here we only verify the proof of the latter. By the Cauchy--Schwarz inequality, Jensen's inequality, the It\^{o} isometry, and the boundedness condition, we observe that\vspace*{-2pt}
\begin{equation*}
\begin{aligned}
&\mathbb{E} \Biggl( \Biggl\llvert\frac{1}{T} \sum_{t=1}^{T} \zeta_{3}^{n}(t, \tau_{j}) \Biggr\rrvert^{m} \Biggr) \\
&\quad\leq\frac{n^{m}}{Tk_{n}} \sum_{t=1}^{T} \sum_{\ell= (j-1)k_{n}+1}^{jk_{n}} \mathbb{E} \biggl( \biggl\llvert\int_{i_{t, \ell}}^{i_{t, \ell+1}} a_{s}^{Y} \,\mathrm{d}s \cdot\int_{i_{t, \ell}}^{i_{t, \ell+1}} \sigma_{s}^{X} \,\mathrm{d}W_{s}^{X} \biggr\rrvert^{m} \biggr) \\
&\quad\leq\frac{n^{m}}{Tk_{n}} \sum_{t=1}^{T} \sum_{\ell=(j-1)k_{n}+1}^{jk_{n}} \biggl[ \mathbb{E} \biggl( \biggl\llvert\int_{i_{t, \ell}}^{i_{t, \ell+1}} a_{s}^{Y} \,\mathrm{d}s \biggr\rrvert^{2m} \biggr) \cdot\mathbb{E} \biggl( \biggl\llvert\int_{i_{t, \ell}}^{i_{t, \ell+1}} \sigma_{s}^{X} \,\mathrm{d}W_{s}^{X} \biggr\rrvert^{2m} \biggr) \biggr]^{1/2} \\
&\quad\leq\frac{C}{Tk_{n}n^{m/2-1}} \sum_{t=1}^{T} \sum_{\ell=(j-1)k_{n}+1}^{jk_{n}} \biggl( \int_{i_{t, \ell}}^{i_{t, \ell+1}} \mathbb{E} \bigl(a_{s}^{Y} \bigr)^{2m} \,\mathrm{d}s \cdot\int_{i_{t, \ell}}^{i_{t, \ell+1}} \mathbb{E} \bigl( \sigma_{s}^{X} \bigr)^{2m} \,\mathrm{d}s \biggr)^{1/2} \\
&\quad\leq\frac{C}{n^{m/2}}.%
\end{aligned} %
\end{equation*}
\upqed\end{proof}

\begin{lemma}%
\label{Lemma2}
Suppose the boundedness condition in Assumption~\ref{ass(C4)} holds. Then\vspace*{-2pt}
\begin{equation*}
\begin{aligned}
\mathbb{E} \Biggl( \Biggl\llvert\frac{1}{T} \sum_{t=1}^{T} \zeta_{4}^{n}(t, \tau_{j}) \Biggr\rrvert\Biggr) &\leq\frac{C}{n^{1/2}}, \qquad\mathbb{E} \Biggl( \Biggl\llvert\frac{1}{T} \sum_{t=1}^{T} \zeta_{5}^{n}(t, \tau_{j}) \Biggr\rrvert\Biggr) \leq\frac{C}{n^{1/2}}, \quad\text{and} \\
\mathbb{E} \Biggl( \Biggl\llvert\frac{1}{T} \sum_{t=1}^{T} \zeta_{6}^{n}(t, \tau_{j}) \Biggr\rrvert\Biggr) &\leq\frac{C}{n}.
\end{aligned}
\end{equation*}
\end{lemma}

\begin{proof}
By the $C_{r}$ inequality, Burkholder--Davis--Gundy inequality, Assumption~(V), and the boundedness condition, for $s \in[i_{t, j}, i_{t, j+1}]$, we find that
\begin{equation*}
\begin{aligned} \mathbb{E} \bigl( \bigl
\llvert\sigma_{s}^{X}- \sigma_{t,j}^{X} \bigr\rrvert^{2} \bigr) & \leq C \biggl[ \mathbb{E} \biggl( \int_{i_{t,j}}^{s} \tilde{a}_{u}^{X} \,\mathrm{d}u \biggr)^{2} + \mathbb{E} \biggl( \int_{i_{t,j}}^{s} \tilde{\sigma}_{u}^{X} \,\mathrm{d}W_{u}^{X} \biggr)^{2} + \mathbb{E} \biggl( \int_{i_{t,j}}^{s} \tilde{\sigma}_{u}^{Y} \,\mathrm{d}W_{u}^{Y} \biggr)^{2} \\
&\quad{}+ \mathbb{E} \biggl( \int_{i_{t,j}}^{s} \tilde{\nu}_{u}^{X} \,\mathrm{d} \tilde{W}_{u}^{X} \biggr)^{2} + \mathbb{E} \biggl( \int_{i_{t,j}}^{s} \tilde{\nu}_{u}^{Y} \,\mathrm{d} \tilde{W}_{u}^{Y} \biggr)^{2} + \mathbb{E} \biggl( \int_{i_{t,j}}^{s} \int_{\mathbb{R}}x \tilde{F}^{X} \,(\mathrm{d}x) \,\mathrm{d}u \biggr)^{2} \biggr] \\
&\leq\frac{C}{n}.%
\end{aligned} %
\end{equation*}
After another round with the Cauchy--Schwarz inequality, Jensen's inequality, the It\^{o} isometry, and the boundedness condition, we arrive at the conclusion that
\begin{equation*}
\begin{aligned} \mathbb{E} \Biggl( \Biggl
\llvert\frac{1}{T} \sum_{t=1}^{T} \zeta_{4}^{n}(t, \tau_{j}) \Biggr\rrvert\Biggr) &\leq\frac{Cn}{Tk_{n}} \sum_{t=1}^{T} \sum_{\ell=(j-1)k_{n}+1}^{jk_{n}} \biggl( \int_{i_{t, \ell}}^{i_{t, \ell+1}} \noxml{\!}\mathbb{E} \bigl( \sigma_{s}^{X}- \sigma_{t,j}^{X} \bigr)^{2} \,\mathrm{d}s \cdot\int_{i_{t, \ell}}^{i_{t, \ell+1}} \mathbb{E} \bigl( \sigma_{s}^{Y} \bigr)^{2} \,\mathrm{d}s \biggr)^{1/2} \\
&\leq\frac{C}{n^{1/2}}.%
\end{aligned} %
\end{equation*}
The proofs of the other inequalities follow in the same footsteps.
\end{proof}

\begin{proof}[Proof of Theorem~\ref{theorem:consistency}]
It suffices to prove the convergence for the covariance term, $\hat{c}_{u, \tau}^{XY}$, for $\tau\in[0, 1]$. We begin with a decomposition of the continuous part of $\hat{c}_{t, \tau_{j}}^{XY}$, that is, $\hat{c}_{t, \tau_{j}}^{X^{c}Y^{c}}$:
\begin{align*}
\hat{c}_{t, \tau_{j}}^{X^{c}Y^{c}} &= \frac{n}{k_{n}} \sum_{\ell=(j-1)k_{n}+1}^{jk_{n}} \Delta_{(t-1)n+ \ell}^{n}X^{c} \Delta_{(t-1)n+ \ell}^{n}Y^{c} \\
&= \sum_{m=1}^{6} \zeta_{m}^{n}(t, \tau_{j}) + \sigma_{t,j}^{X} \sigma_{t,j}^{Y} \rho_{t,j} \cdot\frac{n}{k_{n}} \sum_{\ell=(j-1)k_{n}+1}^{jk_{n}}\bigl( \Delta_{(t-1)n+ \ell}^{n}W^{X} \bigr)^{2} \\
&\quad{}+ \sigma_{t,j}^{X} \sigma_{t,j}^{Y} \sqrt{1- \rho_{t,j}^{2}} \cdot\frac{n}{k_{n}} \sum_{\ell=(j-1)k_{n}+1}^{jk_{n}} \Delta_{(t-1)n+ \ell}^{n}W^{X} \Delta_{(t-1)n+ \ell}^{n}W^{Y} \\
&= \sum_{m=1}^{6} \zeta_{m}^{n}(t, \tau_{j})+ \sigma_{t,j}^{X} \sigma_{t,j}^{Y} \rho_{t,j} \cdot\alpha_{t,j}^{n} + \sigma_{t,j}^{X} \sigma_{t,j}^{Y} \sqrt{1- \rho_{t,j}^{2}} \cdot\beta_{t,j}^{n},
\end{align*}
where
\begin{align*}
\alpha_{t,j}^{n} &= \frac{n}{k_{n}} \sum_{\ell=(j-1)k_{n}+1}^{jk_{n}} \bigl( \Delta_{(t-1)n+ \ell}^{n}W^{X} \bigr)^{2}, \\
\beta_{t,j}^{n} &= \frac{n}{k_{n}} \sum_{\ell=(j-1)k_{n}+1}^{jk_{n}} \Delta_{(t-1)n+ \ell}^{n}W^{X} \Delta_{(t-1)n+ \ell}^{n}W^{Y},
\end{align*}
and $\zeta_{m}^{n}(t, \tau_{j})$ is defined in the preparation step at the beginning of this Appendix. Thus, according to Assumption~\ref{ass(C1)},
\begin{equation*}
\begin{aligned} \hat{c}_{t, \tau_{j}}^{XY}
&= \sigma_{u,j}^{X} \sigma_{u,j}^{Y} k_{u,j} \cdot\sigma_{sv,t,j}^{X} \sigma_{sv,t,j}^{Y} \rho_{sc,t,j} \cdot\alpha_{t,j}^{n} + \sigma_{t,j}^{X} \sigma_{t,j}^{Y} \sqrt{1- \rho_{t,j}^{2}} \cdot\beta_{t,j}^{n} + \sum_{m=1}^{6} \zeta_{m}^{n}(t, \tau_{j}) \\
&= \sigma_{u,j}^{X} \sigma_{u,j}^{Y} k_{u,j} \cdot\sigma_{sv,t,j}^{X} \sigma_{sv,t,j}^{Y} \rho_{sc,t,j} \\
&\quad{}+ \sigma_{u,j}^{X} \sigma_{u,j}^{Y} k_{u,j} \cdot\sigma_{sv,t,j}^{X} \sigma_{sv,t,j}^{Y} \rho_{sc,t,j} \cdot\bigl( \alpha_{t,j}^{n}-1\bigr)+ \sigma_{t,j}^{X} \sigma_{t,j}^{Y} \sqrt{1- \rho_{t,j}^{2}} \cdot\beta_{t,j}^{n} \\
&\quad{}+ \sum_{m=1}^{6} \zeta_{m}^{n}(t, \tau_{j}).%
\end{aligned} %
\end{equation*}
Therefore,
\begin{align*}
\tilde{c}_{u, \tau_{j}}^{XY} &= \frac{1}{T} \sum_{t=1}^{T} \hat{c}_{t, \tau_{j}}^{X^{c}Y^{c}} + \frac{1}{T} \sum_{t=1}^{T}e^{XY}_{t, \tau_{j}} \\
&= \sigma_{u,j}^{X} \sigma_{u,j}^{Y} k_{u,j} \cdot\frac{1}{T} \sum_{t=1}^{T} \sigma_{sv,t,j}^{X} \sigma_{sv,t,j}^{Y} \rho_{sc,t,j} + \frac{1}{T} \sum_{t=1}^{T} e^{XY}_{t, \tau_{j}} + \sum_{m=1}^{6} \Biggl( \frac{1}{T} \sum_{t=1}^{T} \zeta_{m}^{n}(t, \tau_{j}) \Biggr) \\
&\quad{}+ \frac{1}{T} \sum_{t=1}^{T} \bigl( \sigma_{u,j}^{X} \sigma_{u,j}^{Y} k_{u,j} \cdot\sigma_{sv,t,j}^{X} \sigma_{sv,t,j}^{Y} \rho_{sc,t,j} \cdot\bigl( \alpha_{t,j}^{n}-1\bigr) + \sigma_{t,j}^{X} \sigma_{t,j}^{Y} \sqrt{1- \rho_{t,j}^{2}} \cdot\beta_{t,j}^{n} \bigr) \\
&\equiv\sigma_{u,j}^{X} \sigma_{u,j}^{Y} k_{u,j} \cdot\frac{1}{T} \sum_{t=1}^{T} \sigma_{sv,t,j}^{X} \sigma_{sv,t,j}^{Y} \rho_{sc,t,j} + \mathrm{I}_{j,n,T} + \mathrm{II}_{j,n,T} + \mathrm{III}_{j,n,T}.
\end{align*}
First, we observe that by the polarization identity and Lemma~12 of \citet{andersen-su-todorov-zhang:24a}, it readily holds that
\begin{equation*}
\mathbb{E} \bigl( \llvert\mathrm{I}_{j,n,T} \rrvert\bigr) = \mathbb{E} \Biggl( \Biggl\llvert\frac{1}{T} \sum_{t=1}^{T} e^{XY}_{t, \tau_{j}} \Biggr\rrvert\Biggr) \leq C n^{-2 \varpi},
\end{equation*}
where $ \varpi\in(0,1/2)$. Hence, $\mathbb{E} (| \mathrm{I}_{j,n,T}| ) \rightarrow0$. Second, the convergence $\mathbb{E} (| \mathrm{II}_{j,n, T}| ) \rightarrow0$ is a direct consequence of Lemmas~\ref{lem1}--\ref{Lemma2}. Third, it is straightforward to deduce that
\begin{equation*}
\mathbb{E} \bigl[\bigl( \alpha_{t,j}^{n}-1 \bigr)^{2} \bigr] \leq\frac{C}{k_{n}} \quad\text{and} \quad\mathbb{E} \bigl[\bigl( \beta_{t,j}^{n} \bigr)^{2} \bigr] \leq\frac{C}{k_{n}},
\end{equation*}
uniformly in $t$ and $j$. Thus, by the boundedness condition $\mathbb{E} (| \mathrm{III}_{j,n,T}| ) \rightarrow0$.

We write
\begin{equation*}
\eta\equiv\mathbb{E} \bigl( \sigma_{sv,t,j}^{X} \sigma_{sv,t,j}^{Y} \rho_{sc,t,j} \bigr) = \mathbb{E} \bigl( c_{sv, 1}^{XY} \bigr).
\end{equation*}
By applying the law of iterated expectations, H\"{o}lder's inequality and the mixing property in Assumption~\ref{ass(C5)}, for any $\omega> (1+ \ell)/ \ell$,
\begin{equation*}
\begin{aligned}
&\mathbb{E} \Biggl[ \Biggl( \frac{1}{T} \sum_{t=1}^{T} \sigma_{sv,t,j}^{X} \sigma_{sv,t,j}^{Y} \rho_{sc,t,j} - \eta\Biggr)^{2} \Biggr] \\
&\quad= \frac{2}{T^{2}} \sum_{t=1}^{T} \sum_{v=t+1}^{T} \mathbb{E} \bigl[ \bigl( \sigma_{sv,t,j}^{X} \sigma_{sv,t,j}^{Y} \rho_{sc,t,j} - \eta\bigr) \mathbb{E}_{t} \bigl( \sigma_{sv,v,j}^{X} \sigma_{sv,v,j}^{Y} \rho_{sc,v,j} - \eta\bigr) \bigr] \\
&\quad\quad{} + \frac{1}{T^{2}} \sum_{t=1}^{T} \mathbb{E} \bigl( \sigma_{sv,t,j}^{X} \sigma_{sv,t,j}^{Y} \rho_{sc,t,j} - \eta\bigr)^{2} \\
&\quad\leq\frac{2}{T^{2}} \sum_{t=1}^{T} \sum_{v=t+1}^{T} \bigl( \mathbb{E} \bigl[ \bigl\llvert\sigma_{sv,t,j}^{X} \sigma_{sv,t,j}^{Y} \rho_{sc,t,j} - \eta\bigr\rrvert^{\omega} \bigr] \bigr)^{1/ \omega} \\
&\quad\quad{}\cdot\bigl( \mathbb{E} \bigl[ \bigl\llvert\mathbb{E}_{t} \bigl( \sigma_{sv,v,j}^{X} \sigma_{sv,v,j}^{Y} \rho_{sc,v,j} - \eta\bigr) \bigr\rrvert^{\omega/( \omega-1)} \bigr] \bigr)^{1-1/ \omega} + \frac{C}{T} \\
&\quad\leq\frac{C}{T^{2}} \sum_{t=1}^{T} \sum_{v = t+1}^{T} \alpha_{v-t}^{1-2/ \omega} + \frac{C}{T} \\
&\quad\leq\frac{C}{T},%
\end{aligned} %
\end{equation*}
where $\mathbb{E}_{t} ( \cdot) \equiv\mathbb{E}( \cdot\mid\mathcal{G}_{t})$ denotes the conditional expectation with respect to the $\sigma$-algebra $\mathcal{G}_{t} = \sigma(M_{u} \mid u \leq t)$ from Assumption~\ref{ass(C5)}. Therefore,
\begin{equation*}
\tilde{c}_{u, \tau_{j}}^{XY} - \sigma_{u, \tau_{j}}^{X} \sigma_{u, \tau_{j}}^{Y} k_{u, \tau_{j}} \cdot\mathbb{E} \bigl( \sigma_{sv,t,j}^{X} \sigma_{sv,t,j}^{Y} \rho_{sc,t,j} \bigr) = \tilde{c}_{u, \tau_{j}}^{XY} - c_{u, \tau_{j}}^{XY} \mathbb{E} \bigl(c_{sv, 1}^{XY} \bigr) \overset{p}{\longrightarrow} 0.
\end{equation*}
Now, we turn to $\bar{c}_{sv}^{XY}$. By Assumptions \ref{ass(C1)}--\ref{ass(C2)}:
\begin{equation*}
\begin{aligned} \bar{c}_{sv}^{XY}
&= \frac{1}{n/k_{n}} \sum_{j=1}^{n/k_{n}} \tilde{c}_{u, \tau_{j}}^{XY} \\
&= \mathbb{E} \bigl(c^{XY}_{sv, 1} \bigr) \frac{1}{n/k_{n}} \sum_{j=1}^{n/k_{n}} \sigma_{u,j}^{X} \sigma_{u,j}^{Y} k_{u,j} \\
&\quad{}+ \frac{1}{n/k_{n}} \sum_{j=1}^{n/k_{n}} \sigma_{u,j}^{X} \sigma_{u,j}^{Y} k_{u,j} \cdot\frac{1}{T} \sum_{t=1}^{T} \bigl( \sigma_{sv,t,j}^{X} \sigma_{sv,t,j}^{Y} \rho_{sc,t,j} - \mathbb{E} \bigl[c_{sv, 1}^{XY} \bigr] \bigr) \\
&+ \frac{1}{n/k_{n}} \sum_{j=1}^{n/k_{n}} ( \mathrm{I}_{j,n,T}+ \mathrm{II}_{j,n,T}+ \mathrm{III}_{j,n,T} ) \\
&\overset{p}{\longrightarrow} \mathbb{E} \bigl(c_{sv, 1}^{XY} \bigr) \int_{0}^{1} \sigma_{u,s}^{X} \sigma_{u,s}^{Y} k_{u,s} \,\mathrm{d}s = \mathbb{E} \bigl(c_{sv,1}^{XY} \bigr).%
\end{aligned} %
\end{equation*}
Hence,
\begin{equation*}
\tilde{c}_{u, \tau_{j}}^{X} \overset{p}{\longrightarrow} c_{u, \tau_{j}}^{X} \mathbb{E} \bigl(c_{sv, 1}^{X} \bigr) \quad\text{and} \quad\tilde{c}_{u, \tau_{j}}^{Y} \overset{p}{\longrightarrow} c_{u, \tau_{j}}^{Y} \mathbb{E} \bigl(c_{sv,1}^{Y} \bigr).
\end{equation*}
Moreover,
\begin{equation*}
\bar{c}_{sv}^{X} \overset{p}{\longrightarrow} \mathbb{E} \bigl[\bigl( \sigma_{sv,1}^{X} \bigr)^{2} \bigr] \int_{0}^{1}\bigl( \sigma_{u,s}^{X}\bigr)^{2} \,\mathrm{d}s \quad\text{and} \quad\bar{c}_{sv}^{Y} \overset{p}{\longrightarrow} \mathbb{E} \bigl[\bigl( \sigma_{sv,1}^{Y} \bigr)^{2} \bigr] \int_{0}^{1}\bigl( \sigma_{u,s}^{Y}\bigr)^{2} \,\mathrm{d}s.
\end{equation*}
According to Assumption~\ref{ass(C2)}, $\int_{0}^{1} \sigma_{u,s}^{X} \sigma_{u,s}^{Y} k_{u,s} \,\mathrm{d}s = 1$. This implies that $\int_{0}^{1}( \sigma_{u,s}^{X})^{2} \,\mathrm{d}s = 1$ and $\int_{0}^{1}( \sigma_{u,s}^{Y})^{2} \,\mathrm{d}s = 1$ for $X = Y$, where $k_{u,t} = 1$, so by the continuous mapping theorem
\begin{equation*}
\hat{c}_{u, \tau_{j}} \overset{p}{\longrightarrow} c_{u, \tau_{j}}, \qquad\hat{k}_{u, \tau_{j}} \overset{p}{\longrightarrow} k_{u, \tau_{j}},
\end{equation*}
and
\begin{equation*}
\bar{\rho}_{sc} = \frac{\bar{c}_{sv}^{XY}}{\sqrt{\bar{c}_{sv}^{X}} \sqrt{\bar{c}_{sv}^{Y}}} \overset{p}{\longrightarrow} \frac{\mathbb{E}\bigl(c_{sv,1}^{XY} \bigr)}{\sqrt{\mathbb{E}\bigl(c_{sv,1}^{X}\bigr)} \sqrt{\mathbb{E}\bigl(c_{sv,1}^{Y}\bigr)}} = E_{\bar{\rho}_{sc}}.
\end{equation*}
Finally, by Assumption~\ref{ass(C2)}, it follows trivially that $\mathbb{E} [ \lVert c_{u, \tau_{j}} - c_{u, \tau} \rVert] \rightarrow0$. This concludes the proof of Theorem~\ref{theorem:consistency}.
\end{proof}

\begin{proof}[Proof of Theorem~\ref{theorem:functional-clt}]
We adopt the strategy from the proof of Theorem~2 in \citet{andersen-su-todorov-zhang:24a}. Recall that for $Z \in\{ X, XY, Y \}$,
\begin{equation*}
A_{t, \tau}^{Z} = c_{t-1+ \tau}^{Z} -c_{u, \tau}^{Z} \int_{t-1}^{t}c_{s}^{Z} \,\mathrm{d}s.
\end{equation*}
Suppose $Z = X$ and note that for $\tau\in[0,1]$,
\begin{equation*}
\begin{aligned} \hat{c}_{u, \tau}^{X}
- c_{u, \tau}^{X} &= \frac{\tilde{c}_{u, \tau}^{X}}{\bar{c}_{sv}^{X}} - c_{u, \tau}^{X} \\
&= \frac{1}{\bar{c}_{sv}^{X}} \bigl( \tilde{c}_{u, \tau}^{X} - c_{u, \tau}^{X} \bar{c}_{sv}^{X} \bigr) \\
&= \frac{1}{\bar{c}_{sv}^{X}} \Biggl( \frac{1}{T} \sum_{t=1}^{T} \bigl( \hat{c}_{t, \tau}^{X} - c_{u, \tau}^{X} \operatorname{RV}_{t}^{n}(X) \bigr) \Biggr) \\
&= \frac{1}{\bar{c}_{sv}^{X}} \Biggl( \frac{1}{T} \sum_{t=1}^{T} \bigl( \hat{c}_{t, \tau}^{X} - c_{t-1+ \tau}^{X}\bigr) - c_{u, \tau}^{X} \frac{1}{T} \sum_{t=1}^{T} \biggl(\operatorname{RV}_{t}^{n}(X) - \int_{t-1}^{t}c_{s}^{X} \,\mathrm{d}s \biggr)+ \frac{1}{T} \sum_{t=1}^{T} A_{t, \tau}^{X} \Biggr),%
\end{aligned}
\end{equation*}
where $\operatorname{RV}_{t}^{n}(X) \equiv\frac{1}{n/k_{n}} \sum_{j=1}^{n/k_{n}} \hat{c}_{t, \tau_{j}}^{X}$, and $\hat{c}_{t, \tau_{j}}^{X}$ is the (1,1) element of \eqref{equation:realized-covariance}.

By Theorem~\ref{theorem:consistency}, $\bar{c}_{sv}^{X} \overset{p}{\longrightarrow} \mathbb{E} (c_{sv,1}^{X} )$. Furthermore, the proof of Theorem~2 in \citet{andersen-su-todorov-zhang:24a} implies that
\begin{equation*}
\sqrt{T} \Biggl( \frac{1}{T} \sum_{t=1}^{T} \bigl( \hat{c}_{t, \tau}^{X} - c_{t-1+ \tau}^{X} \bigr) \Biggr) \overset{p}{\longrightarrow} 0 \quad\text{and} \quad\sqrt{T} \biggl( \frac{1}{T} \sum_{t=1}^{T} \Bigl( \operatorname{RV}_{t}^{n}(X) - \int_{t-1}^{t} c_{s}^{X} \,\mathrm{d}s \Bigr) \biggr) \overset{p} \longrightarrow0.
\end{equation*}
An analogous result holds for other selections of $Z$. Thus, it suffices to show that
\begin{equation*}
\frac{1}{\sqrt{T}} \sum_{t=1}^{T}
\begin{pmatrix} %
\begin{bmatrix}
A_{t, \tau}^{X} \\[0.10cm]
A_{t, \tau}^{XY} \\[0.10cm]
A_{t, \tau}^{Y} \end{bmatrix} %
\oslash %
\begin{bmatrix} \bar{c}_{sv}^{X}
\\[0.10cm]
\bar{c}_{sv}^{XY} \\[0.10cm]
\bar{c}_{sv}^{Y} \end{bmatrix} %
\end{pmatrix}
\overset{d}{\longrightarrow} \mathcal{W}_{\tau}.
\end{equation*}
We denote the process
\begin{equation*}
\tilde{A}_{t, \tau}^{Z} = \sum_{j=1}^{\infty} \bigl( \mathbb{E}_{t} \bigl(A_{t+j, \tau}^{Z} \bigr) - \mathbb{E}_{t-1} \bigl(A_{t+j, \tau}^{Z} \bigr) \bigr),
\end{equation*}
where $\mathbb{E}_{t}( \cdot)$ is defined as in the proof of Theorem~\ref{theorem:consistency}.

Following Lemma~14 in the Supplementary Appendix of \citet{andersen-su-todorov-zhang:24a}, it follows that $\tilde{A}_{t, \tau}^{Z}$ is well-defined and
\begin{equation*}
\frac{1}{\sqrt{T}} \sum_{t=1}^{T} \bigl(A_{t, \tau}^{Z} - \tilde{A}_{t, \tau}^{Z} \bigr) \overset{p}{\longrightarrow} 0, \qquad\frac{1}{T^{3/2}} \sum_{t=1}^{T} \mathbb{E} \bigl( \bigl\llvert\tilde{A}_{t, \tau}^{Z} \bigr\rrvert^{3} \bigr) \overset{p}{\longrightarrow} 0,
\end{equation*}
and
\begin{equation*}
\frac{1}{T} \sum_{t=1}^{T} \mathbb{E}_{t-1} \bigl( \tilde{A}_{t, \kappa}^{Z} \tilde{A}_{t, \tau}^{Z} \bigr) \overset{p}{\longrightarrow} \mathbb{E} \bigl( \tilde{A}_{1, \kappa}^{Z} \tilde{A}_{1, \tau}^{Z} \bigr) = \sum_{h=-\infty}^{\infty} v_{\kappa, \tau}^{Z}(h),
\end{equation*}
with $v_{\kappa, \tau}^{Z}(h) = \operatorname{cov}(A_{1, \kappa}^{Z}, A_{1, \tau+h}^{Z})$. It also follows for the finite-dimensional covariance that
\begin{equation*}
\mathbb{E} \bigl( \tilde{A}_{1, \kappa}^{X} \tilde{A}_{1, \tau}^{Y} \bigr) = \sum_{j=0}^{\infty} \mathbb{E} \bigl(A_{j+1, \kappa}^{X} A_{1, \tau}^{Y} \bigr) + \sum_{j=1}^{\infty} \mathbb{E} \bigl(A_{1, \kappa}^{X} A_{j+1, \tau}^{Y} \bigr) = \sum_{h=- \infty}^{\infty}v_{\kappa, \tau}^{X,Y}(h),
\end{equation*}
since the expectation of $A_{t, \tau}^{Z}$ is zero for all $Z \in\{ X, XY, Y \}$. Repeating the computation for other cross-products, we conclude that
\begin{align*}
&\frac{1}{T} \sum_{t=1}^{T} \mathbb{E}_{t-1} %
\begin{bmatrix}
\tilde{A}_{t, \kappa}^{X} \tilde{A}^{X}_{t, \tau} & \tilde{A}_{t, \kappa}^{X} \tilde{A}_{t, \tau}^{XY} & \tilde{A}_{t, \kappa}^{X} \tilde{A}_{t, \tau}^{Y} \\[6pt]
\tilde{A}_{t, \kappa}^{XY} \tilde{A}_{t, \tau}^{X} & \tilde{A}_{t, \kappa}^{XY} \tilde{A}_{t, \tau}^{XY} & \tilde{A}_{t, \kappa}^{XY} \tilde{A}_{t, \tau}^{Y} \\[6pt]
\tilde{A}_{t, \kappa}^{Y} \tilde{A}_{t, \tau}^{X} & \tilde{A}_{t, \kappa}^{Y} \tilde{A}_{t, \tau}^{XY} & \tilde{A}_{t, \kappa}^{Y} \tilde{A}_{t, \tau}^{Y}
\end{bmatrix} %
\\
&\quad\overset{p}\longrightarrow\sum_{h=- \infty}^{\infty} %
\begin{bmatrix} v_{\kappa, \tau}^{X}(h) & v_{\kappa, \tau}^{X,XY}(h)
& v_{\kappa, \tau}^{X,Y}(h) \\[6pt]
v_{\kappa, \tau}^{XY,X}(h) & v_{\kappa, \tau}^{XY}(h) & v_{\kappa, \tau}^{XY,Y}(h) \\[6pt]
v_{\kappa, \tau}^{Y,X}(h) & v_{\kappa, \tau}^{Y,XY}(h) & v_{\kappa, \tau}^{Y}(h) \end{bmatrix} %
.
\end{align*}
Hence, finite-dimensional convergence follows by Slutsky's theorem.

To establish the functional convergence in law, we follow the proof of Theorem~2 in the Supplementary Appendix of \citet{andersen-su-todorov-zhang:24a} by verifying three sufficient conditions (for the multivariate version of the problem). To begin with, we write the entries of the covariance operator matrix as follows:\vspace*{-2pt}
\begin{equation*}
\mathcal{K}^{ij} y( \tau) = \int_{0}^{1} \Gamma_{\kappa, \tau}^{ij} y( \kappa) \,\mathrm{d} \kappa,
\end{equation*}
for any $y \in\mathcal{L}^{2}$ and $i, j=1, \dots, 3$.

First, note that for $i=j$:\vspace*{-2pt}
\begin{align*}
\frac{1}{\mathbb{E}\bigl(c_{sv,1}^{X} \bigr)^{2}} \frac{1}{T} \sum_{t=1}^{T} \mathbb{E}_{t-1} \bigl( \bigl\lVert\tilde{A}_{t, \tau}^{X} \bigr\rVert^{2} \bigr) \overset{p}{\longrightarrow} \int_{0}^{1} \Gamma_{\tau, \tau}^{11} \,\mathrm{d} \tau&= \Trace\bigl( \mathcal{K}^{11} \bigr), \\
\frac{1}{\mathbb{E}\bigl(c_{sv,1}^{XY} \bigr)^{2}} \frac{1}{T} \sum_{t=1}^{T} \mathbb{E}_{t-1} \bigl( \bigl\lVert\tilde{A}_{t, \tau}^{XY} \bigr\rVert^{2} \bigr) \overset{p}{\longrightarrow} \int_{0}^{1} \Gamma_{\tau, \tau}^{22} \,\mathrm{d} \tau&= \Trace\bigl( \mathcal{K}^{22} \bigr), \\
\frac{1}{\mathbb{E}\bigl(c_{sv,1}^{Y} \bigr)^{2}} \frac{1}{T} \sum_{t=1}^{T} \mathbb{E}_{t-1} \bigl( \bigl\lVert\tilde{A}_{t, \tau}^{Y} \bigr\rVert^{2} \bigr) \overset{p}{\longrightarrow} \int_{0}^{1} \Gamma_{\tau, \tau}^{33} \,\mathrm{d} \tau&= \Trace\bigl( \mathcal{K}^{33} \bigr).
\end{align*}
The other cases can be handled individually. For example, for $i=1$ and $j=3$,\vspace*{-2pt}
\begin{equation*}
\frac{1}{\mathbb{E}\bigl(c_{sv,1}^{X}\bigr) \mathbb{E} \bigl(c_{sv,1}^{Y}\bigr)} \frac{1}{T} \sum_{t=1}^{T} \mathbb{E}_{t-1} \bigl( \bigl\langle\tilde{A}_{t, \tau}^{X}, \tilde{A}_{t, \tau}^{Y} \bigr\rangle\bigr) \overset{p}{\longrightarrow} \int_{0}^{1} \Gamma_{\tau, \tau}^{13} \,\mathrm{d} \tau= \Trace\bigl( \mathcal{K}^{13}\bigr).
\end{equation*}
Second, it is straightforward to show that\vspace*{-2pt}
\begin{equation*}
\frac{1}{\mathbb{E} \bigl(c_{sv,1}^{Z} \bigr)^{3}T^{3/2}} \sum_{t=1}^{T} \mathbb{E}_{t-1} \bigl( \bigl\lVert\tilde{A}_{t, \tau}^{Z} \bigr\rVert^{3} \bigr) \overset{p}{\longrightarrow} 0,
\end{equation*}
and, therefore, the conditional Lyapunov condition follows immediately from the conditional Cauchy--Schwarz inequality.

Third, for an orthonormal basis $\{e_{i} \}_{i \in\mathbb{N}^{+}}$ in $\mathcal{L}^{2}$,\vspace*{-2pt}
\begin{equation*}
\begin{aligned} \frac{1}{
\mathbb{E} \bigl(c_{sv,1}^{X}\bigr)^{2}T} \sum_{t=1}^{T} \mathbb{E}_{t-1} \bigl( \bigl\langle\tilde{A}_{t, \tau}^{X}, e_{j} \bigr\rangle\bigl\langle\tilde{A}_{t, \tau}^{X}, e_{k} \bigr\rangle\bigr) &\overset{p}{\longrightarrow} \int_{0}^{1} \int_{0}^{1} \Gamma_{\kappa, \tau}^{11}e_{j}( \kappa)e_{k}( \tau) \,\mathrm{d} \kappa\,\mathrm{d} \tau= \bigl\langle\mathcal{K}^{11}e_{j}, e_{k} \bigr\rangle, \\
\frac{1}{\mathbb{E} \bigl(c_{sv,1}^{XY}\bigr) ^{2}T} \sum_{t=1}^{T} \mathbb{E}_{t-1} \bigl( \bigl\langle\tilde{A}^{XY}_{t, \tau}, e_{j} \bigr\rangle\bigl\langle\tilde{A}^{XY}_{t, \tau}, e_{k} \bigr\rangle\bigr) &\overset{p}{\longrightarrow} \int_{0}^{1} \int_{0}^{1} \Gamma_{\kappa, \tau}^{22}e_{j}( \kappa)e_{k}( \tau) \,\mathrm{d} \kappa\,\mathrm{d} \tau= \bigl\langle\mathcal{K}^{22}e_{j}, e_{k} \bigr\rangle, \\
\frac{1}{\mathbb{E} \bigl(c_{sv,1}^{Y}\bigr)^{2}T} \sum_{t=1}^{T} \mathbb{E}_{t-1} \bigl( \bigl\langle\tilde{A}_{t, \tau}^{Y}, e_{j} \bigr\rangle\bigl\langle\tilde{A}_{t, \tau}^{Y}, e_{k} \bigr\rangle\bigr) &\overset{p}{\longrightarrow} \int_{0}^{1} \int_{0}^{1} \Gamma_{\kappa, \tau}^{33}e_{j}( \kappa)e_{k}( \tau) \,\mathrm{d} \kappa\,\mathrm{d} \tau= \bigl\langle\mathcal{K}^{33}e_{j}, e_{k} \bigr\rangle.%
\end{aligned} %
\end{equation*}
As above, the other cases are handled on a stand-alone basis, such as $i=1$ and $j=3$,\vspace*{-2pt}
\begin{align*}
&\frac{1}{\mathbb{E}\bigl(c_{sv,1}^{X} \bigr) \mathbb{E}\bigl(c_{sv,1}^{Y}\bigr)T} \sum_{t=1}^{T} \mathbb{E}_{t-1} \bigl( \bigl\langle\tilde{A}_{t, \tau}^{X}, e_{j} \bigr\rangle\bigl\langle\tilde{A}_{t, \tau}^{Y}, e_{k} \bigr\rangle\bigr) \\
&\quad\overset{p}{\longrightarrow} \int_{0}^{1} \int_{0}^{1} \Gamma_{\kappa, \tau}^{13}e_{j}( \kappa)e_{k}( \tau) \,\mathrm{d} \kappa\,\mathrm{d} \tau= \bigl\langle{\mathcal{K}}^{13}e_{j}, e_{k} \bigr\rangle.
\end{align*}
Hence, the functional convergence follows and the proof is complete.\vadjust{\goodbreak}
\end{proof}

\begin{proof}[Proof of Proposition~\ref{proposition:hac}]
We define
\begin{align}
V_{\tau}^{X} &= \sum_{h=- \infty}^{\infty} \operatorname{cov}\bigl(A_{\tau}^{X},A_{\tau+h}^{X} \bigr), \tag{P.1} \label{eqP.1} \\
V_{\tau}^{XY} &= \sum_{h=- \infty}^{\infty} \operatorname{cov}\bigl(A_{\tau}^{XY},A_{\tau+h}^{XY} \bigr), \tag{P.2} \label{eqP.2} \\
V_{\tau}^{Y} &= \sum_{h=- \infty}^{\infty} \operatorname{cov}\bigl(A_{\tau}^{Y},A_{\tau+h}^{Y} \bigr), \tag{P.3} \label{eqP.3} \\
V_{\tau}^{X,XY} &= \sum_{h=- \infty}^{\infty} \operatorname{cov}\bigl(A_{\tau}^{X},A_{\tau+h}^{XY} \bigr), \tag{P.4} \label{eqP.4} \\
V_{\tau}^{XY,Y} &= \sum_{h=- \infty}^{\infty} \operatorname{cov}\bigl(A_{\tau}^{XY},A_{\tau+h}^{Y} \bigr), \tag{P.5} \label{eqP.5} \\
V_{\tau}^{X,Y} &= \sum_{h=- \infty}^{\infty} \operatorname{cov}\bigl(A_{\tau}^{X},A_{\tau+h}^{Y} \bigr). \tag{P.6} \label{eqP.6}
\end{align}
We also set
\begin{align}
\hat{V}_{\tau}^{X} &= \hat{\nu}_{\tau,0}^{X} + \sum_{h=1}^{H_{T}} \omega\biggl( \frac{h}{H_{T}} \biggr) \bigl( \hat{\nu}_{\tau,h}^{X} + \hat{\nu}_{\tau,-h}^{X} \bigr), \tag{E.1}
\label{eqE.1}
\\
\hat{V}_{\tau}^{X,XY} &= \hat{\nu}_{\tau,0}^{X,XY} + \sum_{h=1}^{H_{T}} \omega\biggl( \frac{h}{H_{T}} \biggr) \bigl( \hat{\nu}_{\tau,h}^{X,XY} + \hat{\nu}_{\tau,-h}^{X,XY} \bigr), \tag{E.2}
\label{eqE.2}
\\
\hat{V}_{\tau}^{XY} &= \hat{\nu}_{\tau,0}^{XY} + \sum_{h=1}^{H_{T}} \omega\biggl( \frac{h}{H_{T}} \biggr) \bigl( \hat{\nu}_{\tau,h}^{XY} + \hat{\nu}_{\tau,-h}^{XY} \bigr), \tag{E.3}
\label{eqE.3}
\\
\hat{V}_{\tau}^{X,Y} &= \hat{\nu}_{\tau,0}^{X,Y} + \sum_{h=1}^{H_{T}} \omega\biggl( \frac{h}{H_{T}} \biggr) \bigl( \hat{\nu}_{\tau,h}^{X,Y} + \hat{\nu}_{\tau,-h}^{X,Y} \bigr), \tag{E.4}
\label{eqE.4}
\\
\hat{V}_{\tau}^{XY,Y} &= \hat{\nu}_{\tau,0}^{XY,Y} + \sum_{h=1}^{H_{T}} \omega\biggl( \frac{h}{H_{T}} \biggr) \bigl( \hat{\nu}_{\tau,h}^{XY,Y} + \hat{\nu}_{\tau,-h}^{XY,Y} \bigr), \tag{E.5}
\label{eqE.5}
\\
\hat{V}_{\tau}^{Y} &= \hat{\nu}_{\tau,0}^{Y} + \sum_{h=1}^{H_{T}} \omega\biggl( \frac{h}{H_{T}} \biggr) \bigl( \hat{\nu}_{\tau,h}^{Y} + \hat{\nu}_{\tau,-h}^{Y} \bigr), \tag{E.6}
\label{eqE.6}
\end{align}
where
\begin{align*}
\hat{\nu}_{\tau,h}^{X} &= \frac{1}{T} \sum_{t=h+1}^{T} \hat{A}_{t, \tau}^{X} \hat{A}_{t-h, \tau}^{X}, \\
\hat{\nu}_{\tau,h}^{XY} &= \frac{1}{T} \sum_{t=h+1}^{T} \hat{A}_{t, \tau}^{XY} \hat{A}_{t-h, \tau}^{XY}, \\
\hat{\nu}_{\tau,h}^{Y} &= \frac{1}{T} \sum_{t=h+1}^{T} \hat{A}_{t, \tau}^{Y} \hat{A}_{t-h, \tau}^{Y}, \\
\hat{\nu}_{\tau,h}^{X,XY} &= \frac{1}{T} \sum_{t=h+1}^{T} \hat{A}_{t, \tau}^{X} \hat{A}_{t-h, \tau}^{XY}, \\
\hat{\nu}_{\tau,h}^{XY,Y} &= \frac{1}{T} \sum_{t=h+1}^{T} \hat{A}_{t, \tau}^{XY} \hat{A}_{t-h, \tau}^{Y}, \\
\hat{\nu}_{\tau,h}^{X,Y} &= \frac{1}{T} \sum_{t=h+1}^{T} \hat{A}_{t, \tau}^{X} \hat{A}_{t-h, \tau}^{Y},
\end{align*}
with $\widehat{A}_{t, \kappa}^{X}$, $\widehat{A}_{t, \kappa}^{XY}$ and $\widehat{A}_{t, \kappa}^{Y}$ as defined in \eqref{eq45}, and $\omega( h, H_{T}) \equiv\omega( h/H_{T} )$ a kernel function upholding basic regularity conditions (e.g., \citet{andrews:91a}). Then, the required results follow from the next proposition.
\end{proof}

\begin{proposition}
\label{proposition:hac-component}
Let $H_{T}$ be a deterministic sequence of integers such that $H_{T} / \sqrt{T} \rightarrow0$, $H_{T}/k_{n} \rightarrow0$, $k_{n}/ \sqrt{n} \rightarrow0$, and $H_{T}/n^{2 \varpi} \rightarrow0$. Then, it holds that\vspace*{-2pt}
\begin{equation*}
\text{(E.I)} \overset{p}{\longrightarrow} \text{(P.I)},
\end{equation*}
for $I = 1, \dots, 6$.
\end{proposition}

\begin{proof}[Proof of Proposition~\ref{proposition:hac-component}]
First, we show $\hat{V}_{\tau}^{X} \overset{p}{\longrightarrow} V_{\tau}^{X}$. To this end, we define\vspace*{-2pt}
\begin{align*}
\nu_{\tau,h}^{X,T} &= \frac{1}{T} \sum_{t=h+1}^{T} A_{t, \tau}^{X} A_{t-h, \tau}^{X} \quad\text{and} \\
V_{\tau}^{X,T} &= \nu_{\tau,0}^{X,T} + \sum_{h=1}^{H_{T}} \omega\biggl( \frac{h}{H_{T}} \biggr) \bigl( \nu_{\tau,h}^{X,T} + \nu_{\tau,-h}^{X,T} \bigr).
\end{align*}
By a standard argument for HAC estimators (see, e.g., \cite{andrews:91a}),\vspace*{-2pt}
\begin{equation*}
V_{\tau}^{X,T} \overset{p}{\longrightarrow} \sum_{h=- \infty}^{\infty} \operatorname{cov} \bigl(A_{t, \tau}^{X},A_{t+h, \tau}^{X} \bigr) = V_{\tau}^{X}.
\end{equation*}
Thus, it suffices to show $\hat{V}_{\tau}^{X} - V_{\tau}^{X,T} \overset{p}{\longrightarrow} 0$. Note that
\begin{equation*}
\hat{\nu}_{\tau,h}^{X} - \nu_{\tau,h}^{X,T} = \frac{1}{T} \sum_{t=h+1}^{T} \bigl( \hat{A}_{t, \tau}^{X} \hat{A}_{t-h, \tau}^{X} - A_{t, \tau}^{X} A_{t-h, \tau}^{X} \bigr) \equiv A_{n,T}.
\end{equation*}
By the proof of Theorem~\ref{theorem:consistency}, we deduce that
\begin{equation*}
\mathbb{E} \bigl( \llvert A_{n,T} \rrvert\bigr) = O \biggl( \Delta_{n}^{2 \varpi} \vee\frac{1}{k_{n}} \vee k_{n}^{2} \Delta_{n} \vee\frac{1}{\sqrt{T}} \biggr),
\end{equation*}
and the result follows from the rate conditions imposed a priori, that is, $H_{T}/ \sqrt{T} \rightarrow0$, $H_{T}/k_{n} \rightarrow0$, $k_{n}/ \sqrt{n} \rightarrow0$, and $H_{T}/n^{2 \varpi} \rightarrow0$.

The proofs for $\hat{V}_{\tau}^{XY}$ and $\hat{V}_{\tau}^{Y}$ follow the outline above. The last three terms can be dealt with using polarization identity for covariance. Hence, because
\begin{equation*}
\Gamma_{\tau} = %
\begin{bmatrix}
V_{\tau}^{X} & V_{\tau}^{X,XY} & V_{\tau}^{X,Y} \\
V_{\tau}^{X,XY} & V_{\tau}^{XY} & V_{\tau}^{XY,Y} \\
V_{\tau}^{X,Y} & V_{\tau}^{XY,Y} & V_{\tau}^{Y} \end{bmatrix} %
,
\end{equation*}
Proposition~\ref{proposition:hac-component} follows upon noting that $H_{T} / \sqrt{T} \rightarrow0$ and $T / n^{4 \varpi} \rightarrow0$ lead to $H_{T} / n^{2 \varpi} \rightarrow0$.
\end{proof}

The functional CLT for the correlation process is derived by applying the functional delta method to \eqref{equation:functional-clt}. On the one hand, the mapping $g(x,y,z) = y(xz)^{-1/2}$ is not Hadamard differentiable on the unrestricted product space $\mathcal{L}^{2}([0,1])^{3}$, as shown in \citet{andersen-tan-tan-todorov-zhang:26a}. On the other hand, the local covariance estimators employed in this paper are positive semidefinite by construction and, hence, their correlation ratios are bounded. As we show in Lemma~\ref{lemma:covariance-cone} below, Hadamard differentiability of $g$ can be established on a proper subset of $\mathcal{L}^{2}([0,1])^{3}$ under an additional nondegeneracy condition.

To accomplish this, we define a ``covariance cone'':
\begin{equation*}
\mathcal{D}_{M} = \left\{(x,y,z) \in\mathcal{L}^{2}([0,1])^{3} : x \geq0, z \geq0, |y| \leq M \sqrt{xz} \ \text{a.e.} \right\},
\tag{A.1}
\end{equation*}
for a fixed constant $0 < M < \infty$.

$\mathcal{D}_{M}$ is a closed convex set, because $(x,y,z) \in\mathcal{D}_M$ if and only if
\begin{equation*}
\begin{pmatrix}
x & y/M \\ y/M & z
\end{pmatrix}
\end{equation*}
is positive semidefinite a.e.

In our framework, the generic element $(x,y,z)$ of $\mathcal{D}_{M}$ is identified with the triplet of diurnal covariance components. Writing $c_{u} = \big(c_{u, \cdot}^{X}, c_{u, \cdot}^{XY}, c_{u, \cdot}^{Y} \big)$ and $\widehat{c}_{u} = \big( \widehat{c}_{u, \cdot}^{X}, \widehat{c}_{u, \cdot}^{XY}, \widehat{c}_{u, \cdot}^{Y} \big)$, we regard $c_{u}$ as a deterministic element and $\widehat{c}_{u}$ as a random element of $\mathcal L^2([0,1])^3$, so that \eqref{equation:functional-clt} reads $\sqrt{T}( \widehat{c}_{u} - c_{u}) \overset{d}{\longrightarrow} \mathcal{W}$, where $\mathcal{W} = ( \mathcal{W}_{1}, \mathcal{W}_{2}, \mathcal{W}_{3})^{\top}$ is the $\mathcal{L}^{2}([0,1])^{3}$-valued Gaussian limit of Theorem~\ref{theorem:functional-clt}.

Our estimator in \eqref{equation:realized-covariance} has the required structure. In particular, each return increment contributes the outer product of $\Big(\Delta_{(t-1)n+\ell}^{n}X \mathbbm{1}_{\mathcal{A}_{t, \tau_{j}}^{X,n}}, \Delta_{(t-1)n+\ell}^{n}Y \mathbbm{1}_{\mathcal{A}_{t, \tau_{j}}^{Y,n}} \Big)$, so every blockwise covariance matrix estimator $\widehat{c}_{t, \tau_{j}}$ is positive semidefinite. Averaging over $t$ preserves this property. Hence,
\begin{equation*}
\big| \widetilde{c}^{XY}_{u, \tau} \big|^{2} \leq\widetilde{c}_{u, \tau}^{X} \widetilde c_{u,\tau}^{Y} \quad\text{and} \quad|\widetilde k_{u,\tau}| \leq1,
\tag{A.2}
\end{equation*}
where $\widetilde{k}_{u,\tau}$ is the correlation ratio for $\widetilde{c}^{X}_{u, \tau}\widetilde{c}^{Y}_{u, \tau} > 0$ and zero otherwise.

Subsequently, \eqref{eq25}--\eqref{eq26} imply
\begin{equation}
\widehat{k}_{u, \tau} = \frac{\widetilde{k}_{u, \tau}}{\bar{\rho}_{sc}}.
\tag{A.3}
\end{equation}
We set
\begin{equation*}
\rho_{0} \equiv E_{\bar{\rho}_{sc}} = \frac{\mathbb{E}(c^{XY}_{sv,1})}{\sqrt{\mathbb{E}(c^X_{sv,1}) \mathbb{E}(c^Y_{sv,1})}},
\tag{A.4}
\end{equation*}
and suppose that $\rho_{0} \neq0$, which follows from Assumption~\ref{ass(ND)}. Theorem~\ref{theorem:consistency} then gives $\bar{\rho}_{sc} \overset{p}{\longrightarrow} \rho_{0}$. Therefore, for every fixed $M > 2/| \rho_{0}|$,
\begin{equation*}
P\left( \esssup_{\tau\in[0,1]} | \widehat{k}_{u, \tau}| \leq M \right) \rightarrow1.
\tag{A.5}
\end{equation*}
Thus, for a suitably large $M$, the estimated covariance component triplet $\widehat{c}_{u}$ belongs to $\mathcal{D}_{M}$ with probability tending to one. The population triplet belongs to $\mathcal{D}_{M}$ deterministically: $c_{u, \tau}^{X}$ and $c_{u, \tau}^{Y}$ are nonnegative, and by \eqref{eq12} the ratio $c_{u, \tau}^{XY} / \sqrt{c_{u, \tau}^{X} c_{u, \tau}^{Y}}$ equals $k_{u, \tau}$. Since $| \widetilde{k}_{u, \tau}| \leq1$ for every $\tau$ by (A.2), while $\widetilde{k}_{u, \tau} \overset{p}{\longrightarrow} k_{u, \tau} \rho_{0}$ by \eqref{eq28}, the deterministic limit obeys $|k_{u, \tau} \rho_{0}| \leq1$, that is, $|k_{u, \tau}| \leq1/| \rho_{0}|$. As $M > 2/| \rho_{0}|$, it follows that $|c_{u, \tau}^{XY}| \leq M \sqrt{c_{u, \tau}^{X} c_{u, \tau}^{Y}}$ for Lebesgue-a.e.\ $\tau$. This restriction controls the problematic shrinking-set perturbations of \citet{andersen-tan-tan-todorov-zhang:26a} by imposing a uniform bound on the correlation ratio.

We adopt a local nondegeneracy condition of the form:

\def\theassumptionnd{(ND)}
\begin{assumptionnd}\label{ass(ND)}
We assume that
\begin{equation*}
\essinf_{\tau\in[0,1]} c_{u, \tau}^{X} > 0, \quad\essinf_{\tau\in[0,1]} c_{u, \tau}^{Y} > 0,
\tag{ND.1}
\end{equation*}
and
\begin{equation*}
\mathbb{E} \big(c^{XY}_{sv,1} \big) \neq0.
\tag{ND.2}
\end{equation*}
\end{assumptionnd}

As elements of $\mathcal{L}^{2}$ are equivalence classes defined up to null sets, the lower-bound condition is stated using the essential infimum. Assumption~(ND.1) concerns only the diurnal variance components and strengthens Assumption~\ref{ass(C3)}. The latter requires $\sigma_{u,t}^{\wp} > 0$ a.e. and does not bound the curves away from zero. Without a uniform lower bound, the coefficients in $g_{\theta}'$ may be unbounded. However, we do not require $|k_{u,\tau}|$ to be bounded away from zero, so the correlation curve can cross zero. Assumption~(ND.2) is already implicit in Theorem~\ref{theorem:functional-clt}, because the covariance matrix in \eqref{eq33} contains $\mathbb{E}(c^{XY}_{sv,1})$ in the denominator, and \eqref{eq25}--\eqref{eq29} normalize by the average correlation. Assumption~(ND.1) is equivalent to the existence of $c_{L}>0$ such that $c_{u,\tau}^{X} \geq c_{L}$ and $c_{u,\tau}^{Y} \geq c_{L}$ for Lebesgue-a.e. $\tau\in[0,1]$. Assumption~(ND.2), separately, is equivalent to $\rho_{0} \neq0$, because the variance expectations in \eqref{eq29} are strictly positive.

We define the mapping:
\begin{equation*}
g(x,y,z) =
\begin{cases}
\displaystyle\frac{y}{\sqrt{xz}}, & xz>0, \\
0, & xz=0.
\end{cases}
\end{equation*}
We denote by $g_{M}$ the restriction of $g$ to $\mathcal{D}_{M}$. With the above identification, $k_{u, \cdot} = g_{M}(c_{u})$ and $\widehat{k}_{u, \cdot} = g( \widehat{c}_{u})$. We remark that, on $\{\widehat{c}_{u} \in\mathcal{D}_{M} \}$, $\widehat{k}_{u, \cdot} = g_{M}( \widehat{c}_{u})$. The boundary convention is immaterial at $c_{u}$ under Assumption~\ref{ass(ND)}, and $|g_{M}| \leq M$ everywhere on $\mathcal{D}_{M}$.

\begin{lemma} \label{lemma:covariance-cone}
Take an arbitrary $\theta= ( \theta_{1}, \theta_{2}, \theta_{3})^{\top} \in\mathcal{D}_{M}$ with $\essinf_{\tau\in[0,1]} \theta_{1}( \tau) \geq c_{L}$ and $\essinf_{\tau\in[0,1]} \theta_{3}( \tau) \geq c_{L}$, for some $c_{L} > 0$. Let $t_{n} \downarrow0$ and $h_{n} \rightarrow h$ in $\mathcal{L}^{2}([0,1])^{3}$, such that
\begin{equation*}
\theta_{n} \equiv\theta+ t_{n} h_{n} \in\mathcal{D}_{M}
\end{equation*}
for every $n$. Then,
\begin{equation*}
\left\| \frac{g_{M}( \theta_{n}) - g_{M}( \theta)}{t_{n}} - g_{\theta}'(h) \right\|_{2} \rightarrow0,
\tag{A.6}
\end{equation*}
where
\begin{equation*}
g_{\theta}'(h) = - \frac{\theta_{2}}{2 \theta_{1}^{3/2} \theta_{3}^{1/2}}h_{1} + \frac{1}{\theta_{1}^{1/2} \theta_{3}^{1/2}}h_{2} - \frac{\theta_{2}}{2 \theta_{1}^{1/2} \theta_{3}^{3/2}}h_{3}.
\tag{A.7}
\end{equation*}
\end{lemma}
Lemma~\ref{lemma:covariance-cone} shows there exists a continuous linear map $g_{\theta}' : \mathcal{L}^{2}([0,1])^{3} \rightarrow\mathcal{L}^{2}([0,1])$, so $g_{M}$ is Hadamard differentiable at $\theta$, relative to the domain $\mathcal{D}_{M}$, tangentially to $\mathcal{L}^{2}([0,1])^{3}$.

\begin{proof}[Proof of Lemma~\ref{lemma:covariance-cone}]

The coefficients in (A.7) are essentially bounded. In fact, because $|\theta_{2}| \leq M \sqrt{\theta_{1} \theta_{3}}$ and $\theta_{1}, \theta_{3} \geq c_{L}$ a.e.,
\begin{equation*}
\left| \frac{\theta_{2}}{2 \theta_{1}^{3/2} \theta_{3}^{1/2}} \right| \leq\frac{M}{2c_{L}}, \qquad\frac{1}{\sqrt{\theta_{1} \theta_{3}}} \leq\frac{1}{c_{L}}, \qquad\left| \frac{\theta_{2}}{2 \theta_{1}^{1/2} \theta_{3}^{3/2}} \right| \leq\frac{M}{2c_{L}}.
\end{equation*}
Thus, $g_{\theta}' : \mathcal{L}^{2}([0,1])^{3} \rightarrow\mathcal{L}^{2}([0,1])$ is a bounded linear operator.

We define a ``bad'' set
\begin{equation*}
B_{n} = \left\{\theta_{1,n} < \frac{c_{L}}{2} \right\} \cup\left\{\theta_{3,n} < \frac{c_{L}}{2} \right\}.
\end{equation*}
It must hold that either $|h_{1,n}| \geq c_{L}/(2t_{n})$ or $|h_{3,n}| \geq c_{L}/(2t_{n})$ on $B_{n}$. As $h_{n} \rightarrow h$ in $\mathcal{L}^{2}([0,1])^{3}$, the sequence $(h_{n})_{n=1}^{\infty}$ is bounded in $\mathcal{L}^{2}([0,1])^{3}$. Moreover, $\bigl\||h_{n}|^{2} - |h|^{2}\bigr\|_{1} \leq\norm{h_{n}-h}_{2} \bigl( \norm{h_{n}}_{2} + \norm{h}_{2} \bigr) \rightarrow0$, so $|h_{n}|^{2} \rightarrow|h|^{2}$ in $\mathcal{L}^{1}$, which implies that $(|h_{n}|^{2})_{n=1}^{\infty}$ is uniformly integrable. Furthermore,
\begin{equation*}
\lambda(B_{n}) \leq\frac{4t_{n}^{2}}{c_{L}^{2}} \int_{0}^{1}(h_{1,n}^{2} + h_{3,n}^{2}) \mathrm{d} \tau\rightarrow0.
\end{equation*}
The uniform integrability of $(|h_{n}|^{2})_{n=1}^{\infty}$ also renders
\begin{equation*}
\int_{B_{n}}(h_{1,n}^{2} + h_{3,n}^{2}) \mathrm{d}\tau\rightarrow0.
\end{equation*}
Consequently,
\begin{equation*}
\frac{\lambda(B_{n})}{t_{n}^{2}} \leq\frac{4}{c_{L}^{2}} \int_{B_{n}}(h_{1,n}^{2} + h_{3,n}^{2}) \mathrm{d}\tau\rightarrow0.
\end{equation*}
Since $|g_{M}| \leq M$ on $\mathcal{D}_{M}$,
\begin{equation}
\left\| \frac{g_{M}( \theta_{n}) - g_{M}( \theta)}{t_{n}} \mathbbm{1}_{B_{n}} \right\|_{2} \leq\frac{2M}{t_{n}} \lambda(B_{n})^{1/2} \rightarrow0.
\tag{A.8}
\end{equation}
Moreover, $\norm{g_{\theta}'(h_{n}) \mathbbm{1}_{B_{n}}}_{2} \rightarrow0$, as the coefficients of $g_{\theta}'$ are bounded and $(|h_{n}|^{2})_{n=1}^{\infty}$ is uniformly integrable.

It remains to study the ``good'' set $B_{n}^{\complement}$. For $s \in[0,1]$, we set $\theta_{n}^{(s)} = \theta+ s( \theta_{n} - \theta)$. On $B_{n}^{\complement}$ the first and third coordinates of every $\theta_{n}^{(s)}$ are at least $c_{L}/2$. And since $\mathcal{D}_{M}$ is convex, $\theta_{n}^{(s)} \in\mathcal{D}_{M}$. Furthermore, on
\begin{equation*}
\mathcal{K}_{M,c_{L}} = \left\{(x,y,z) : x,z \geq\frac{c_{L}}{2}, |y| \leq M \sqrt{xz} \right\},
\end{equation*}
the gradient
\begin{equation*}
\nabla g(x,y,z) = \left(- \frac{g(x,y,z)}{2x}, \frac{1}{\sqrt{xz}}, -\frac{g(x,y,z)}{2z} \right)
\end{equation*}
is bounded. The Hessian is also bounded on $\mathcal{K}_{M,c_{L}}$, because each second derivative is a product of a bounded power of $x^{-1}$ and $z^{-1}$ with either $1$ or $g$, while $x,z \geq c_{L}/2$ and $|g| \leq M$. Since $\mathcal{K}_{M,c_{L}}$ is convex, boundedness of the Hessian implies that $\nabla g$ is uniformly Lipschitz, and hence uniformly continuous, on $\mathcal{K}_{M,c_{L}}$.

The fundamental theorem of calculus implies that, at a.e. $\tau\in B_{n}^{\complement}$,
\begin{equation}
\frac{g_{M}( \theta_{n}) - g_{M}( \theta)}{t_{n}} - g_{\theta}'(h_{n}) = \int_{0}^{1} \Bigl[ \bigl( \nabla g \bigl( \theta_{n}^{(s)} \bigr) - \nabla g( \theta) \bigr) \cdot h_{n} \Bigr] \mathrm{d}s.
\tag{A.9}
\end{equation}
Write
\begin{equation*}
R_{n} = \sup_{0 \leq s \leq1} \bigl| \nabla g \bigl( \theta_{n}^{(s)} \bigr) - \nabla g( \theta) \bigr| \mathbbm{1}_{B_{n}^{\complement}}.
\end{equation*}
As $t_{n}h_{n} \rightarrow0$ in $\mathcal{L}^{2}$, it converges to zero in measure. Uniform continuity of $\nabla g$ implies $R_{n} \rightarrow0$ in measure, while boundedness of $\nabla g$ implies $0 \leq R_{n} \leq C < \infty$.

To conclude, we shall establish that
\begin{equation}
\int_{0}^{1} R_{n}^{2}|h_{n}|^{2} \mathrm{d} \tau\rightarrow0.
\tag{A.10}
\end{equation}
We fix $\varepsilon> 0$. Now, uniform integrability of $(|h_{n}|^{2})_{n=1}^{\infty}$ means that we can choose a constant $K < \infty$ so that
\begin{equation*}
\sup_{n} \int_{\{|h_{n}|^{2} > K \}}|h_{n}|^{2} \mathrm{d} \tau< \frac{\varepsilon}{2C^{2}}.
\end{equation*}
On the complementary set,
\begin{equation*}
\int_{\{|h_{n}|^{2} \leq K \}}R_{n}^{2}|h_{n}|^{2} \mathrm{d} \tau\leq K \int_{0}^{1} R_{n}^{2} \mathrm{d} \tau.
\end{equation*}
As $R_{n}$ is uniformly bounded and converges to zero in measure on $[0,1]$, $\int_{0}^{1} R_{n}^{2} \mathrm{d} \tau\rightarrow0$. So the last display is eventually smaller than $\varepsilon/2$, while the contribution from $\{|h_{n}|^{2} > K \}$ is at most $C^{2} \varepsilon/(2C^{2}) = \varepsilon/2$. This proves (A.10). By Jensen's inequality (in $s$) and the Cauchy--Schwarz inequality (in $\mathbb{R}^{3}$), (A.10) implies that the nonlinear remainder converges to zero in $\mathcal{L}^{2}([0,1])$ on $B_{n}^{\complement}$. Combining this with (A.8), the derivative bound on $B_{n}$, and $g_{\theta}'(h_{n}-h) \rightarrow0$ in $\mathcal{L}^{2}$, proves (A.6).
\end{proof}

Taking $\theta= c_{u}$ in (A.7), the derivative $g_{c_{u}}'$ has exactly the form of the gradient displayed in \eqref{eq37}:
\begin{equation*}
\nabla g(c_{u, \tau}^{X}, c_{u, \tau}^{XY}, c_{u,\tau}^{Y}) = (4c_{u, \tau}^{X} c_{u,\tau}^{Y})^{-1/2} \left( -\frac{c_{u, \tau}^{XY}}{c_{u, \tau}^{X}}, 2,- \frac{c_{u, \tau}^{XY}}{c_{u, \tau}^{Y}} \right).
\end{equation*}

\begin{corollary}
\label{cor:stoch}
Suppose Theorem~\ref{theorem:functional-clt} and Assumption~\ref{ass(ND)} hold and that, for some fixed $0 < M < \infty$,
\begin{equation*}
P( \widehat{c}_{u} \in\mathcal{D}_{M}) \rightarrow1 \qquad\text{and} \qquad c_{u} \in\mathcal{D}_{M}.
\tag{A.11}
\end{equation*}
Then
\begin{equation*}
\sqrt{T} \big(g( \widehat{c}_u) - g(c_{u}) \big) = g_{c_{u}}' \big( \sqrt{T}( \widehat{c}_{u} - c_{u}) \big) + o_{p}(1) \quad\text{in } \mathcal{L}^{2}([0,1]),
\tag{A.12}
\end{equation*}
and, therefore,
\begin{equation*}
\sqrt{T}( \widehat{k}_{u} - k_{u}) \overset{d}{\longrightarrow} g_{c_{u}}'( \mathcal{W}) \quad\text{in } \mathcal{L}^{2}([0,1]).
\tag{A.13}
\end{equation*}
\end{corollary}

\begin{proof}[Proof of Corollary~\ref{cor:stoch}]
Let $M$ be the fixed constant in (A.11), and define the event $A_{T} \equiv\{\widehat{c}_{u} \in\mathcal{D}_{M} \}$. By (A.11), $P(A_{T}) \rightarrow1$ and $c_{u} \in\mathcal{D}_{M}$. Now, define a modified estimator
\begin{equation*}
\widehat{c}_{u}^{(M)} = \widehat{c}_{u} \mathbbm{1}_{A_{T}} + c_{u} \mathbbm{1}_{A_{T}^{\complement}}.
\end{equation*}
By construction, $\widehat{c}_{u}^{(M)} \in\mathcal{D}_{M}$ for every $T$, almost surely, and
\begin{equation*}
\sqrt{T} \bigl( \widehat{c}_{u}^{(M)} - c_{u} \bigr) - \sqrt{T} \bigl( \widehat{c}_{u} - c_{u} \bigr) \overset{p}{\longrightarrow} 0,
\end{equation*}
because the random elements differ only on the asymptotically negligible set $A_{T}^{\complement}$. Indeed, for every $\varepsilon>0$,
\begin{equation*}
P \biggl( \left\| \sqrt{T} \bigl( \widehat{c}_{u}^{(M)} - \widehat{c}_{u} \bigr) \right\|_{2} > \varepsilon\biggr) \leq P \bigl(A_{T}^{\complement} \bigr) \rightarrow0,
\end{equation*}
so Theorem~\ref{theorem:functional-clt} holds with $\widehat{c}_{u}$ replaced by $\widehat{c}_{u}^{(M)}$.

To verify the asymptotic linearization, we define
\begin{equation*}
R_{T} = \sqrt{T} \bigl( g_{M} \bigl( \widehat{c}_{u}^{(M)} \bigr) - g_{M}(c_{u}) \bigr) - g_{c_{u}}' \left( \sqrt{T} \bigl( \widehat{c}_{u}^{(M)} - c_{u} \bigr) \right).
\end{equation*}
Since $\mathcal{L}^{2}([0,1])^{3}$ is a separable Hilbert space, for an arbitrary subsequence, the Skorokhod representation theorem provides a further subsequence and coupled versions, defined on a common probability space, such that
\begin{equation*}
\sqrt{T} \bigl( \widehat{c}_{u}^{(M)} - c_{u} \bigr) \rightarrow\mathcal{W}
\end{equation*}
almost surely in $\mathcal{L}^{2}([0,1])^{3}$.

At each sample point in this probability-one event, we apply Lemma~\ref{lemma:covariance-cone} with $\theta= c_{u}$, $t_{T} = T^{-1/2}$, and $h_{T} = \sqrt{T} \bigl( \widehat{c}_{u}^{(M)} - c_{u} \bigr)$. Its hypotheses are met: $c_{u} \in\mathcal{D}_{M}$ by (A.11); Assumption~(ND.1) delivers a constant $c_{L} > 0$ with $\essinf_{\tau\in[0,1]} \min\{ c_{u, \tau}^{X}, c_{u, \tau}^{Y} \} \geq c_{L}$; $t_{T} \downarrow0$ and $h_{T} \rightarrow\mathcal{W}$ in $\mathcal{L}^{2}([0,1])^{3}$ along the coupled subsequence; and $\theta+ t_{T} h_{T} = \widehat{c}_{u}^{(M)} \in\mathcal{D}_{M}$ for every $T$, by construction of $\widehat{c}_{u}^{(M)}$. Therefore, $R_{T} \rightarrow0$, almost surely, along the coupled subsequence, since $R_{T}$ is a measurable functional of $\sqrt{T} \bigl( \widehat{c}_{u}^{(M)} - c_{u} \bigr)$. Hence, along every subsequence, there exists a further subsequence such that $R_{T} \overset{p}{\longrightarrow} 0$. By the subsequence characterization of convergence in probability,
\begin{equation*}
R_{T} \overset{p}{\longrightarrow} 0
\end{equation*}
along the original sequence.

Since the modified and original estimators, and their correlation transforms, differ only on $A_{T}^{\complement}$, whose probability tends to zero, the asymptotic linearization in (A.12) also holds with $\widehat{c}_{u}^{(M)}$ replaced by $\widehat{c}_{u}$. Combining (A.12) with Theorem~\ref{theorem:functional-clt} and continuity of the bounded linear operator $g_{c_{u}}'$ verifies (A.13).

\end{proof}

Thus, under the local nondegeneracy condition in Assumption~\ref{ass(ND)}, \eqref{eq36} follows without requiring unrestricted Hadamard differentiability on $\mathcal{L}^{2}([0,1])^{3}$.
\begin{proof}[Proof of Theorem~\ref{theorem:functional-clt-correlation}]
(a) Since $k_{u,t}$ is a bounded function, this is a direct consequence of Theorem~\ref{theorem:consistency} and Riemann integrability.

(b) In restriction to $\mathcal{H}_{0}$, $k_{u, \tau} \equiv1$, Corollary~\ref{cor:stoch} shows that $\sqrt{T} ( \widehat{k}_{u, \cdot} - 1) \overset{d}{\longrightarrow} \mathcal{H}$ in $\mathcal{L}^{2}([0,1])$. Since $\widehat{k}_{u, \tau}$ is constant on each of the $m = n/k_{n}$ blocks, $\mathcal{N} = \lVert\sqrt{T} ( \widehat{k}_{u, \cdot} - 1) \rVert^{2}$, and the result follows by the continuous mapping theorem.

\end{proof}

\begin{proof}[Proof of Theorem~\ref{theorem:process-estimator}]
The consistency of the long-run covariance matrix estimator can be shown as in Proposition~\ref{proposition:hac-component} above. Moreover, following the proof of Theorem~6 in \citet{andersen-su-todorov-zhang:24a}, we can further show that
\begin{equation*}
\widehat{\mathcal{H}}_{\tau} \overset{d}{\longrightarrow} \mathcal{H}_{\tau},
\end{equation*}
so the result follows from Slutsky's theorem and the continuous mapping theorem.
\end{proof}

\begin{proof}[Proof of Theorem~\ref{theorem:functional-clt-random}]

The result follows from the proof of Theorem~\ref{theorem:functional-clt} and Theorem~\ref{theorem:functional-clt-correlation} without considering the estimator $\widebar{c}_{sv}$.
\end{proof}

\begin{proof}[Proof of Theorem~\ref{theorem:functional-clt-correlation-random}]
(a) Following the idea in the proof of Theorem~\ref{theorem:consistency}, we can show that $\hat{k}_{u, \tau} \overset{p}{\longrightarrow} k_{u, \tau}$ for $\tau\in[0,1]$ uniformly (because $k_{u,t}$ is bounded). Hence, the result again follows by Riemann integrability.

(b) The result follows from Theorem~\ref{theorem:functional-clt-random} and the proof of Theorem~\ref{theorem:functional-clt-correlation}.
\end{proof}

\clearpage

\section{Additional Monte Carlo analysis} \label{appendix:monte-carlo}

This Appendix contains the results (see Tables~\ref{table:sim-diurnal-correlation-t=N-r=0.60-q=0.10.tex}--\ref{table:sim-diurnal-correlation-t=N-r=0.20-q=0.05.tex}) for the Monte Carlo analysis with rejection rates of the test statistic at the $\alpha= 0.10$ and $\alpha= 0.05$ significance levels.

\begin{table}[h!]
\tabcolsep=0pt

\caption{Rejection rate of the test statistic for diurnal variation in the
correlation process at the $\alpha= 0.10$ significance level and with $\rho= 0.60$.}
\label{table:sim-diurnal-correlation-t=N-r=0.60-q=0.10.tex}
\begin{tabular*}{\textwidth}{@{\extracolsep{\fill}}D{.}{.}{5.0}D{.}{.}{3.0}cccccccccc@{}}
\hline
\multicolumn{1}{l}{\multirow{3}{*}[-17pt]{$n$}}& \multicolumn{1}{c}{\multirow{3}{*}[-17pt]{$k_{n}$}} &
\multicolumn{5}{c}{Equidistant Sampling} & \multicolumn{5}{c}{Irregular Sampling}
\\
\cline{3-7}\cline{8-12}
& & \multicolumn{5}{c}{$a = $} & \multicolumn{5}{c}{$a = $} \\
\cline{3-7}\cline{8-12}
& & 1.000 & 0.950 & 0.900 & 0.850 & 0.800 & 1.000 & 0.950 & 0.900 & 0.850 & 0.800 \\
\hline
\multicolumn{12}{l}{Panel A: $T = 5$} \\
26 & 13 & 0.232 & 0.254 & 0.322 & 0.399 & 0.488 & 0.229 & 0.251 & 0.309 & 0.390 & 0.462 \\
39 & 13 & 0.232 & 0.261 & 0.335 & 0.425 & 0.521 & 0.228 & 0.266 & 0.336 & 0.417 & 0.512 \\
78 & 26 & 0.225 & 0.288 & 0.432 & 0.584 & 0.692 & 0.229 & 0.292 & 0.428 & 0.556 & 0.674 \\
390 & 130 & 0.226 & 0.511 & 0.787 & 0.908 & 0.958 & 0.226 & 0.436 & 0.720 & 0.874 & 0.941 \\
780 & 195 & 0.217 & 0.622 & 0.880 & 0.957 & 0.983 & 0.225 & 0.456 & 0.766 & 0.915 & 0.964 \\
1{,}560 & 390 & 0.222 & 0.759 & 0.947 & 0.987 & 0.995 & 0.226 & 0.453 & 0.786 & 0.937 & 0.976 \\
4{,}680 & 936 & 0.215 & 0.909 & 0.990 & 0.997 & 0.999 & 0.210 & 0.334 & 0.628 & 0.853 & 0.947 \\[6pt]
\multicolumn{12}{l}{Panel B: $T = 22$}\\
26 & 13 & 0.141 & 0.204 & 0.348 & 0.513 & 0.655 & 0.146 & 0.204 & 0.344 & 0.519 & 0.655 \\
39 & 13 & 0.144 & 0.213 & 0.386 & 0.558 & 0.717 & 0.139 & 0.216 & 0.380 & 0.557 & 0.708 \\
78 & 26 & 0.134 & 0.320 & 0.626 & 0.820 & 0.918 & 0.135 & 0.316 & 0.616 & 0.808 & 0.913 \\
390 & 130 & 0.137 & 0.764 & 0.957 & 0.988 & 0.996 & 0.136 & 0.688 & 0.940 & 0.983 & 0.995 \\
780 & 195 & 0.131 & 0.868 & 0.983 & 0.996 & 0.998 & 0.132 & 0.747 & 0.965 & 0.990 & 0.998 \\
1{,}560 & 390 & 0.132 & 0.947 & 0.995 & 0.998 & 0.999 & 0.132 & 0.772 & 0.976 & 0.995 & 0.998 \\
4{,}680 & 936 & 0.126 & 0.988 & 0.998 & 1.000 & 1.000 & 0.127 & 0.582 & 0.952 & 0.992 & 0.997 \\[6pt]
\multicolumn{12}{l}{Panel C: $T = 66$}\\
26 & 13 & 0.119 & 0.256 & 0.519 & 0.772 & 0.889 & 0.120 & 0.253 & 0.537 & 0.764 & 0.892 \\
39 & 13 & 0.119 & 0.262 & 0.564 & 0.813 & 0.925 & 0.118 & 0.253 & 0.578 & 0.808 & 0.923 \\
78 & 26 & 0.114 & 0.512 & 0.866 & 0.966 & 0.989 & 0.112 & 0.495 & 0.871 & 0.966 & 0.990 \\
390 & 130 & 0.108 & 0.933 & 0.994 & 0.999 & 0.999 & 0.114 & 0.903 & 0.992 & 0.998 & 0.999 \\
780 & 195 & 0.105 & 0.974 & 0.998 & 1.000 & 1.000 & 0.117 & 0.941 & 0.996 & 0.999 & 1.000 \\
1{,}560 & 390 & 0.114 & 0.993 & 0.999 & 1.000 & 1.000 & 0.111 & 0.960 & 0.998 & 1.000 & 1.000 \\
4{,}680 & 936 & 0.109 & 0.999 & 1.000 & 1.000 & 1.000 & 0.106 & 0.912 & 0.997 & 0.999 & 0.999 \\
\hline
\end{tabular*}
\begin{tabnotes}
\legend{We simulate a bivariate jump-diffusion model with diurnal variation in the correlation coefficient, such that
$\rho_{t} = \rho_{sc,t} k_{u,t}$, where $\rho_{sc,t}$ is a stochastic process and $k_{u,t} = a + bt$ with $b = 2(1-a)$ captures the deterministic component. The hypothesis $\mathcal{H}_{0}: \int_{0}^{1} ( k_{u,t} - 1)^{2} \,\mathrm{d}t = 0$ is tested against $\mathcal{H}_{a}: \int_{0}^{1} ( k_{u,t} - 1)^{2} \,\mathrm{d}t \neq0$. In the model, the null is equivalent to $a = 1$, whereas the alternative corresponds to $a \neq1$. The table reports rejection rates of the test statistic from \eqref{equation:test-statistic} at significance level $\alpha= 0.10$. $n$ is the number of intraday observations over a sample period of $T$ days, while $k_{n}$ is the number of log-price increments used to compute the blockwise realized covariance estimator.}
\end{tabnotes}%
\end{table}

\newpage

\begin{table}
\tabcolsep=0pt

\caption{Rejection rate of the test statistic for diurnal variation in the
correlation process at the $\alpha= 0.10$ significance level and with $\rho= 0.40$.}
\label{table:sim-diurnal-correlation-t=N-r=0.40-q=0.10.tex}
\begin{tabular*}{\textwidth}{@{\extracolsep{\fill}}D{.}{.}{5.0}D{.}{.}{3.0}cccccccccc@{}}
\hline
\multicolumn{1}{l}{\multirow{3}{*}[-17pt]{$n$}}& \multicolumn{1}{c}{\multirow{3}{*}[-17pt]{$k_{n}$}} &
\multicolumn{5}{c}{Equidistant Sampling} & \multicolumn{5}{c}{Irregular Sampling}
\\
\cline{3-7}\cline{8-12}
& & \multicolumn{5}{c}{$a = $} & \multicolumn{5}{c}{$a = $} \\
\cline{3-7}\cline{8-12}
& & 1.000 & 0.950 & 0.900 & 0.850 & 0.800 & 1.000 & 0.950 & 0.900 & 0.850 & 0.800 \\
\hline
\multicolumn{12}{l}{Panel A: $T = 5$} \\
26 & 13 & 0.209 & 0.211 & 0.241 & 0.259 & 0.302 & 0.204 & 0.215 & 0.232 & 0.259 & 0.294 \\
39 & 13 & 0.208 & 0.217 & 0.242 & 0.275 & 0.322 & 0.211 & 0.215 & 0.247 & 0.275 & 0.312 \\
78 & 26 & 0.217 & 0.230 & 0.280 & 0.355 & 0.421 & 0.216 & 0.240 & 0.281 & 0.342 & 0.409 \\
390 & 130 & 0.224 & 0.308 & 0.482 & 0.643 & 0.748 & 0.222 & 0.293 & 0.444 & 0.588 & 0.717 \\
780 & 195 & 0.214 & 0.356 & 0.583 & 0.741 & 0.836 & 0.222 & 0.309 & 0.486 & 0.657 & 0.773 \\
1{,}560 & 390 & 0.217 & 0.451 & 0.719 & 0.850 & 0.913 & 0.228 & 0.324 & 0.523 & 0.706 & 0.820 \\
4{,}680 & 936 & 0.210 & 0.624 & 0.863 & 0.939 & 0.962 & 0.206 & 0.265 & 0.427 & 0.604 & 0.757 \\[6pt]
\multicolumn{12}{l}{Panel B: $T = 22$} \\
26 & 13 & 0.130 & 0.152 & 0.215 & 0.300 & 0.399 & 0.133 & 0.157 & 0.208 & 0.301 & 0.385 \\
39 & 13 & 0.136 & 0.161 & 0.228 & 0.329 & 0.434 & 0.137 & 0.160 & 0.225 & 0.316 & 0.423 \\
78 & 26 & 0.134 & 0.196 & 0.345 & 0.502 & 0.650 & 0.138 & 0.198 & 0.332 & 0.499 & 0.633 \\
390 & 130 & 0.135 & 0.425 & 0.735 & 0.863 & 0.920 & 0.139 & 0.374 & 0.681 & 0.829 & 0.904 \\
780 & 195 & 0.131 & 0.538 & 0.827 & 0.918 & 0.955 & 0.135 & 0.425 & 0.756 & 0.879 & 0.938 \\
1{,}560 & 390 & 0.132 & 0.689 & 0.910 & 0.955 & 0.975 & 0.134 & 0.470 & 0.809 & 0.908 & 0.955 \\
4{,}680 & 936 & 0.124 & 0.860 & 0.963 & 0.978 & 0.986 & 0.122 & 0.361 & 0.733 & 0.879 & 0.938 \\[6pt]
\multicolumn{12}{l}{Panel C: $T = 66$} \\
26 & 13 & 0.108 & 0.170 & 0.303 & 0.482 & 0.626 & 0.115 & 0.168 & 0.324 & 0.484 & 0.629 \\
39 & 13 & 0.112 & 0.176 & 0.335 & 0.519 & 0.676 & 0.115 & 0.167 & 0.332 & 0.527 & 0.671 \\
78 & 26 & 0.108 & 0.274 & 0.556 & 0.754 & 0.855 & 0.108 & 0.258 & 0.548 & 0.748 & 0.856 \\
390 & 130 & 0.104 & 0.654 & 0.894 & 0.953 & 0.974 & 0.112 & 0.591 & 0.879 & 0.946 & 0.971 \\
780 & 195 & 0.106 & 0.776 & 0.942 & 0.970 & 0.985 & 0.116 & 0.663 & 0.919 & 0.962 & 0.978 \\
1{,}560 & 390 & 0.110 & 0.880 & 0.968 & 0.984 & 0.991 & 0.108 & 0.737 & 0.942 & 0.971 & 0.982 \\
4{,}680 & 936 & 0.108 & 0.952 & 0.985 & 0.990 & 0.994 & 0.108 & 0.643 & 0.917 & 0.967 & 0.979 \\
\hline
\end{tabular*}
\begin{tabnotes}
\legend{We simulate a bivariate jump-diffusion model with diurnal variation in the correlation coefficient, such that
$\rho_{t} = \rho_{sc,t} k_{u,t}$, where $\rho_{sc,t}$ is a stochastic process and $k_{u,t} = a + bt$ with $b = 2(1-a)$ captures the deterministic component. The hypothesis $\mathcal{H}_{0}: \int_{0}^{1} ( k_{u,t} - 1)^{2} \,\mathrm{d}t = 0$ is tested against $\mathcal{H}_{a}: \int_{0}^{1} ( k_{u,t} - 1)^{2} \,\mathrm{d}t \neq0$. In the model, the null is equivalent to $a = 1$, whereas the alternative corresponds to $a \neq1$. The table reports rejection rates of the test statistic from \eqref{equation:test-statistic} at significance level $\alpha= 0.10$. $n$ is the number of intraday observations over a sample period of $T$ days, while $k_{n}$ is the number of log-price increments used to compute the blockwise realized covariance estimator.}
\end{tabnotes}%
\end{table}

\clearpage

\begin{table}
\tabcolsep=0pt

\caption{Rejection rate of the test statistic for diurnal variation in the
correlation process at the $\alpha= 0.10$ significance level and with $\rho= 0.20$.}
\label{table:sim-diurnal-correlation-t=N-r=0.20-q=0.10.tex}
\begin{tabular*}{\textwidth}{@{\extracolsep{\fill}}D{.}{.}{5.0}D{.}{.}{3.0}cccccccccc@{}}
\hline
\multicolumn{1}{l}{\multirow{3}{*}[-17pt]{$n$}}& \multicolumn{1}{c}{\multirow{3}{*}[-17pt]{$k_{n}$}} &
\multicolumn{5}{c}{Equidistant Sampling} & \multicolumn{5}{c}{Irregular Sampling}
\\
\cline{3-7}\cline{8-12}
& & \multicolumn{5}{c}{$a = $} & \multicolumn{5}{c}{$a = $} \\
\cline{3-7}\cline{8-12}
& & 1.000 & 0.950 & 0.900 & 0.850 & 0.800 & 1.000 & 0.950 & 0.900 & 0.850 & 0.800 \\
\hline
\multicolumn{12}{l}{Panel A: $T = 5$} \\
26 & 13 & 0.152 & 0.158 & 0.158 & 0.165 & 0.182 & 0.146 & 0.155 & 0.160 & 0.167 & 0.175 \\
39 & 13 & 0.151 & 0.159 & 0.163 & 0.172 & 0.191 & 0.155 & 0.157 & 0.166 & 0.180 & 0.187 \\
78 & 26 & 0.167 & 0.176 & 0.191 & 0.214 & 0.234 & 0.175 & 0.180 & 0.192 & 0.215 & 0.235 \\
390 & 130 & 0.202 & 0.218 & 0.288 & 0.353 & 0.435 & 0.190 & 0.220 & 0.276 & 0.333 & 0.411 \\
780 & 195 & 0.194 & 0.238 & 0.330 & 0.434 & 0.520 & 0.201 & 0.221 & 0.288 & 0.374 & 0.452 \\
1{,}560 & 390 & 0.203 & 0.277 & 0.418 & 0.539 & 0.631 & 0.205 & 0.239 & 0.309 & 0.414 & 0.516 \\
4{,}680 & 936 & 0.197 & 0.356 & 0.558 & 0.676 & 0.758 & 0.184 & 0.205 & 0.269 & 0.350 & 0.447 \\[6pt]
\multicolumn{12}{l}{Panel B: $T = 22$} \\
26 & 13 & 0.099 & 0.106 & 0.116 & 0.151 & 0.193 & 0.098 & 0.104 & 0.124 & 0.153 & 0.180 \\
39 & 13 & 0.104 & 0.107 & 0.126 & 0.165 & 0.204 & 0.108 & 0.112 & 0.132 & 0.154 & 0.193 \\
78 & 26 & 0.117 & 0.132 & 0.172 & 0.240 & 0.315 & 0.116 & 0.132 & 0.170 & 0.239 & 0.304 \\
390 & 130 & 0.128 & 0.205 & 0.384 & 0.519 & 0.633 & 0.129 & 0.189 & 0.351 & 0.489 & 0.598 \\
780 & 195 & 0.117 & 0.253 & 0.476 & 0.615 & 0.715 & 0.124 & 0.205 & 0.398 & 0.552 & 0.655 \\
1{,}560 & 390 & 0.118 & 0.347 & 0.599 & 0.724 & 0.793 & 0.121 & 0.238 & 0.462 & 0.609 & 0.707 \\
4{,}680 & 936 & 0.110 & 0.513 & 0.739 & 0.822 & 0.871 & 0.111 & 0.198 & 0.398 & 0.560 & 0.664 \\[6pt]
\multicolumn{12}{l}{Panel C: $T = 66$} \\
26 & 13 & 0.086 & 0.108 & 0.152 & 0.229 & 0.300 & 0.088 & 0.106 & 0.160 & 0.228 & 0.303 \\
39 & 13 & 0.089 & 0.116 & 0.166 & 0.238 & 0.330 & 0.093 & 0.107 & 0.167 & 0.240 & 0.330 \\
78 & 26 & 0.091 & 0.145 & 0.253 & 0.380 & 0.488 & 0.092 & 0.141 & 0.249 & 0.378 & 0.492 \\
390 & 130 & 0.094 & 0.308 & 0.558 & 0.692 & 0.764 & 0.106 & 0.272 & 0.525 & 0.668 & 0.749 \\
780 & 195 & 0.098 & 0.398 & 0.654 & 0.766 & 0.821 & 0.104 & 0.324 & 0.601 & 0.732 & 0.794 \\
1{,}560 & 390 & 0.100 & 0.533 & 0.750 & 0.834 & 0.871 & 0.099 & 0.374 & 0.656 & 0.770 & 0.826 \\
4{,}680 & 936 & 0.096 & 0.698 & 0.844 & 0.894 & 0.913 & 0.100 & 0.318 & 0.607 & 0.739 & 0.807 \\
\hline
\end{tabular*}
\begin{tabnotes}
\legend{We simulate a bivariate jump-diffusion model with diurnal variation in the correlation coefficient, such that
$\rho_{t} = \rho_{sc,t} k_{u,t}$, where $\rho_{sc,t}$ is a stochastic process and $k_{u,t} = a + bt$ with $b = 2(1-a)$ captures the deterministic component. The hypothesis $\mathcal{H}_{0}: \int_{0}^{1} ( k_{u,t} - 1)^{2} \,\mathrm{d}t = 0$ is tested against $\mathcal{H}_{a}: \int_{0}^{1} ( k_{u,t} - 1)^{2} \,\mathrm{d}t \neq0$. In the model, the null is equivalent to $a = 1$, whereas the alternative corresponds to $a \neq1$. The table reports rejection rates of the test statistic from \eqref{equation:test-statistic} at significance level $\alpha= 0.10$. $n$ is the number of intraday observations over a sample period of $T$ days, while $k_{n}$ is the number of log-price increments used to compute the blockwise realized covariance estimator.}
\end{tabnotes}%
\end{table}

\clearpage

\begin{table}
\tabcolsep=0pt

\caption{Rejection rate of the test statistic for diurnal variation in the
correlation process at the $\alpha= 0.05$ significance level and with $\rho= 0.60$.}
\label{table:sim-diurnal-correlation-t=N-r=0.60-q=0.05.tex}
\begin{tabular*}{\textwidth}{@{\extracolsep{\fill}}D{.}{.}{5.0}D{.}{.}{3.0}cccccccccc@{}}
\hline
\multicolumn{1}{l}{\multirow{3}{*}[-17pt]{$n$}}& \multicolumn{1}{c}{\multirow{3}{*}[-17pt]{$k_{n}$}} &
\multicolumn{5}{c}{Equidistant Sampling} & \multicolumn{5}{c}{Irregular Sampling}
\\
\cline{3-7}\cline{8-12}
& & \multicolumn{5}{c}{$a = $} & \multicolumn{5}{c}{$a = $} \\
\cline{3-7}\cline{8-12}
& & 1.000 & 0.950 & 0.900 & 0.850 & 0.800 & 1.000 & 0.950 & 0.900 & 0.850 & 0.800 \\
\hline
\multicolumn{12}{l}{Panel A: $T = 5$}\\
26 & 13 & 0.160 & 0.177 & 0.241 & 0.309 & 0.396 & 0.159 & 0.175 & 0.228 & 0.298 & 0.378 \\
39 & 13 & 0.143 & 0.165 & 0.228 & 0.310 & 0.411 & 0.143 & 0.164 & 0.227 & 0.303 & 0.392 \\
78 & 26 & 0.146 & 0.194 & 0.327 & 0.474 & 0.601 & 0.145 & 0.194 & 0.320 & 0.446 & 0.577 \\
390 & 130 & 0.149 & 0.404 & 0.715 & 0.867 & 0.938 & 0.145 & 0.328 & 0.633 & 0.819 & 0.911 \\
780 & 195 & 0.135 & 0.514 & 0.826 & 0.933 & 0.973 & 0.137 & 0.331 & 0.673 & 0.870 & 0.942 \\
1{,}560 & 390 & 0.134 & 0.672 & 0.921 & 0.977 & 0.992 & 0.140 & 0.327 & 0.692 & 0.891 & 0.960 \\
4{,}680 & 936 & 0.117 & 0.860 & 0.981 & 0.995 & 0.999 & 0.118 & 0.215 & 0.484 & 0.760 & 0.909 \\[6pt]
\multicolumn{12}{l}{Panel B: $T = 22$}\\
26 & 13 & 0.079 & 0.128 & 0.257 & 0.417 & 0.563 & 0.081 & 0.131 & 0.250 & 0.421 & 0.566 \\
39 & 13 & 0.078 & 0.125 & 0.267 & 0.447 & 0.621 & 0.072 & 0.122 & 0.261 & 0.444 & 0.612 \\
78 & 26 & 0.067 & 0.219 & 0.520 & 0.751 & 0.880 & 0.071 & 0.215 & 0.510 & 0.735 & 0.874 \\
390 & 130 & 0.077 & 0.690 & 0.935 & 0.983 & 0.995 & 0.076 & 0.601 & 0.910 & 0.975 & 0.992 \\
780 & 195 & 0.070 & 0.818 & 0.976 & 0.994 & 0.997 & 0.069 & 0.657 & 0.947 & 0.987 & 0.996 \\
1{,}560 & 390 & 0.068 & 0.923 & 0.993 & 0.998 & 0.999 & 0.070 & 0.679 & 0.963 & 0.993 & 0.997 \\
4{,}680 & 936 & 0.064 & 0.981 & 0.998 & 0.999 & 0.999 & 0.067 & 0.450 & 0.923 & 0.987 & 0.997 \\[6pt]
\multicolumn{12}{l}{Panel C: $T = 66$}\\
26 & 13 & 0.063 & 0.174 & 0.418 & 0.687 & 0.840 & 0.064 & 0.167 & 0.434 & 0.679 & 0.842 \\
39 & 13 & 0.057 & 0.164 & 0.446 & 0.737 & 0.884 & 0.055 & 0.154 & 0.460 & 0.727 & 0.881 \\
78 & 26 & 0.055 & 0.404 & 0.812 & 0.946 & 0.984 & 0.057 & 0.384 & 0.816 & 0.947 & 0.983 \\
390 & 130 & 0.056 & 0.907 & 0.990 & 0.997 & 0.999 & 0.060 & 0.861 & 0.987 & 0.997 & 0.999 \\
780 & 195 & 0.053 & 0.961 & 0.997 & 0.999 & 1.000 & 0.058 & 0.914 & 0.993 & 0.999 & 0.999 \\
1{,}560 & 390 & 0.059 & 0.988 & 0.999 & 1.000 & 1.000 & 0.056 & 0.938 & 0.997 & 0.999 & 1.000 \\
4{,}680 & 936 & 0.054 & 0.998 & 1.000 & 1.000 & 1.000 & 0.050 & 0.864 & 0.995 & 0.999 & 0.999 \\
\hline
\end{tabular*}
\begin{tabnotes}
\legend{We simulate a bivariate jump-diffusion model with diurnal variation in the correlation coefficient, such that
$\rho_{t} = \rho_{sc,t} k_{u,t}$, where $\rho_{sc,t}$ is a stochastic process and $k_{u,t} = a + bt$ with $b = 2(1-a)$ captures the deterministic component. The hypothesis $\mathcal{H}_{0}: \int_{0}^{1} ( k_{u,t} - 1)^{2} \,\mathrm{d}t = 0$ is tested against $\mathcal{H}_{a}: \int_{0}^{1} ( k_{u,t} - 1)^{2} \,\mathrm{d}t \neq0$. In the model, the null is equivalent to $a = 1$, whereas the alternative corresponds to $a \neq1$. The table reports rejection rates of the test statistic from \eqref{equation:test-statistic} at significance level $\alpha= 0.05$. $n$ is the number of intraday observations over a sample period of $T$ days, while $k_{n}$ is the number of log-price increments used to compute the blockwise realized covariance estimator.}
\end{tabnotes}%
\end{table}

\clearpage

\begin{table}
\tabcolsep=0pt

\caption{Rejection rate of the test statistic for diurnal variation in the
correlation process at the $\alpha= 0.05$ significance level and with $\rho= 0.40$.}
\label{table:sim-diurnal-correlation-t=N-r=0.40-q=0.05.tex}
\begin{tabular*}{\textwidth}{@{\extracolsep{\fill}}D{.}{.}{5.0}D{.}{.}{3.0}cccccccccc@{}}
\hline
\multicolumn{1}{l}{\multirow{3}{*}[-17pt]{$n$}}& \multicolumn{1}{c}{\multirow{3}{*}[-17pt]{$k_{n}$}} &
\multicolumn{5}{c}{Equidistant Sampling} & \multicolumn{5}{c}{Irregular Sampling}
\\
\cline{3-7}\cline{8-12}
& & \multicolumn{5}{c}{$a = $} & \multicolumn{5}{c}{$a = $} \\
\cline{3-7}\cline{8-12}
& & 1.000 & 0.950 & 0.900 & 0.850 & 0.800 & 1.000 & 0.950 & 0.900 & 0.850 & 0.800 \\
\hline
\multicolumn{12}{l}{Panel A: $T = 5$}\\
26 & 13 & 0.145 & 0.147 & 0.169 & 0.185 & 0.225 & 0.141 & 0.145 & 0.158 & 0.186 & 0.220 \\
39 & 13 & 0.133 & 0.135 & 0.156 & 0.185 & 0.225 & 0.127 & 0.136 & 0.159 & 0.183 & 0.217 \\
78 & 26 & 0.139 & 0.153 & 0.190 & 0.252 & 0.318 & 0.137 & 0.154 & 0.190 & 0.246 & 0.307 \\
390 & 130 & 0.146 & 0.214 & 0.382 & 0.552 & 0.677 & 0.141 & 0.202 & 0.346 & 0.488 & 0.632 \\
780 & 195 & 0.128 & 0.247 & 0.482 & 0.659 & 0.775 & 0.131 & 0.199 & 0.373 & 0.559 & 0.696 \\
1{,}560 & 390 & 0.133 & 0.339 & 0.633 & 0.794 & 0.880 & 0.139 & 0.214 & 0.405 & 0.615 & 0.749 \\
4{,}680 & 936 & 0.116 & 0.519 & 0.809 & 0.912 & 0.948 & 0.115 & 0.160 & 0.288 & 0.477 & 0.657 \\[6pt]
\multicolumn{12}{l}{Panel B: $T = 22$}\\
26 & 13 & 0.071 & 0.092 & 0.139 & 0.215 & 0.305 & 0.075 & 0.094 & 0.131 & 0.216 & 0.287 \\
39 & 13 & 0.072 & 0.085 & 0.141 & 0.228 & 0.324 & 0.074 & 0.089 & 0.139 & 0.218 & 0.315 \\
78 & 26 & 0.071 & 0.118 & 0.241 & 0.397 & 0.557 & 0.077 & 0.121 & 0.232 & 0.398 & 0.545 \\
390 & 130 & 0.073 & 0.322 & 0.661 & 0.818 & 0.895 & 0.075 & 0.277 & 0.602 & 0.781 & 0.873 \\
780 & 195 & 0.069 & 0.431 & 0.775 & 0.892 & 0.939 & 0.068 & 0.316 & 0.681 & 0.839 & 0.915 \\
1{,}560 & 390 & 0.070 & 0.605 & 0.879 & 0.943 & 0.969 & 0.069 & 0.358 & 0.744 & 0.878 & 0.938 \\
4{,}680 & 936 & 0.062 & 0.814 & 0.949 & 0.972 & 0.983 & 0.063 & 0.240 & 0.646 & 0.838 & 0.916 \\[6pt]
\multicolumn{12}{l}{Panel C: $T = 66$}\\
26 & 13 & 0.056 & 0.097 & 0.218 & 0.377 & 0.537 & 0.059 & 0.100 & 0.228 & 0.381 & 0.536 \\
39 & 13 & 0.054 & 0.102 & 0.231 & 0.410 & 0.584 & 0.058 & 0.093 & 0.230 & 0.422 & 0.583 \\
78 & 26 & 0.051 & 0.178 & 0.458 & 0.681 & 0.808 & 0.052 & 0.168 & 0.446 & 0.676 & 0.808 \\
390 & 130 & 0.052 & 0.569 & 0.863 & 0.939 & 0.966 & 0.060 & 0.495 & 0.839 & 0.928 & 0.962 \\
780 & 195 & 0.053 & 0.709 & 0.924 & 0.963 & 0.981 & 0.060 & 0.580 & 0.894 & 0.953 & 0.972 \\
1{,}560 & 390 & 0.057 & 0.845 & 0.960 & 0.980 & 0.989 & 0.055 & 0.653 & 0.923 & 0.964 & 0.979 \\
4{,}680 & 936 & 0.054 & 0.938 & 0.982 & 0.988 & 0.993 & 0.052 & 0.541 & 0.891 & 0.957 & 0.974 \\
\hline
\end{tabular*}
\begin{tabnotes}
\legend{We simulate a bivariate jump-diffusion model with diurnal variation in the correlation coefficient, such that
$\rho_{t} = \rho_{sc,t} k_{u,t}$, where $\rho_{sc,t}$ is a stochastic process and $k_{u,t} = a + bt$ with $b = 2(1-a)$ captures the deterministic component. The hypothesis $\mathcal{H}_{0}: \int_{0}^{1} ( k_{u,t} - 1)^{2} \,\mathrm{d}t = 0$ is tested against $\mathcal{H}_{a}: \int_{0}^{1} ( k_{u,t} - 1)^{2} \,\mathrm{d}t \neq0$. In the model, the null is equivalent to $a = 1$, whereas the alternative corresponds to $a \neq1$. The table reports rejection rates of the test statistic from \eqref{equation:test-statistic} at significance level $\alpha= 0.05$. $n$ is the number of intraday observations over a sample period of $T$ days, while $k_{n}$ is the number of log-price increments used to compute the blockwise realized covariance estimator.}
\end{tabnotes}%
\end{table}

\clearpage

\begin{table}
\tabcolsep=0pt

\caption{Rejection rate of the test statistic for diurnal variation in the
correlation process at the $\alpha= 0.05$ significance level and with $\rho= 0.20$.}
\label{table:sim-diurnal-correlation-t=N-r=0.20-q=0.05.tex}
\begin{tabular*}{\textwidth}{@{\extracolsep{\fill}}D{.}{.}{5.0}D{.}{.}{3.0}cccccccccc@{}}
\hline
\multicolumn{1}{l}{\multirow{3}{*}[-17pt]{$n$}}& \multicolumn{1}{c}{\multirow{3}{*}[-17pt]{$k_{n}$}} &
\multicolumn{5}{c}{Equidistant Sampling} & \multicolumn{5}{c}{Irregular Sampling}
\\
\cline{3-7}\cline{8-12}
& & \multicolumn{5}{c}{$a = $} & \multicolumn{5}{c}{$a = $} \\
\cline{3-7}\cline{8-12}
& & 1.000 & 0.950 & 0.900 & 0.850 & 0.800 & 1.000 & 0.950 & 0.900 & 0.850 & 0.800 \\
\hline
\multicolumn{12}{l}{Panel A: $T = 5$}\\
26 & 13 & 0.102 & 0.107 & 0.104 & 0.112 & 0.124 & 0.098 & 0.104 & 0.109 & 0.113 & 0.121 \\
39 & 13 & 0.090 & 0.094 & 0.101 & 0.107 & 0.121 & 0.094 & 0.091 & 0.103 & 0.111 & 0.116 \\
78 & 26 & 0.103 & 0.110 & 0.124 & 0.139 & 0.160 & 0.108 & 0.110 & 0.122 & 0.140 & 0.156 \\
390 & 130 & 0.131 & 0.144 & 0.197 & 0.265 & 0.346 & 0.123 & 0.144 & 0.190 & 0.241 & 0.317 \\
780 & 195 & 0.113 & 0.147 & 0.231 & 0.333 & 0.429 & 0.119 & 0.132 & 0.190 & 0.275 & 0.353 \\
1{,}560 & 390 & 0.121 & 0.177 & 0.320 & 0.448 & 0.552 & 0.122 & 0.146 & 0.212 & 0.306 & 0.414 \\
4{,}680 & 936 & 0.112 & 0.246 & 0.463 & 0.596 & 0.700 & 0.104 & 0.115 & 0.164 & 0.244 & 0.334 \\[6pt]
\multicolumn{12}{l}{Panel B: $T = 22$}\\
26 & 13 & 0.053 & 0.055 & 0.067 & 0.089 & 0.125 & 0.053 & 0.058 & 0.071 & 0.093 & 0.116 \\
39 & 13 & 0.052 & 0.056 & 0.072 & 0.096 & 0.129 & 0.055 & 0.057 & 0.074 & 0.090 & 0.117 \\
78 & 26 & 0.062 & 0.071 & 0.102 & 0.166 & 0.229 & 0.062 & 0.071 & 0.101 & 0.161 & 0.219 \\
390 & 130 & 0.068 & 0.132 & 0.297 & 0.440 & 0.562 & 0.068 & 0.116 & 0.260 & 0.408 & 0.518 \\
780 & 195 & 0.057 & 0.165 & 0.389 & 0.545 & 0.662 & 0.062 & 0.124 & 0.302 & 0.472 & 0.590 \\
1{,}560 & 390 & 0.060 & 0.255 & 0.525 & 0.669 & 0.754 & 0.063 & 0.150 & 0.373 & 0.538 & 0.649 \\
4{,}680 & 936 & 0.050 & 0.417 & 0.691 & 0.787 & 0.846 & 0.056 & 0.115 & 0.296 & 0.471 & 0.599 \\[6pt]
\multicolumn{12}{l}{Panel C: $T = 66$}\\
26 & 13 & 0.042 & 0.057 & 0.093 & 0.154 & 0.220 & 0.045 & 0.055 & 0.095 & 0.151 & 0.224 \\
39 & 13 & 0.044 & 0.061 & 0.097 & 0.157 & 0.244 & 0.046 & 0.056 & 0.101 & 0.165 & 0.240 \\
78 & 26 & 0.045 & 0.081 & 0.171 & 0.295 & 0.411 & 0.047 & 0.080 & 0.167 & 0.294 & 0.410 \\
390 & 130 & 0.049 & 0.221 & 0.482 & 0.634 & 0.724 & 0.053 & 0.190 & 0.445 & 0.607 & 0.705 \\
780 & 195 & 0.048 & 0.313 & 0.592 & 0.724 & 0.790 & 0.049 & 0.237 & 0.526 & 0.683 & 0.761 \\
1{,}560 & 390 & 0.048 & 0.450 & 0.707 & 0.806 & 0.850 & 0.046 & 0.287 & 0.593 & 0.729 & 0.796 \\
4{,}680 & 936 & 0.048 & 0.643 & 0.820 & 0.879 & 0.901 & 0.045 & 0.220 & 0.533 & 0.694 & 0.772 \\
\hline
\end{tabular*}
\begin{tabnotes}
\legend{We simulate a bivariate jump-diffusion model with diurnal variation in the correlation coefficient, such that
$\rho_{t} = \rho_{sc,t} k_{u,t}$, where $\rho_{sc,t}$ is a stochastic process and $k_{u,t} = a + bt$ with $b = 2(1-a)$ captures the deterministic component. The hypothesis $\mathcal{H}_{0}: \int_{0}^{1} ( k_{u,t} - 1)^{2} \,\mathrm{d}t = 0$ is tested against $\mathcal{H}_{a}: \int_{0}^{1} ( k_{u,t} - 1)^{2} \,\mathrm{d}t \neq0$. In the model, the null is equivalent to $a = 1$, whereas the alternative corresponds to $a \neq1$. The table reports rejection rates of the test statistic from \eqref{equation:test-statistic} at significance level $\alpha= 0.05$. $n$ is the number of intraday observations over a sample period of $T$ days, while $k_{n}$ is the number of log-price increments used to compute the blockwise realized covariance estimator.}\vspace*{-6pt}
\end{tabnotes}%
\end{table}
\end{appendices}

\clearpage

\renewcommand{\baselinestretch}{1.0} {\small
\bibliographystyle{rfs}
\bibliography{userref}}

\end{document}